\documentclass[12pt, nonatbib]{elsarticle}

\usepackage{graphicx}
\usepackage{subcaption}
\usepackage{amssymb}
\usepackage{float}
\usepackage{booktabs}
\usepackage{graphicx}
\usepackage[table,xcdraw]{xcolor}
\usepackage{amsthm}
\usepackage{amsmath}
\usepackage{dirtytalk}
\usepackage{hyperref}
\hypersetup{
    colorlinks=true,
    linkcolor=blue,
    filecolor=magenta,      
    urlcolor=cyan,
    pdftitle={Overleaf Example},
    pdfpagemode=FullScreen,
    }
\usepackage[left=2.5cm, right=2.5cm, top=2.5cm, bottom=2.5cm, footskip=1cm]{geometry}

\makeatletter
\let\c@author\relax
\makeatother
\usepackage[style=apa]{biblatex}
\begin{document}

\begin{frontmatter}

\title{Migration-Driven Demographic Changes:\\effects on local communities in the canton of Fribourg.\footnote{This project was partially funded by the Conseil d'État de Fribourg under a mandate to the University of Fribourg to study the influence of Fribourg's demographics on future public policies, titled: \textit{\say{Etude démographique et les conséquences sur la politique publique cantonale}}.}}
\date{May 7, 2026}

\author[1]{Emma Bacci}

\address[1]{Department of Economics, University of Fribourg \& Unidistance Suisse}

\begin{abstract}
Migration is reshaping demographic landscapes across Europe, raising urgent questions about how to adapt to rapid population changes. This study examines the case of the canton of Fribourg, Switzerland, which experienced a 30\% population increase over the past 15 years, primarily driven by international and internal migration.
As local governments are often the first to face pressures from demographic shifts, particularly in housing, education, and social services, understanding the causal effects of migration is essential for evidence-based policymaking.
We study how migration reshapes local demographic, educational, and housing outcomes across 112 municipalities in the canton of Fribourg (2010-2021). Using the intertemporal difference-in-differences estimator of De Chaisemartin and D'Haultfoeuille (2024), which accommodates staggered timing and cumulative, non-binary treatment, we identify the effect of a one-percentage-point increase in cumulative migration balance (relative to baseline population). Results show that migration exposure generates modest but persistent adjustments across demographic, educational, and housing dimensions. International and internal migration both reduce the share of elderly residents, and international inflows are associated with higher birth counts. Internal migration increases the number of resident students and alters the composition of compulsory and secondary-school cohorts, while international migration slightly reduces the tertiary-education share. Housing adjustments are gradual and concentrated in household composition and selected dwelling types, with international migration increasing mid-sized households and internal migration reducing mixed-use dwellings. Although yearly effects are small, their persistence over time leads to meaningful cumulative changes. Overall, migration acts as a counterweight to population aging and generates incremental adjustments in service demand, underscoring the importance of incorporating migration exposure into cantonal and municipal planning.

\end{abstract}

\end{frontmatter}

\section{Introduction}

Across Europe, migration has for many years been a central topic as international mobility expands and the free movement of people intensifies. Policymakers face growing pressure to understand not only how migration affects economies, but also how it reshapes the social and demographic fabric of local communities. As migration continues to rise, identifying which aspects of population change are most affected, and how, becomes essential for anticipating future needs and allocating public resources effectively at both national and local levels.

This paper contributes to the literature by providing causal evidence on how migration reshapes local demographic, educational, and housing outcomes at the municipal level. Focusing on the canton of Fribourg, Switzerland, we estimate the effects of incremental changes in cumulative migration exposure on key community-level outcomes using panel data for municipalities between 2010 and 2021. Methodologically, we apply a modern intertemporal difference-in-differences estimator that accommodates staggered timing and non-binary, cumulative treatments. This analysis moves beyond descriptive accounts of population movements by quantifying how migration alters age structure, household composition, and student populations in a highly decentralized institutional setting.

At the fiscal level, research on migration's impact on public finances shows mixed results. Young, skilled, and highly educated migrants often make a positive net contribution to public budgets (\textcite{boffi_decomposing_2024}; \textcite{suari-andreu_intra-eu_2023}), while lower-skilled migrants may generate short-term pressure through increased demand for welfare and basic services (\textcite{razin_tax_1998}; \textcite{rowthorn_fiscal_2008}; \textcite{jofre-monseny_immigration_2016}). Comparable dynamics appear in local contexts: evidence from Germany, Finland, and Italy shows that migration tends to raise local expenditures, particularly in education, childcare, and social assistance, as service demand expands with new arrivals (\textcite{viren_fiscal_2022}; \textcite{makela_migration_2018}; \textcite{maxand_local_2024}). At the same time, younger and healthier migrants can reduce per capita healthcare spending (\textcite{bettin_health_2020}). While these fiscal effects are important, they reflect deeper demographic and social processes set in motion by migration.

Changes in who lives in a municipality , by age, household type, or family status, translate directly into shifts in school enrolment, housing demand, and the use of local infrastructure. In other words, before migration alters budgets, it alters communities. Across Europe, towns and villages are experiencing visible demographic shifts as people move for work, study, or lifestyle reasons, and these transformations are most clearly felt at the local level. Research on studentification in British cities, for instance, shows how large student inflows reshaped neighbourhood age structures and household composition, prompting new local planning measures \parencite{sage_rapidity_2012}. Rural studies in England trace how selective in-migration rejuvenated ageing villages and altered housing markets \parencite{lewis_counter-urbanization_1991}, while Alpine villages in northern Italy have experienced population renewal through amenity migration, leading to integration and service adjustments \parencite{loffler_amenity_2016}.

In Switzerland and continental Europe, empirical work has largely focused on documenting migration patterns and demographic mechanisms rather than evaluating their causal effects on local outcomes. Large-scale studies such as \textcite{lerch_end_2023} for Switzerland and \textcite{ghio_age_2023} for Europe provide detailed mappings of internal and international migration, tracing processes of peri-urbanisation, return migration, and age-selective mobility. However, this literature generally stops short of assessing how these population movements affect concrete local realities such as school enrolment, household composition, or housing demand. As a result, demographic mechanisms are well described, but their tangible consequences at the municipal level remain less understood.

This gap is particularly relevant in Switzerland, where municipalities (\textit{communes}) are directly responsible for many of the services most affected by population change, including education, local infrastructure, housing, and land-use planning. Because these services depend on population composition rather than sheer population size, even modest migration inflows can significantly reshape local demand patterns. Switzerland's highly decentralized and heterogeneous municipal system, with more than 2,000 communes of varying size and growth trajectories, therefore provides a natural setting to study how migration translates into measurable demographic and social change.

The canton of Fribourg illustrates these dynamics clearly. Over the past fifteen years, its population has grown by roughly 30\%, driven primarily by international and internal migration (\cite{statistics}). Rather than analysing governmental reactions to these demographic shifts, this study focuses on estimating their causal effects on local outcomes. By combining detailed municipal panel data with an intertemporal difference-in-differences design, the paper provides new evidence on how cumulative migration exposure reshapes communities over time within the Swiss context.

The remainder of the paper is organized as follows. Section \ref{sec:context} presents the institutional context of the canton of Fribourg. Section \ref{data} describes the data and treatment definition. Section \ref{sec:identification} outlines the identification strategy and empirical framework. Section \ref{results} presents the results, and Section \ref{discussion} discusses their implications for local policy. Section \ref{conclusions} concludes.

\section{Institutional context: the canton of Fribourg}
\label{sec:context}
The canton of Fribourg has a moderately younger population profile than Switzerland as a whole. In recent years, individuals aged 0-19 have represented roughly 22 \% of the population in Fribourg, compared to about 20 \% at the national level, while the share of residents aged 65 and over has been around 17 \% in Fribourg versus approximately 19-20 \% in Switzerland. Consistently with these shares, the median age in Fribourg has been about 2-3 years lower than the Swiss average.\footnote{Source: Office fédéral de la statistique (OFS), Bilan démographique selon l'âge et le canton, 1981-2024, updated August 27 2025.} These differences indicate a relatively higher presence of younger cohorts, including students and young families. At the same time, Fribourg faces the same long-term demographic challenges as the rest of Switzerland, notably population aging and the gradual retirement of the baby-boom generation. These demographic dynamics exert increasing pressure on public revenues and expenditures, particularly in education and health, and have important implications for infrastructure, mobility, and housing.
Between 2007 and 2022, the population of the canton increased by 27\%, rising from 263,000 to 334,000 residents, marking the highest growth rate among Swiss cantons during this period. This consistent population growth, averaging about +4,700 residents per year, is driven mainly by both international immigration and inter-cantonal migration rather than natural increase, especially after the 2002 Agreement on the Free Movement of Persons with the EU.\footnote{Data sources: Office fédéral de la statistique, Compte de l'Etat, Service de la statistique Fribourg.}

This context is suitable to investigate the effect of migration given the relevance of migration increases: the analysis aims at finding causal effects of local factors, given the observed correlations. 

\begin{figure}[H]
    \centering
    \includegraphics[width=1\linewidth]{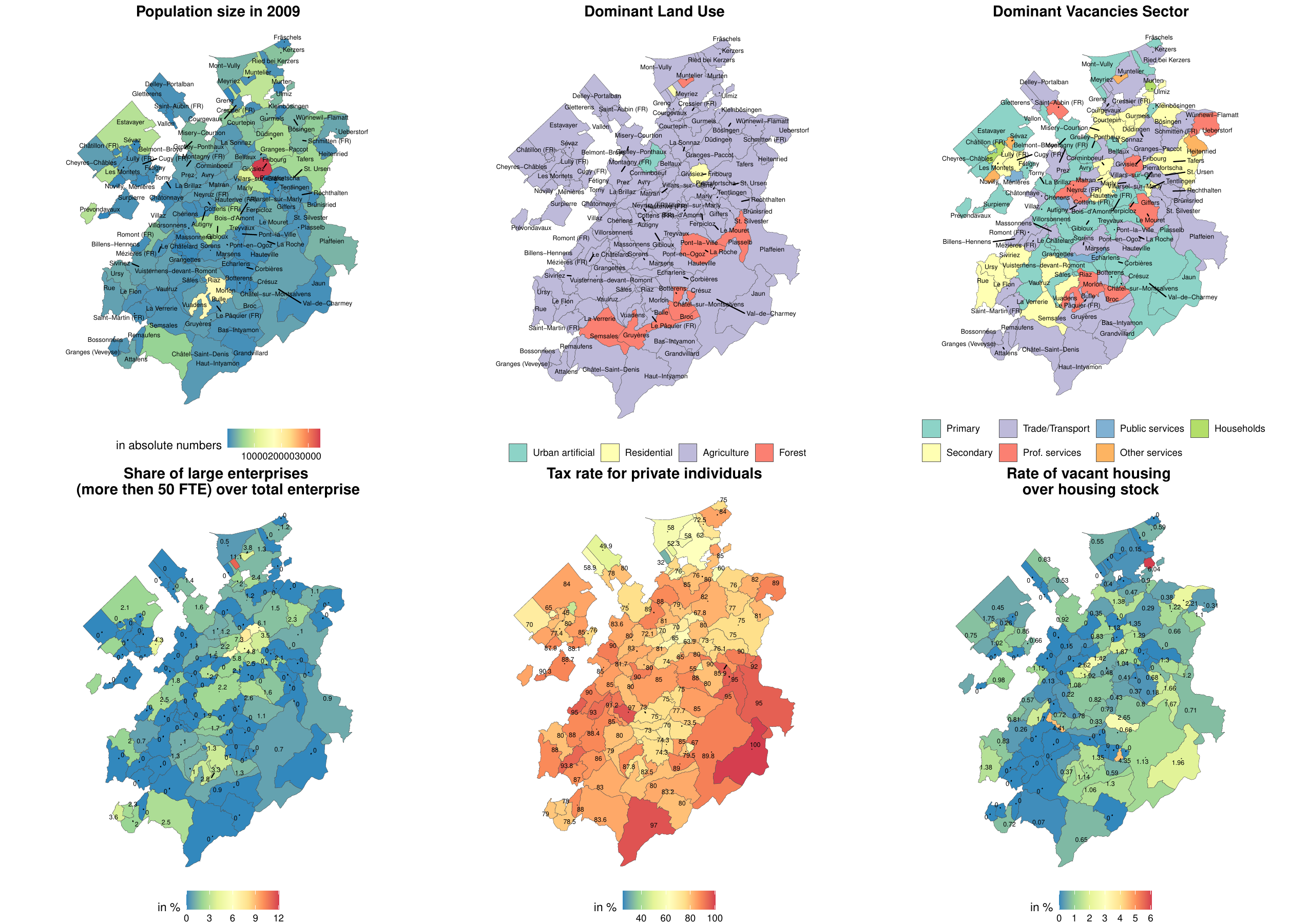}
    \caption{Status for Fribourg municipalities in 2009}
    \label{fig:covariates_map}
\end{figure}

Going into a little bit more detail, focusing on Fribourg's municipalities, we observe that internal geography and economy are heterogeneous. Figure~\ref{fig:covariates_map} provides an overview of this diversity, showing key municipal characteristics in 2009, before the main study period. Population is concentrated in the canton's main centres (Fribourg city, Villars-sur-Glâne,Düdinge, Bulle and Murten) while smaller rural communes dominate the pre-Alpine and southern regions. Land use patterns reveal an urban-residential corridor running along the Sarine valley and the Broye plain, contrasting with agricultural and forested landscapes elsewhere. The map of dominant employment sectors shows a service-oriented economy in urban areas, whereas primary and secondary activities remain prominent in rural communes.
The distribution of large enterprises (over 50 full-time equivalent employees) is similarly clustered around major centres. Fiscal conditions also vary: tax rates for private individuals differ considerably across communes, creating local incentives for residential mobility. Finally, the rate of housing vacancies, generally low across the canton, tends to be higher in peri-urban areas where housing development has expanded most rapidly.
Together, these maps depict a canton characterized by a majority of peri-urban and rural municipalities with markedly different demographic, fiscal, and economic profiles.

\section{Data and variables}
\label{data}
\subsection{Data sources and panel construction}
Our data is a panel composed by municipalities present in the canton of Fribourg observed across the years from 2010 to 2021. The data is all sourced from the portal \href{https://opendata.fr.ch/explore/?disjunctive.theme&disjunctive.keyword&sort=modified}{Opendata Fribourg}, curated by the Service de la statistique Fribourg. To ensure a balanced panel, we harmonize municipal boundaries to the post-merger 2021 map, aggregating pre-merger entities up to their 2021 successors. Moreover, we exclude very small municipalities because they could generate too extreme ratios. This yields 112 municipalities observed for 12 years (N=1,344 municipality-year observations).
\subsection{Treatment definition and descriptive patterns}
We define treatment as a \textbf{1 percentage point increase in the cumulative migration balance as a share of the baseline population}. 
The idea is to capture the effects of the exposure on the local communities of an increased number of migrants.We use cumulative migration as local outcomes (age structure, schooling, housing) respond to stocks of new residents rather than single-year flows; cumulative exposure maps those stock effects.  The migration balance is divided into two components:
\begin{itemize}
    \item {\textbf{Internal migration}}: net migration from other cantons.
    \item {\textbf{International migration}}: net migration from abroad.
\end{itemize}

For each municipality $i$ and year $t$, the treatment intensity is defined as:
\begin{equation}
    D_{i,t} = \frac{\sum_{j=\text{baseline year}}^{t} \text{MigrationBalance}_{i,j}}{\text{Population}_{i, \text{baseline}}}
\end{equation}
We discretize the continuous treatment variable $D_{i,t}$ into categories (or "bins") corresponding to 1 percentage point increases, which allows us to interpret treatment as a discrete change in exposure to migration.

Because of this binning, $D_{i,t}=0$ corresponds to cumulative migration between 0\% and 1\% of baseline population.  
This category therefore includes municipalities with either very small net inflows or outflows, or nearly balanced cumulative migration.  
Negative values of $D_{i,t}$ indicate net outflows (e.g. $D_{i,t}=-2$ means cumulative emigration equal to roughly 2\% of the baseline population), while positive values reflect net inflows of the same magnitude.

Tables \ref{tab:first_SC} and \ref{tab:first_FA} present the distribution of $D_{i,t}$ at the start of the observation period, in 2010, separately for internal and international migration.  
For internal migration, most municipalities (about 65\%) fall in the zero bin, implying little or balanced net migration relative to the baseline population. Around 17\% of municipalities show cumulative in-migration of approximately 1\%, and a few record higher inflows of 2-5\%. Only three municipalities record an initial outflow of migrants.
For international migration, nearly 91\% of municipalities exhibit cumulative balances close to zero, with about 8\% displaying small positive inflows and only one municipality showing a mild outflow.  
These patterns indicate that at baseline, cumulative migration exposure was generally limited in magnitude but not necessarily nonexistent.

\begin{table}[H]\centering
\def\sym#1{\ifmmode^{#1}\else\(^{#1}\)\fi}
\caption{Distribution of initial cumulative internal migration exposure by municipality in 2010. Each entry shows the share of municipalities whose cumulative migration balance (internal/international) equals the given percentage of baseline population. A value of 1 corresponds to net in-migration of 1\% of baseline population; negative values indicate net emigration.}
\begin{tabular}{l*{1}{cc}}
\hline\hline

                    \multicolumn{2}{c}{$D_{i,2010}$ for Internal migration}&          \\
                    &           $\#$ of municipalities&         pct\\
\hline
-2                  &           1&    .89\\
-1                  &           2&    1.78\\
0                   &          73&    65.18\\
1                   &          19&    16.96\\
2                   &          10&    8.93\\
3                   &           3&    2.68\\
4                   &           3&    2.68\\
5                   &           1&    .89\\  \hline 
Total               &        112&         100\\           
        
\hline\hline
\end{tabular}
\label{tab:first_SC}
\end{table}

\begin{table}[H]\centering
\def\sym#1{\ifmmode^{#1}\else\(^{#1}\)\fi}
\caption{Distribution of initial cumulative international migration exposure by municipality in 2010. Each entry shows the share of municipalities whose cumulative migration balance (internal/international) equals the given percentage of baseline population. A value of 1 corresponds to net in-migration of 1\% of baseline population; negative values indicate net emigration.}
\begin{tabular}{l*{1}{ccc}}
\hline\hline
                    \multicolumn{2}{c}{$D_{i,2010}$ for International migration}&          \\
&           $\#$ of municipalities&         pct\\
\hline
-1                  &           1&    .89\\
0                   &         102&    91.07\\
1                   &           8&    7.14\\
2                   &           1&    .89\\
\hline
Total               &        112&         100\\
\hline\hline
\end{tabular}
\label{tab:first_FA}
\end{table}
 
Changes in treatment over time are driven by annual migration flows that accumulate gradually.  
Tables \ref{tab:d_SC} and \ref{tab:d_FA} show that most municipality-year observations display little change from the previous year, typically around one percentage point, but for a few municipalities the differences can take slightly bigger positive and negative values: municipalities can experience both increases and decreases in their level of cumulative migration.  Tables \ref{tab:switches_SC} and \ref{tab:switches_FA} illustrate that many municipalities move across several bins during the sample period. This combination of small annual increments and multiple category changes implies that migration exposure evolves slowly but persistently over time.  
In practice, most municipalities experience year-to-year changes of no more than one percentage point, indicating that exposure increases through gradual accumulation rather than abrupt inflows.  
At the same time, the cumulative nature of $D_{i,t}$ means that even modest but sustained inflows eventually shift municipalities into higher treatment categories.  
As a result, municipalities traverse several bins over the sample period, reflecting meaningful long-term heterogeneity in cumulative migration exposure.

\begin{table}[H]\centering
\def\sym#1{\ifmmode^{#1}\else\(^{#1}\)\fi}
\caption{Descriptive Statistics for first differences of internal migration: measures the annual change in the net migration balance as a share of baseline population, expressed in percentage points.}
\begin{tabular}{l*{1}{ccc}}
\hline\hline
                    \multicolumn{2}{c}{$D_{i,t}$ -  $D_{i,t-1}$ for Internal migration}&              \\
&           $\#$ of municipalities $\times$ year&         pct\\
\hline
-5                  &           1&    .08\\
-4                  &           1&    .08\\
-2                  &          14&    1.14\\
-1                  &          91&    7.39\\
0                   &         583&    47.32\\
1                   &         269&    21.83\\
2                   &         146&    11.85\\
3                   &          76&    6.17\\
4                   &          26&     2.11\\
5                   &          16&    1.30\\
6                   &           4&    .32\\
7                   &           4&    .32\\
9                   &           1&    .08\\
Total               &        1232&         100\\
\hline\hline
\end{tabular}
\label{tab:d_SC}
\end{table}

\begin{table}[H]\centering
\def\sym#1{\ifmmode^{#1}\else\(^{#1}\)\fi}
\caption{Descriptive Statistics for first differences of international migration: measures the annual change in the net migration balance as a share of baseline population, expressed in percentage points.}
\begin{tabular}{l*{1}{ccc}}
\hline\hline
                    \multicolumn{2}{c}{$D_{i,t}$ -  $D_{i,t-1}$ for International migration}&              \\
&           $\#$ of municipalities $\times$ year&         pct\\
\hline
-2                  &           8&    .65\\
-1                  &         129&    10.47\\
0                   &         660&    53.57\\
1                   &         353&     28.65\\
2                   &          71&    5.76\\
3                   &           6&     .49\\
4                   &           3&    .24\\
5                   &           1&    .08\\
8                   &           1&    .08\\
Total               &        1232&         100\\

\hline\hline
\end{tabular}
\label{tab:d_FA}
\end{table}

\begin{table}[H]\centering
\def\sym#1{\ifmmode^{#1}\else\(^{#1}\)\fi}
\caption{Number of times a municipality experiences a change in internal migration over the duration of the panel (2010-2021).}
\begin{tabular}{l*{1}{ccc}}
\hline\hline
                    \multicolumn{2}{c}{Number of Switches (Internal migration)}&              \\
&           $\#$ of municipalities  $\times$ year&         pct\\
\hline
0                   &           4&    3.57\\
1                   &           6&    5.36\\
2                   &          15&    13.39\\
3                   &           6&    5.36\\
4                   &          12&    10.71\\
5                   &          12&    10.71\\
6                   &           9&    8.03\\
7                   &           9&    8.03\\
8                   &           9&    8.03\\
9                   &          11&    9.82\\
10                  &          10&    8.93\\
11                  &           9&    8.03\\
\hline
Total               &        112&         100\\
\hline\hline
\end{tabular}
\label{tab:switches_SC}
\end{table}

\begin{table}[H]\centering
\def\sym#1{\ifmmode^{#1}\else\(^{#1}\)\fi}
\caption{Number of times a municipality experiences a change in international migration over the duration of the panel(2010-2021).}
\begin{tabular}{l*{1}{ccc}}
\hline\hline
                    \multicolumn{2}{c}{Number of Switches (International migration)}&              \\
&           $\#$ of municipalities $\times$ year&         pct\\
\hline
0                   &           2&    1.78\\
1                   &           2&    1.78\\
2                   &           8&    7.14\\
3                   &          15&    13.39\\
4                   &          20&    17.86\\
5                   &          21&       18.75\\
6                   &          13&    11.61\\
7                   &          11&    9.82\\
8                   &          15&    13.39\\
9                   &           2&    1.78\\
10                  &           2&    1.78\\
11                  &           1&    .89\\
\hline
Total               &        112&         100\\
\hline\hline
\end{tabular}
\label{tab:switches_FA}
\end{table}

\subsection{Outcomes}
As mentioned in the introduction, this study focuses on a set of outcomes selected ex ante based on their direct relevance for the demographic and social composition of local communities in the canton of Fribourg. 
We would like to explore how internal and international migration differentially affect the development of the population age pyramid, the composition of the student population, the composition of  household sizes and types of housing and the birth rate over total population in each municipality.  These outcomes define the full set of variables considered in the empirical analysis and are motivated by their importance for municipal planning and service provision. Some descriptive statistics for those selected outcomes in the period before treatment can be found in Table \ref{tab:outcomes}.
The table shows substantial heterogeneity across municipalities. 
On average, youth (below 20) represent 25\% of the population, while older residents (65+) account for 17\%.
The wide variation in logged student counts (mean 5.69, SD 0.87) may capture the mix of small rural communes and larger urban centers with more secondary and tertiary institutions. 
Household composition varies widely, with single-person households ranging from 18 to 47\% of all households.
Housing characteristics display a clear urban-rural divide: small family apartments and mixed-use dwellings are concentrated in urban communes, whereas detached houses dominate rural ones.

\begin{table}[H]\centering
\def\sym#1{\ifmmode^{#1}\else\(^{#1}\)\fi}
\caption{Descriptive Statistics for Selected outcomes}
\label{tab:outcomes}
\resizebox{\linewidth}{!}{
\begin{tabular}{l*{1}{ccccc}}
\hline\hline
                    &\multicolumn{5}{c}{}                                            \\
                    &        mean&          sd&         min&         max&       count\\
\hline
Below 20-year-olds relative to total population&    .2531289&    .0288882&    .1769911&    .3145009&         112\\
Population aged 20 to 64 relative to total population&    .6118107&    .0243126&         .54&    .6930233&         112\\
Population aged above 64 relative to total population&    .1694148&    .0316934&    .1004601&    .2685512&         112\\
Total students resident in the municipality (logged)  &    5.688087&    .8660595&    4.043051&    8.547723&         112\\
Proportion of students in compulsory school      &    .4960943&    .0572121&    .3216374&    .6358024&         112\\
Proportion of students in vocational  secondary education             &    .1743908&    .0444228&    .0666667&    .3171806&         112\\
Proportion of students in general secondary education            &    .0873708&    .0362787&    .0049751&    .1929825&         110\\
Proportion of students in  tertiary education         &    .0173999&    .0088232&    .0038911&    .0512821&         107\\
Proportion of 1-people households relative to total households&    .2538227&     .046781&    .1787709&    .4658605&         112\\
Proportion of 2-people households relative to total households&    .3166137&    .0354179&     .231405&    .4224423&         112\\
Proportion of 3-people households relative to total households&    .1526579&    .0210453&    .1011673&    .2027778&         112\\
Proportion of above 4-people households relative to total households&    .2769057&    .0455584&    .1470013&    .3913043&         112\\
Proportion of 1/2 bedroom apartments over total housing        &    .0840987&    .0306507&     .015544&    .1788321&         112\\
Proportion of small family apartments over total housing      &    .2482523&    .0773483&    .1098266&    .4882353&         112\\
Proportion of large family apartments over total housing       &    .3102221&    .0705014&      .07496&    .5150376&         112\\
Proportion of single-detached dwellings over total housing    &    .1663934&    .0821986&    .0091213&    .5090909&         112\\
Proportion of mixed-use dwellings over total housing          &    .1910335&    .0773393&    .0578035&    .4863014&         112\\
Proportion of births relative to total population             &    .0112345&    .0033391&     .003012&    .0248447&         112\\
Total number of births in the municipality (logged)            &    2.797866&    .9177132&           0&    6.016157&         112\\
\hline
Observations        &         112&            &            &            &            \\
\hline\hline
\end{tabular}
}
\end{table}

\section{Identification and Empirical Framework}
\label{sec:identification}

Our empirical strategy exploits variation in the timing and intensity of migration inflows across municipalities in the canton of Fribourg between 2010 and 2021. We estimate the causal effects of cumulative migration exposure using the \textbf{intertemporal Difference-in-Differences (DiD)} estimator developed by \textcite{de_chaisemartin_difference--differences_2024}. This estimator accommodates staggered treatment timing, non-binary intensity, and heterogeneous effects across units and over time. 

The estimator compares municipalities that experience an increase in treatment ("switchers-in")\footnote{The method originally would allow to explore the comparison also among "switchers-out" or a mix of the two. We decide to focus on "switchers-in" for two main reasons: interpretability of the results and data availability. As mostmunicipalities experience an increase in treatment we have not enough power to estimate the isolated effects of experiencing a decrease in treatment, and estimating both effects together leads to a bigger complication in the interpretation of the results.} with municipalities that have not yet experienced such an increase, conditional on having the same treatment status in the first period (baseline). Identification is based on within-baseline-treatment comparisons between municipalities that differ only in the timing of their first treatment increase. As a result, the parallel-trends assumption is restricted to municipalities that shared the same baseline exposure level but experience treatment changes at different points in time.

\subsection*{Notation and main assumptions}

We follow the notation of \textcite{de_chaisemartin_difference--differences_2024}. The panel consists of a set of municipalities, indexed by $i$, observed over time periods indexed by $t = 1,\dots,T$. For the purpose of the intertemporal DiD design, municipalities are grouped into sets indexed by $g$, where a group $g$ consists of municipalities that share the same baseline treatment level $D_{i,1}$ and belong to the same pre-treatment cluster. Outcomes and treatments are therefore defined at the group level.

Let $Y_{g,t}$ denote the average outcome of municipalities in group $g$ at time $t$, and let $D_{g,t}$ denote the group-level treatment intensity at time $t$. Outcomes may depend not only on the contemporaneous treatment $D_{g,t}$, but also on the history of past treatments.

Let $Y_{g,t}(d_1,\dots,d_t)$ denote the potential outcome of group $g$ at time $t$ if its treatment path from period $1$ to $t$ were equal to $(d_1,\dots,d_t)$. Let $F_g$ denote the first period in which group $g$'s treatment changes relative to its baseline level $D_{g,1}$. We refer to $Y_{g,t}(D_{g,1},\dots,D_{g,1})$ as the \emph{status quo} potential outcome, corresponding to a counterfactual path in which group $g$'s treatment remains fixed at its period-one level in all periods.

We focus on groups that experience at least one increase in treatment relative to their baseline level over the sample period (``switchers-in''). Let $\mathcal{G}$ denote the set of such groups.  For a given group $g \in \mathcal{G}$ and relative time $\ell \geq 1$, the estimand $\delta_{g,\ell}$ compares the outcome evolution of group $g$ between periods $F_g - 1$ and $F_g - 1 + \ell$ to the outcome evolution of groups that share the same baseline treatment level $D_{g,1}$ and whose treatment has not yet changed by period $F_g - 1 + \ell$. These groups serve as not-yet-treated controls for group $g$ at relative time $\ell$.

We rely on the following key assumptions for identification:

\begin{enumerate}
    \item \textbf{No Anticipation}: Municipalities do not adjust their outcomes in anticipation of future treatment changes.  Formally, outcomes $Y_{g,t}$ remain unaffected by future increases in cumulative migration before the first change $F_g$. We could imagine that municipalities have limited advance information about future net migration flows and that migration decisions are largely individual and driven by employment, family, or housing opportunities rather than municipal policies decided years ahead.  Moreover, treatment changes accumulate gradually: annual increments are typically under one percentage point, too small to trigger anticipatory adjustments in education or housing within the same year. Additionally, the inclusion of within-canton migration as a covariate (see Section~\ref{sec:est_cov}) partially absorbs any local planning responses that might otherwise confound the estimate, providing further support for this assumption.
    
    \item \textbf{Parallel Trends for Status-Quo Potential Outcomes.}
For any two groups $g$ and $g'$ such that $D_{g,1} = D_{g',1}$, and for all periods $t \geq 2$,
\[
\mathbb{E}\!\left[
Y_{g,t}(D_{g,1},\dots,D_{g,1})
-
Y_{g,t-1}(D_{g,1},\dots,D_{g,1})
\right]
=
\mathbb{E}\!\left[
Y_{g',t}(D_{g',1},\dots,D_{g',1})
-
Y_{g',t-1}(D_{g',1},\dots,D_{g',1})
\right].
\]
That is, among groups that share the same period-one treatment level, the expected evolution of the status-quo potential outcome is the same.

    \item \textbf{Design Restrictions}:
    \begin{itemize}
        \item There exists at least one pair of municipalities $g$ and $g'$ such that they share the same initial treatment status ($D_{g,1}=D_{g',1}$) but experience a first treatment increase at different times ($F_g\neq F_{g'}$):
        \begin{align*}
            &\exists (g,g') \text{ such that } D_{g,1}=D_{g',1} \text{ and } F_g\neq F_{g'}.
        \end{align*}
        \item We do not require global monotonicity of treatment for all years. Instead, we impose a \textbf{one-sided window condition}: once a municipality's treatment increases above baseline, we focus only on periods $\ell$ where 
        \[
        D_{g,\tau} \geq D_{g,1} \quad \forall \tau \in [F_g, F_g+\ell],
        \]
        ensuring that the estimated contrast reflects exposure levels above baseline.\footnote{The one-sided window condition imposed in our design restricts attention to periods in which treatment exposure remains equal or above its baseline level. Importantly, this restriction rules out \emph{full reversals} of treatment, in the sense that, once a group's cumulative migration exposure increases above its period-one level, we exclude from the estimation sample any group-period cells in which exposure subsequently falls 
        below the baseline level. In contrast, \emph{partial reversals} (temporary decreases in treatment intensity that remain weakly above the baseline) are allowed by construction and do not violate the identifying assumptions of the intertemporal DiD estimator.
This distinction is important in a setting with non-binary and cumulative treatments. Full reversals would imply that a group has experienced both treatment levels strictly below and strictly above its baseline prior to a given period, in which case the group-level DiD contrasts may combine effects of increases and decreases in exposure. Following \textcite{de_chaisemartin_difference--differences_2024}, we therefore discard from the estimation sample all group-period cells for which such full reversals occur, yielding an unbalanced panel on which the one-sided window condition holds by construction. This ensures that the estimated effects can be interpreted as the impact of being exposed to a weakly higher treatment level for $\ell$ consecutive periods. Moreover, because treatment is defined as cumulative net migration exposure at the group level,  full reversals require sustained net outflows sufficient to offset earlier cumulative inflows, which are empirically rare in our setting.
We acknowledge that allowing partial reversals may introduce additional noise if treatment effects are highly non-linear in intensity, but this affects precision rather than identification. We refer to section \ref{sec:treatmentpaths} for an overview of the treatment paths.} The original design restriction would also include the groups g for which $ D_{g,\tau} \leq D_{g,1} \quad \forall \tau \in [F_g, F_g+\ell]$, but as argued above we will consider only the sample of ``switchers-in''. 
    \end{itemize}
\end{enumerate}

These assumptions ensure that estimated differences in outcomes can be attributed to changes in migration exposure rather than to other confounding trends.

\subsection{Estimands and research design}
\label{sec:estimation}

The initial parameter we would like to explore is the true treatment effect for group $g$ at relative time $\ell$:
\begin{equation}
  \delta_{g,\ell}  = 
    \mathbb{E}\!\left[
    Y_{g,F_g -1 + \ell} 
    - 
    \textcolor{blue}{
    Y_{g,F_g -1 + \ell}(D_{g,1}, \dots, D_{g,1})
    }
    \mid D
    \right],
\end{equation}
where the second term (in blue) is the unobserved counterfactual outcome that group $g$ would have experienced had migration exposure remained at its baseline level.

In the sample, we estimate these dynamic treatment effects by comparing how outcomes change for treated municipalities after their treatment time, relative to a control group of municipalities that have not yet received treatment but share the same baseline treatment status.

For a given group $g$ and relative time $\ell$, the estimator takes the form:
\begin{equation}
    \text{DID}_{g,\ell} = \left( Y_{g,F_g -1 + \ell} - Y_{g,F_g -1} \right) - 
    \frac{1}{N} \sum_{g'\in \mathcal{C}(g,\ell)} 
    \left( Y_{g',F_g -1 + \ell} - Y_{g',F_g -1} \right),
\end{equation}
where:
\begin{itemize}
    \item $F_g$ is the first year the treatment increases for municipalities in group $g$,
    \item $Y_{g,t}$ is the observed outcome for municipalities in group $g$ at time $t$,
    \item The second term averages over ``not-yet-increased'' controls $g'$ sharing the same baseline treatment level $D_{g',1}=D_{g,1}$.
\end{itemize}

It can be shown that:
\begin{equation}
      \mathbb{E}[\text{DID}_{g,\ell} \mid D] = \delta_{\ell,g}
\end{equation}

\subsubsection{Definition of Treatment Effects}

The overall effect at relative time $\ell$ is obtained by aggregating group-specific effects among switchers-in:
\begin{equation}
\label{eq:non_normalized}
    \delta_\ell = 
    \frac{1}{N_\ell} 
    \sum_{g: F_g -1 + \ell \leq T_g,\, D_{g,t}>D_{g,1}} 
    \delta_{g,\ell} \rightarrow DID_\ell = \frac{1}{N_\ell} DID_{g,\ell}
\end{equation}
This estimator focuses exclusively on municipalities that experience a discrete increase in cumulative migration exposure relative to their baseline and remain above that level for the event window under consideration. 

\paragraph{Normalized event-study effects}

Because treatment intensity varies across groups and over time, the magnitude of the group- and lag-specific effect $\delta_{g,\ell}$ reflects both the duration of exposure and the cumulative size of the treatment change experienced by group $g$ up to relative time $\ell$. To facilitate interpretation in terms of effects per unit of cumulative exposure, we normalize these effects by the total treatment increment received over the event window.

Formally, for each group $g$ and relative time $\ell \geq 1$, we define the cumulative treatment increment relative to the status quo as
\begin{equation*}
\Delta_{g,\ell}^D = \sum_{k=0}^{\ell-1} \left( D_{g,F_g+k} - D_{g,1} \right),
\end{equation*}
which captures the total deviation of group $g$'s treatment path from its baseline level between periods $F_g$ and $F_g - 1 + \ell$.

We then define the \emph{normalized} ($n$) treatment effect for group $g$ at relative time $\ell$ as
\begin{equation}
\delta_{g,\ell}^n = \frac{\delta_{g,\ell}}{\Delta_{g,\ell}^D}.
\end{equation}
The superscript $n$ denotes normalization by the cumulative treatment increment. The parameter $\delta_{g,\ell}^n$ can therefore be interpreted as the average effect on the outcome per unit of cumulative treatment exposure over the $\ell$-period window.

For every $(g,\ell)$ cell such that $1 \leq \ell \leq T_g - F_g + 1$, it can be shown that the normalized effect admits the representation
\begin{equation}
\delta_{g,\ell}^n = \sum_{k=0}^{\ell-1} w_{g,\ell,k} \, s_{g,\ell,k},
\end{equation}
where $s_{g,\ell,k}$ denotes the slope of the expected potential outcome function of group $g$ with respect to its $k$-th treatment lag at time $F_g - 1 + \ell$, and the weights $w_{g,\ell,k}$ are proportional to the absolute deviation of the $k$-th lagged treatment from its status-quo level. This representation highlights that $\delta_{g,\ell}^n$ aggregates contemporaneous and lagged treatment effects into a single, interpretable measure.

Finally, the overall normalized event-study effect at relative time $\ell$ is obtained by aggregating the group-specific normalized effects among switchers-in. Let
\[
\Delta_\ell^D = \frac{1}{N_\ell} \sum_{g: F_g - 1 + \ell \leq T_g} \left| \Delta_{g,\ell}^D \right|
\]
denote the average cumulative treatment increment at relative time $\ell$. The corresponding normalized estimator is then given by
\begin{equation}
\label{eq:normalized eff}
DID_\ell^n = \frac{DID_\ell}{\Delta_\ell^D},
\end{equation}
which estimates the average effect per unit of cumulative treatment exposure at horizon $\ell$.

\paragraph{Average Total treatment effects}
Lastly, one more parameter we could be interested in is the average total effect per unit of treatment: 
\begin{equation}
    \delta = \frac{\sum_{g:F_g \leq T_g} \sum_{\ell=1}^{T_g -F_g+1} \delta_{\ell,g}}{\sum_{g:F_g \leq T_g} \sum_{\ell=1}^{T_g -F_g+1} (D_{g, F_g -1 +\ell}- D_{g,1})}
\end{equation}

This parameter can be estimated with:
\begin{equation}
\label{eq: Av tot effects}
    \hat{\delta} = \frac{\sum_{g:F_g \leq T_g} \sum_{\ell=1}^{T_g -F_g+1} DID_{\ell,g}}{\sum_{g:F_g \leq T_g} \sum_{\ell=1}^{T_g -F_g+1} (D_{g, F_g -1 +\ell}- D_{g,1})}
\end{equation}

Notice that the horizon over which the total effects are cumulated are conditional on the design and can vary across groups and lags. So, the parameter  $\delta$ can be interpreted as an average total effect per unit of treatment, aggregating the contemporaneous and lagged effects of all treatment increments over the sample period. This object is closely related to what standard two-way fixed effects (TWFE) regressions with treatment intensity would implicitly target in settings with dynamic and cumulative treatments. However, as shown by \textcite{de_chaisemartin_difference--differences_2024}, TWFE estimators generally recover non-transparent weighted averages of dynamic treatment effects, potentially contaminated by heterogeneous timing and lag structures, and may even place negative weight on some effects. In contrast, their estimator recovers $\delta$ under explicit design restrictions, yielding a parameter with a clear cost-benefit interpretation.
\subsection{Clustering}
\label{sec:clustering}
To account for systematic baseline differences across municipalities that could 
correlate with both migration dynamics and subsequent local outcomes, we relax 
the standard parallel-trends assumption and allow untreated potential outcomes 
to evolve differently across sets of municipalities. We partition municipalities 
into a small number of groups defined by similarities in their initial 
migration-population profile, restricting identification to within-group 
comparisons only. This weakens the identifying assumption by requiring parallel 
trends only among municipalities that share the same historical baseline 
characteristics, rather than across all municipalities, thereby mitigating 
concerns that time-varying structural differences induce differential trends 
between treated and control units. We construct these groups using hierarchical 
agglomerative clustering \cite{Koch_2013} based on two pre-treatment 
characteristics: cumulative migration inflows up to 2009 (aggregated over 1991-2009) and population size.\footnote{In  \ref{sec: app_clustering} can be found a graph showing the distribution of municipalities in clusters over these two characteristics. Additionally, we tried to construct an index of ``Attractiveness'' of these municipalities based on some of their characteristics described in Section \ref{sec:context} and to use this index as a third dimension for clustering; results do not vary. This check can also be found in the Appendix.} After standardizing both variables, we compute pairwise Euclidean distances between municipalities and apply complete-linkage hierarchical  clustering, which iteratively merges municipalities into compact and well-separated clusters without requiring the number of groups to be specified  ex ante. Inspection of the resulting dendrogram reveals a clear break suggesting 
the presence of three meaningful clusters, which we retain as our grouping  structure. In the estimation, the effect for municipalities whose treatment  changes is identified by comparing their outcome evolution only to municipalities 
in the same cluster that share the same initial treatment level and have not yet  changed treatment.
\subsection{Estimation with covariate}
\label{sec:est_cov}

One last dimension we would like to control for is potential spillovers of our treatment via within canton migration. We can observe in our data for each municipality the level of net within-canton migration  every year, defined as the difference between immigration from all the other municipalities in the canton of Fribourg and the emigration towards any of the other municipalities in the canton of Fribourg. 

To include the covariate in our estimation we follow the residualization procedure proposed by \textcite{de_chaisemartin_difference--differences_2024}. 
The estimator remains conceptually identical to that without controls, except that the first-difference of the outcome is replaced by residuals from auxiliary regressions that partial out the effect of the covariate and time fixed effects.

Formally, for each baseline treatment level $D_{g,1}$, we estimate the following auxiliary regression in the subsample of ``not-yet-increased'' $(g,t)$ pairs:
\begin{equation}
    \Delta Y_{g,t} 
    = 
    \alpha_{D_{g,1}} 
    + 
    \boldsymbol{\beta}_{D_{g,1}}' \Delta \mathbf{X}_{g,t} 
    + 
    \lambda_t 
    + 
    \varepsilon_{g,t},
\end{equation}
where $\Delta Y_{g,t}$ denotes the first difference of the outcome, $\Delta \mathbf{X}_{g,t}$ the first differences of the control variables, and $\lambda_t$ year fixed effects. 
The residuals $\widehat{\varepsilon}_{g,t}$ replace $\Delta Y_{g,t}$ in the computation of the intertemporal DiD estimator.

This approach ensures that estimators with controls are unbiased even when municipalities follow distinct underlying trends, provided such trends can be explained by a linear model in covariate changes.

The resulting residualized outcome $\widehat{\varepsilon}_{g,t}$ captures changes in outcomes that are orthogonal to both temporal shocks and the evolution of the observed covariate, allowing the estimator to isolate the dynamic effects of migration exposure.

\subsection{Treatment paths}
\label{sec:treatmentpaths}
As mentioned above, our analysis focuses on \textbf{switchers-in}: municipalities whose cumulative migration exposure rises above their baseline bin at least once during the study period. 
Municipalities that have not yet increased serve as \emph{not-yet-increased controls} within the same baseline bin and calendar year. 
This one-sided design aligns with the estimator described in Section~\ref{sec:identification}, which isolates the effect of incremental increases in migration exposure relative to the baseline level.
This section visualizes the treatment variation.
The heatmaps illustrate how staggered and gradual treatment timing generates the quasi-experimental contrasts exploited by the intertemporal DiD.

\begin{figure}[H]
    \centering
    \includegraphics[width=0.95\linewidth]{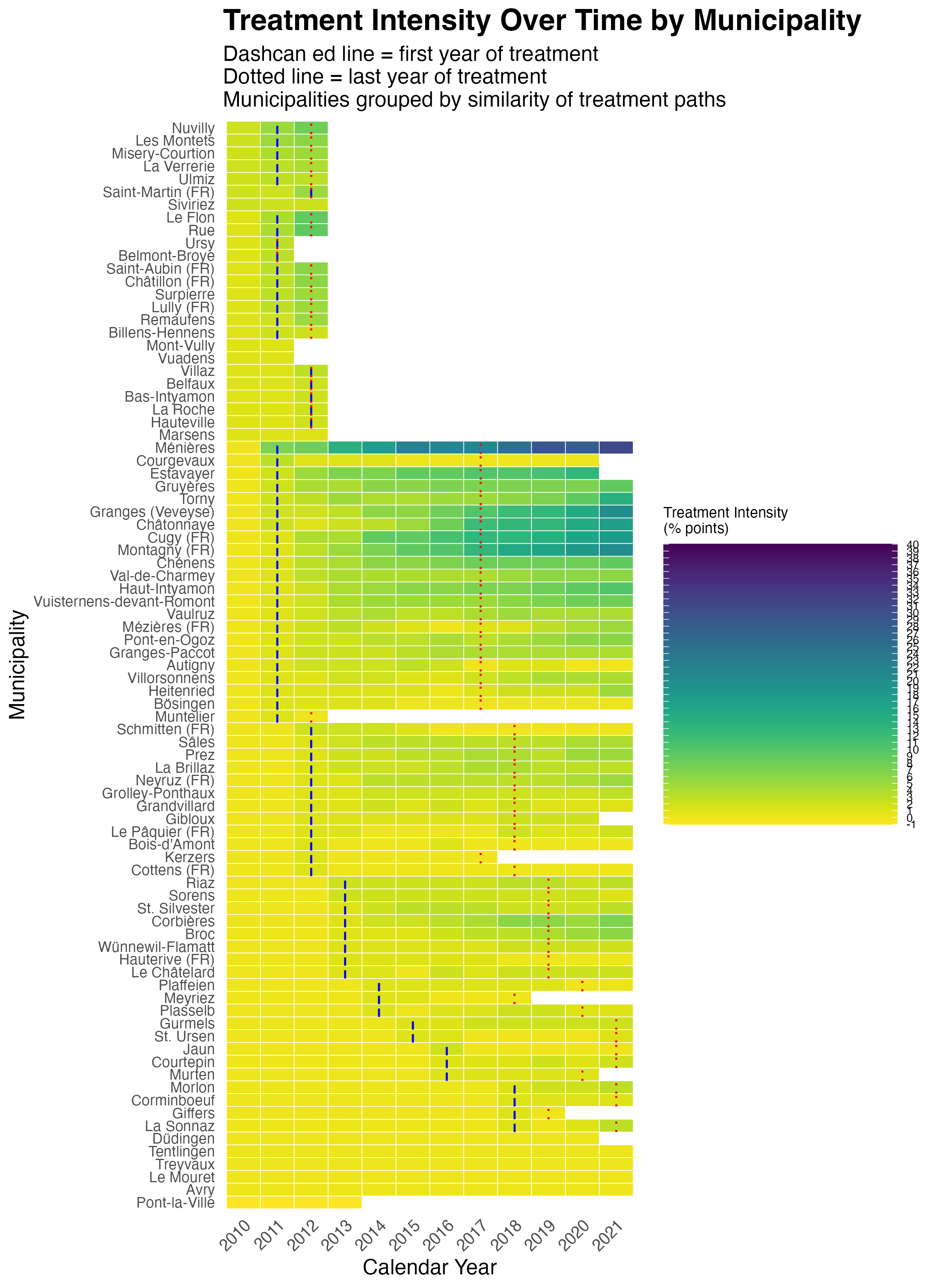}
     \caption{Cumulative internal migration exposure (1 p.p. bins) by municipality (rows) and year (columns), 2010-2021.
Blue ticks = first treatment increase.
Red ticks =last treatment period. 
Darker cells = higher cumulative inflows. 
     Number of municipalities involved in estimating the effect of internal migration: 85/112, with 10 never-treated. 27 municipalities were excluded either because they were "switchers-out" or because no comparable controls were found. }
    \label{fig:SC_1pp_intensity}
\end{figure}

\begin{figure}[H]
    \centering
    \includegraphics[width=0.95\linewidth]{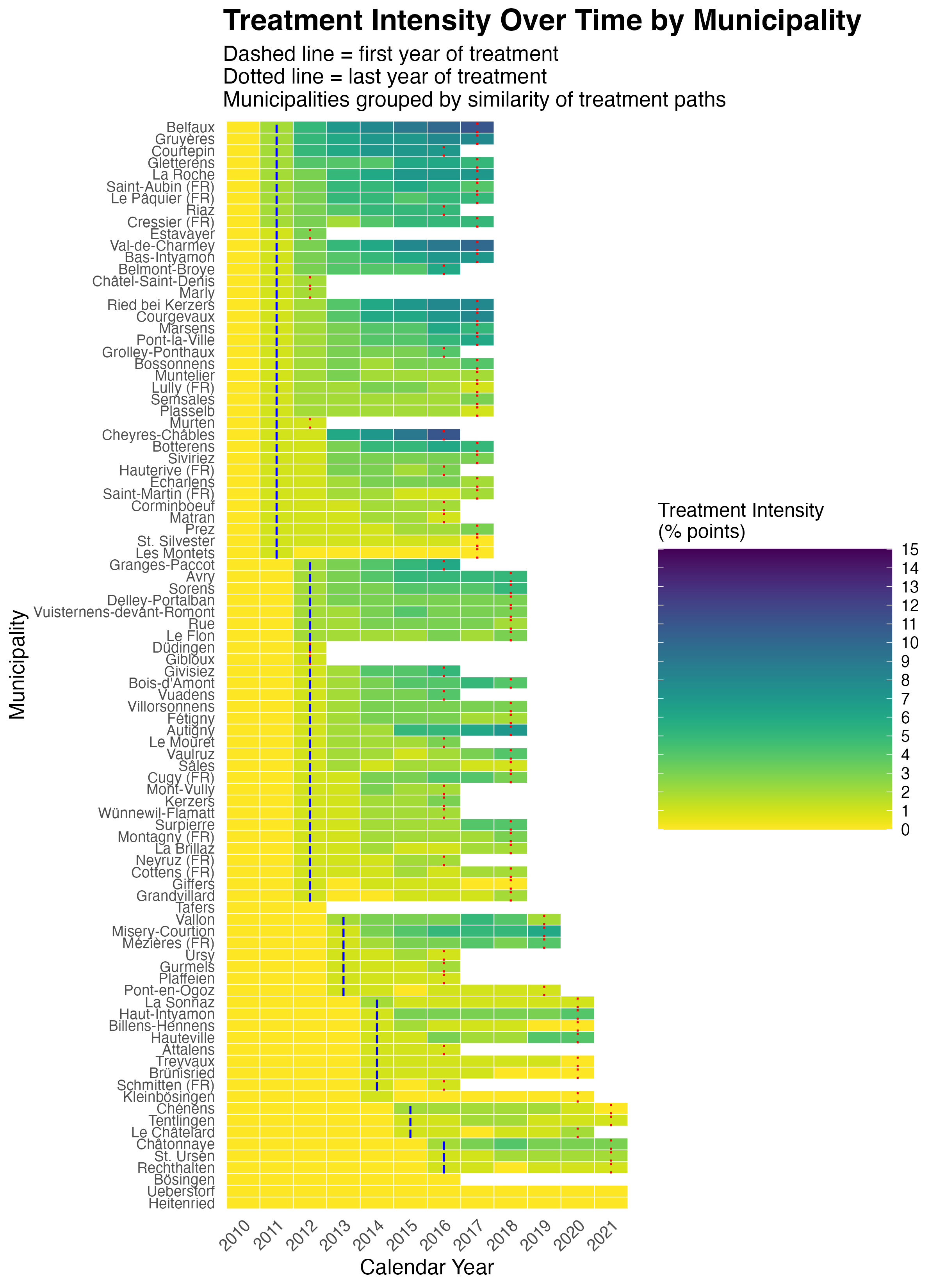}
    \caption{Cumulative international migration exposure (1 p.p. bins) by municipality (rows) and year (columns), 2010-2021.
Blue ticks = first treatment increase.
Red ticks =last treatment period. 
Darker cells = higher cumulative inflows. 
    Number of municipalities involved in estimating the effect of international migration: 89/112, with 4 never-treated. 20 municipalities were excluded either because they were "switchers-out" or because no comparable controls were found.}
    \label{fig:FA_1pp_intensity}
\end{figure}

\section{Results}
\label{results}
 We report in the following sections the results of the model that controls for within migration and that is performed on clusters given by the pre-treatment values of migration following \textcite{de_chaisemartin_difference--differences_2024} as described in Section \ref{sec:estimation}. Most of the following results are robust to specifications of the models where we don't control for spillovers or where we don't cluster for initial levels of migration. A brief discussion on these model specifications can be found in \ref{app:model specifications}.
In the discussion below, we focus primarily on outcomes for which the identifying assumptions are most credible, as assessed by placebo tests of pre-treatment dynamics. We will not discuss treatment effects for those outcomes for which placebo tests fail as we cannot argue the identifying assumptions would hold. 
 Placebo leads are used to test for pre-treatment effects, and high placebo $p$-values indicate that we cannot reject the null hypothesis of no effects, suggesting the absence of significant pre-trends. 
Placebo tests serve as a diagnostic for the plausibility of the identification strategy, and we therefore emphasize outcomes exhibiting stable pre-treatment trajectories.

\textbf{Significance and inference.}
95\% confidence intervals are shown alongside $p$-values for the treatment effects.
Standard errors are computed following \textcite{de_chaisemartin_difference--differences_2024} and are clustered at the pre-treatment migration cluster level, 
using their group-level asymptotic variance formula and the \texttt{less\_conservative\_se} option.\footnote{Standard errors are clustered across three pre-treatment groups. With a small number of clusters, cluster-robust inference can be imprecise and the reported confidence intervals should be interpreted with some caution. Reassuringly, the key results are stable across all model specifications reported in Appendix~\ref{app:model specifications}, including specifications without clustering, which supports the robustness of the findings.}

\subsection{Dynamic effects: event studies}

\subsubsection{Non-normalized treatment effects}

\begin{figure}[H]
    \centering
    \begin{subfigure}[b]{0.45\textwidth}
        \centering
        \includegraphics[width=\textwidth]{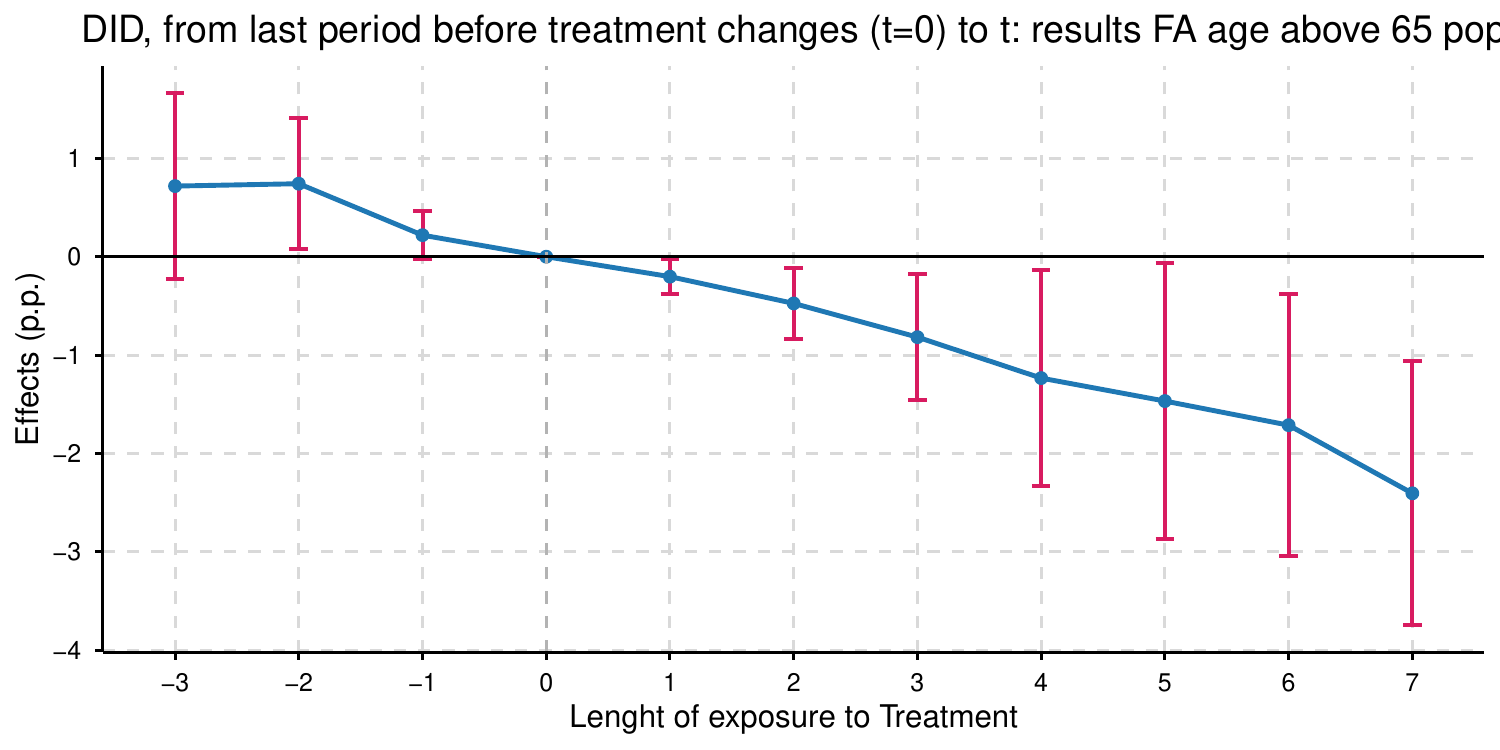}
        \caption{Proportion of population aged $>65$}
        \label{fig:age_less_20}
    \end{subfigure}
    \hfill
    \begin{subfigure}[b]{0.45\textwidth}
        \centering
        \includegraphics[width=\textwidth]{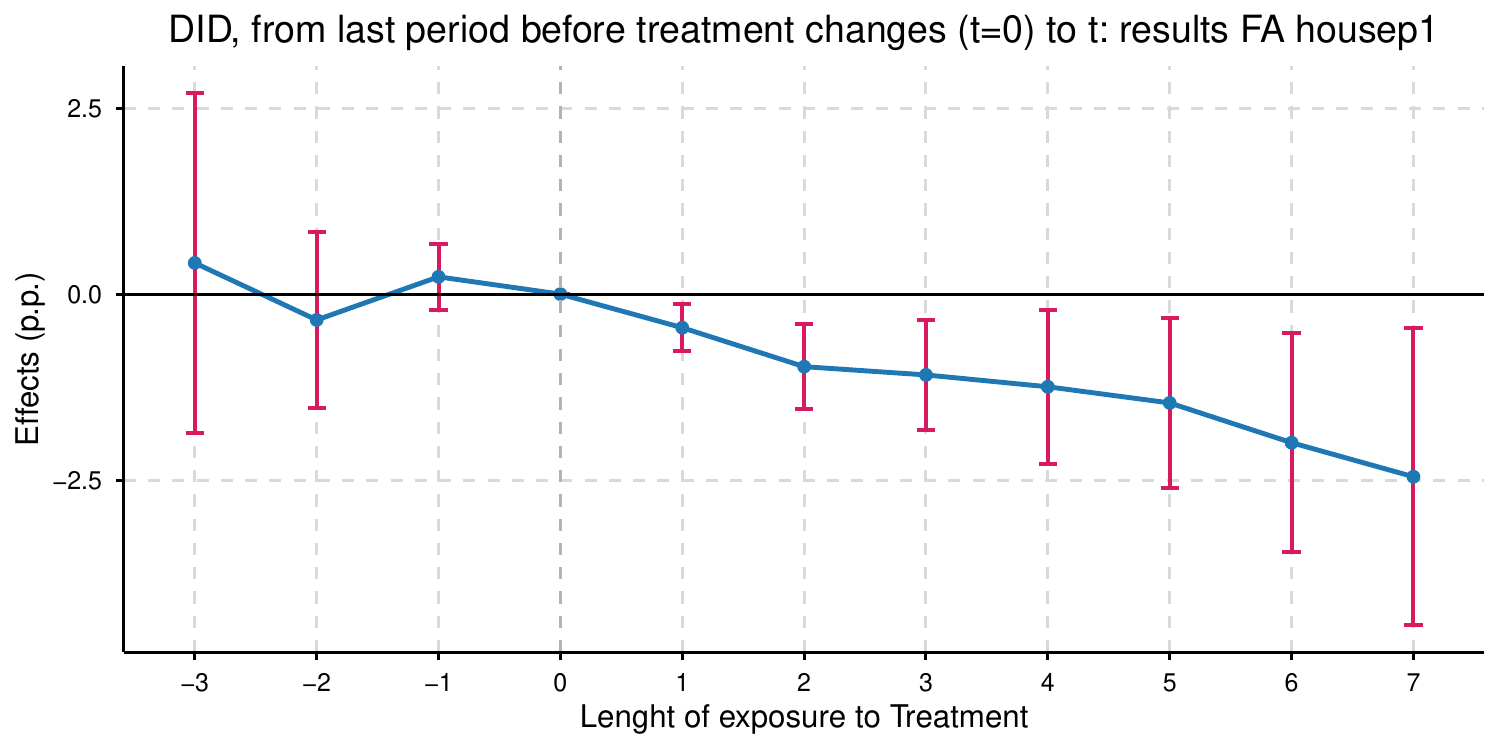}
        \caption{Proportion of single households}
        \label{fig:housep1}
    \end{subfigure}
    \vskip\baselineskip
    \begin{subfigure}[b]{0.45\textwidth}
        \centering
        \includegraphics[width=\textwidth]{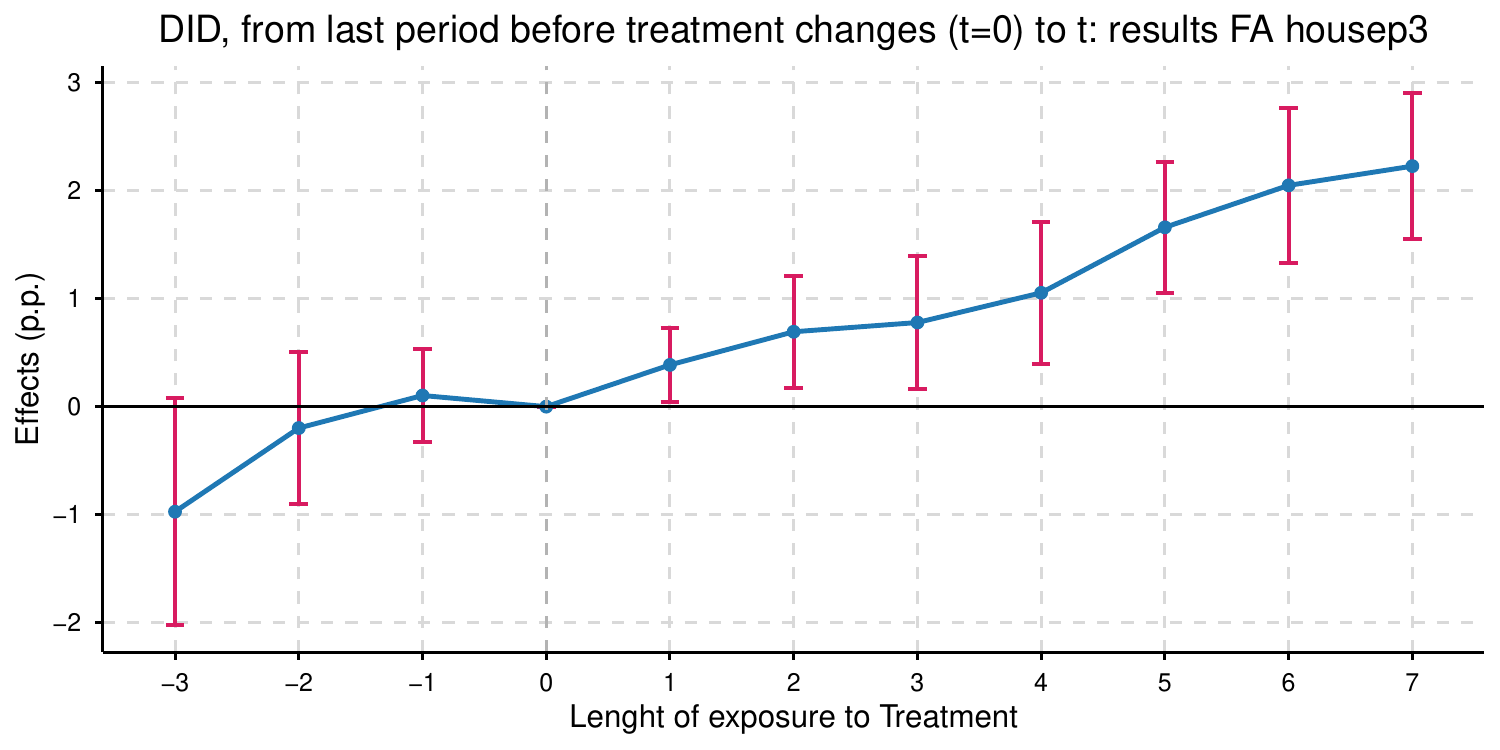}
        \caption{Proportion of 3-people households}
        \label{fig:housep3}
    \end{subfigure}
    \hfill
    \begin{subfigure}[b]{0.45\textwidth}
        \centering
        \includegraphics[width=\textwidth]{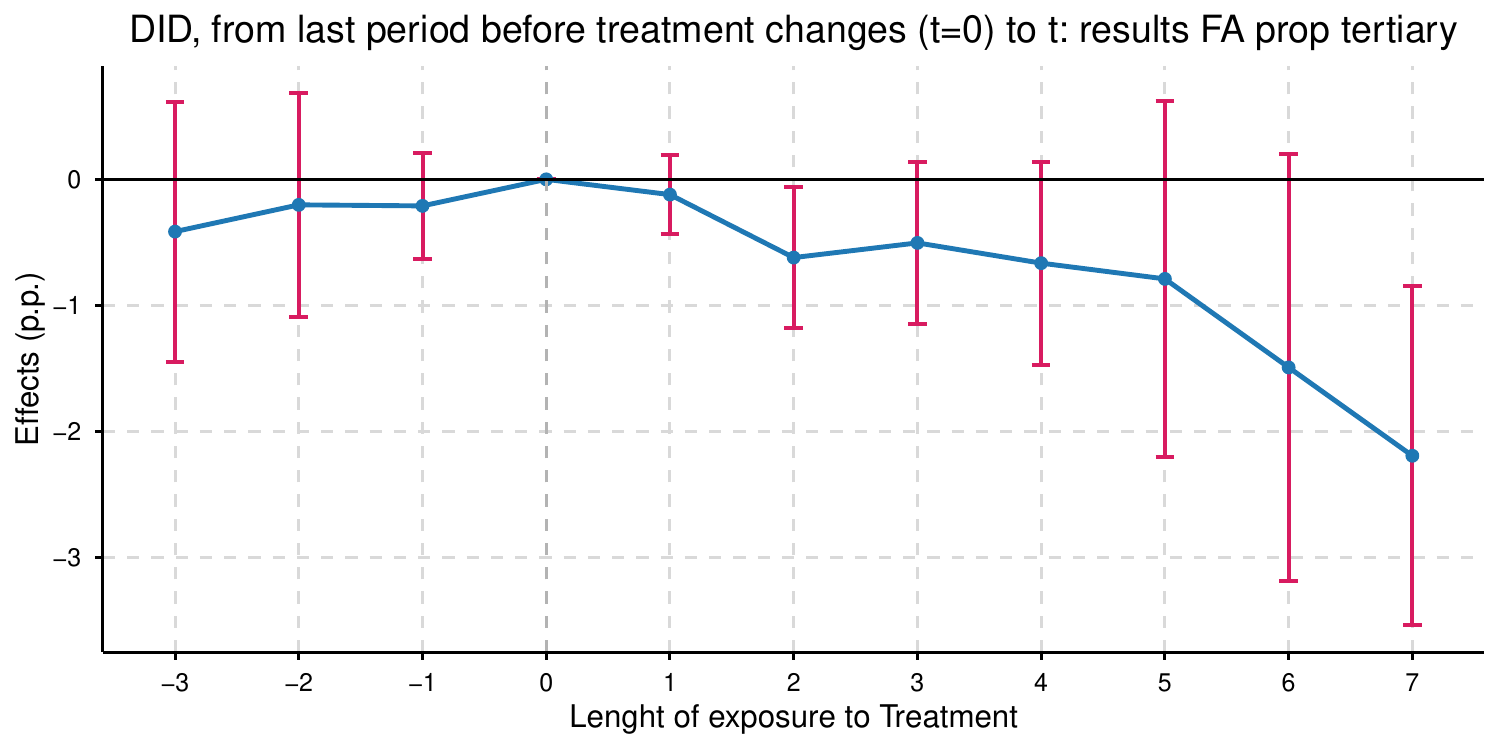}
        \caption{Proportion of students in tertiary education}
        \label{fig:prop_tertiary}
    \end{subfigure}
    \vskip\baselineskip
    \begin{subfigure}[b]{0.45\textwidth}
        \centering
        \includegraphics[width=\textwidth]{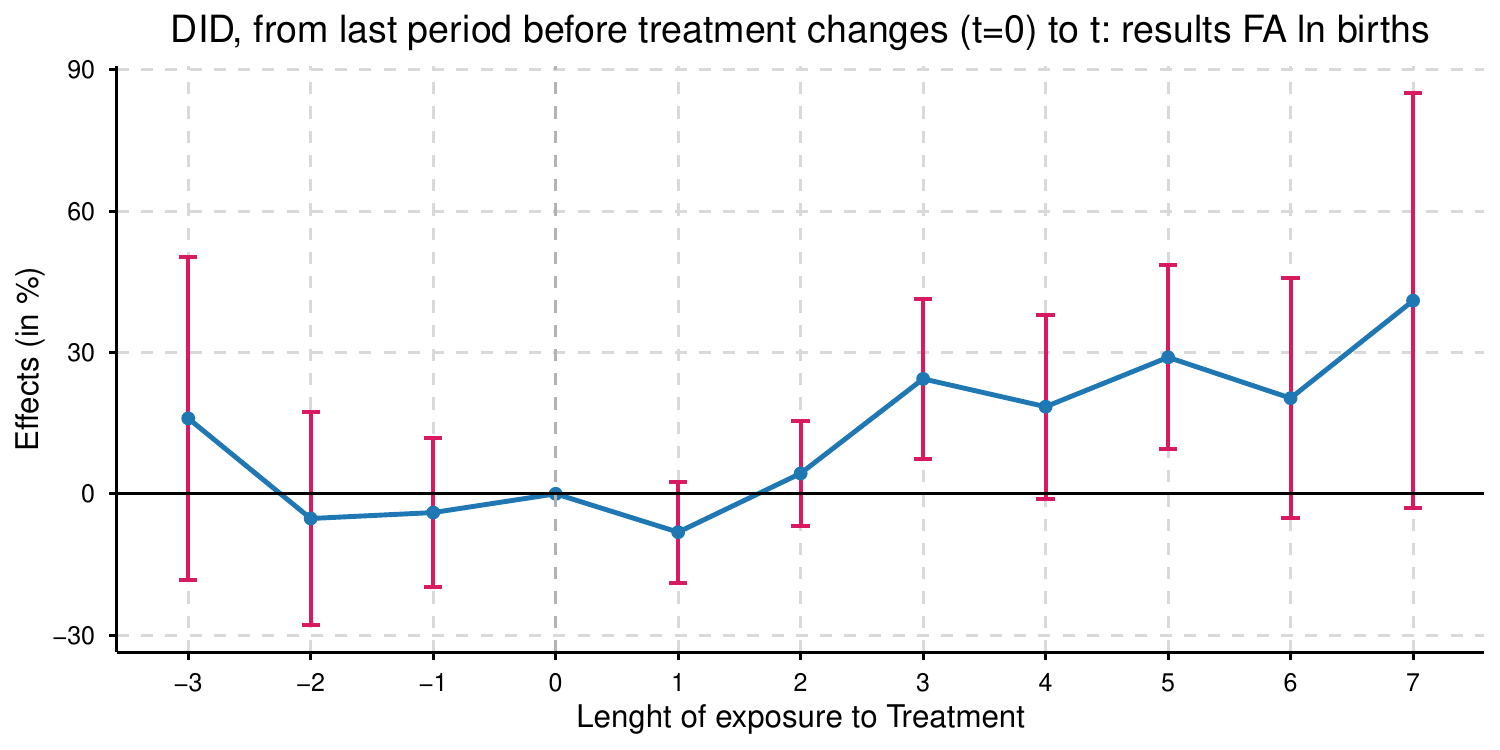}
        \caption{Total births }
        \label{fig:FA_births}
    \end{subfigure}
    \caption{Event studies: effects over time of international migration. Lines plot estimated dynamic treatment effects ($DID_\ell$) relative to the year before treatment; red bars = 95\% CI.}
    \label{fig:FA_event}
\end{figure}

\begin{figure}[H]
    \centering
    \begin{subfigure}[b]{0.45\textwidth}
        \centering
        \includegraphics[width=\textwidth]{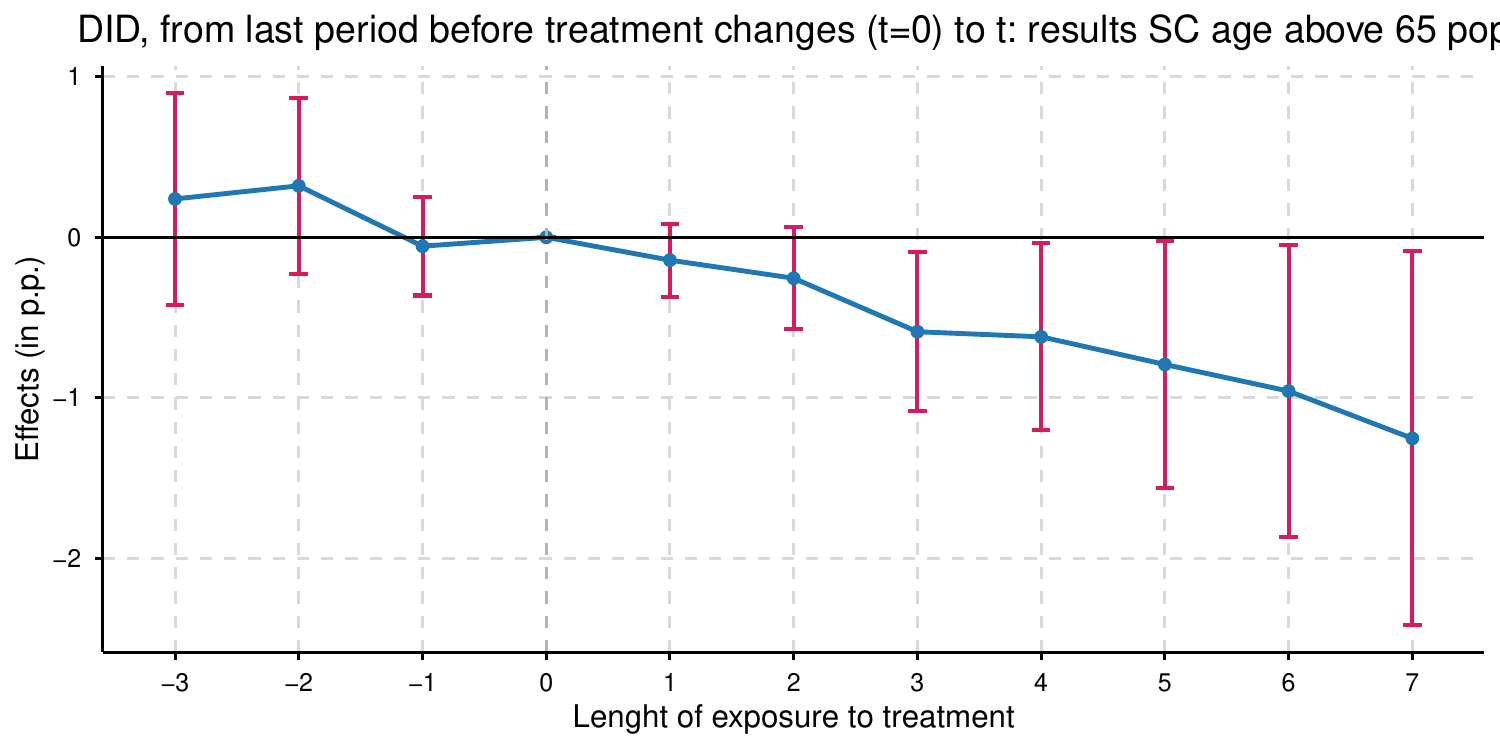}
        \caption{Proportion of population aged $>$ 65}
        \label{fig:age65_SC}
    \end{subfigure}
    \hfill
    \begin{subfigure}[b]{0.45\textwidth}
        \centering
        \includegraphics[width=\textwidth]{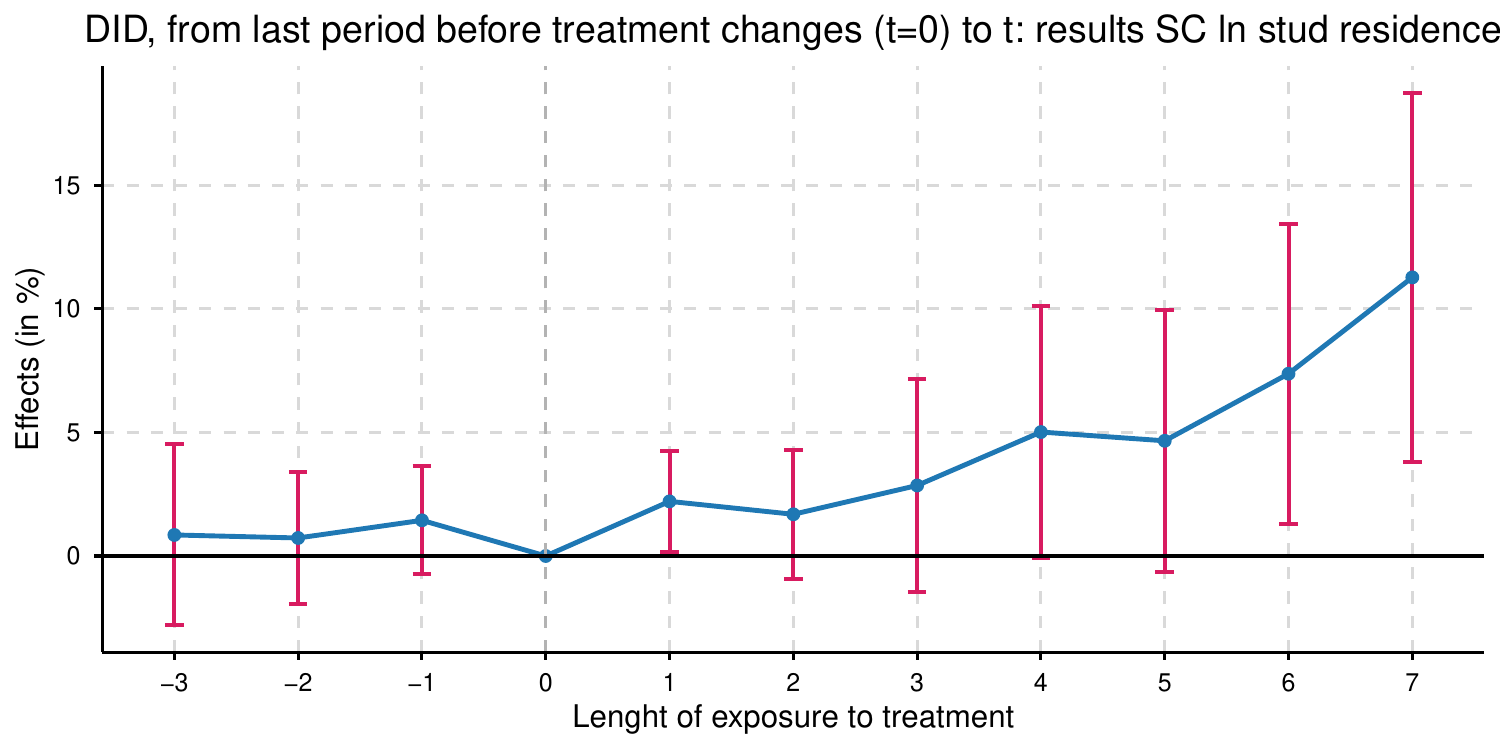}
        \caption{Total students at residence}
        \label{fig:sub2}
    \end{subfigure}
    \vskip\baselineskip
    \begin{subfigure}[b]{0.45\textwidth}
        \centering
        \includegraphics[width=\textwidth]{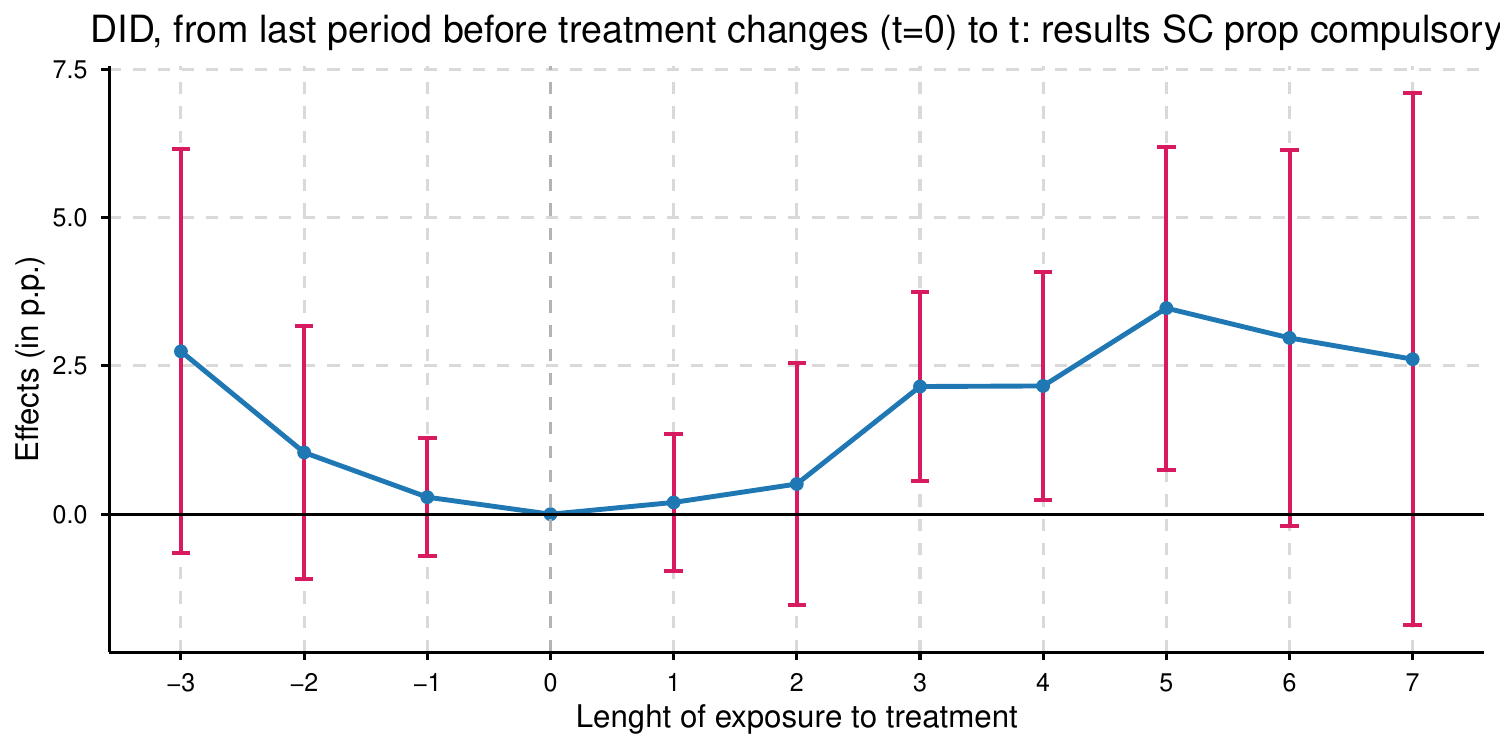}
        \caption{Proportion of students in compulsory education}
        \label{fig:compuls}
    \end{subfigure}
    \hfill
    \begin{subfigure}[b]{0.45\textwidth}
        \centering
        \includegraphics[width=\textwidth]{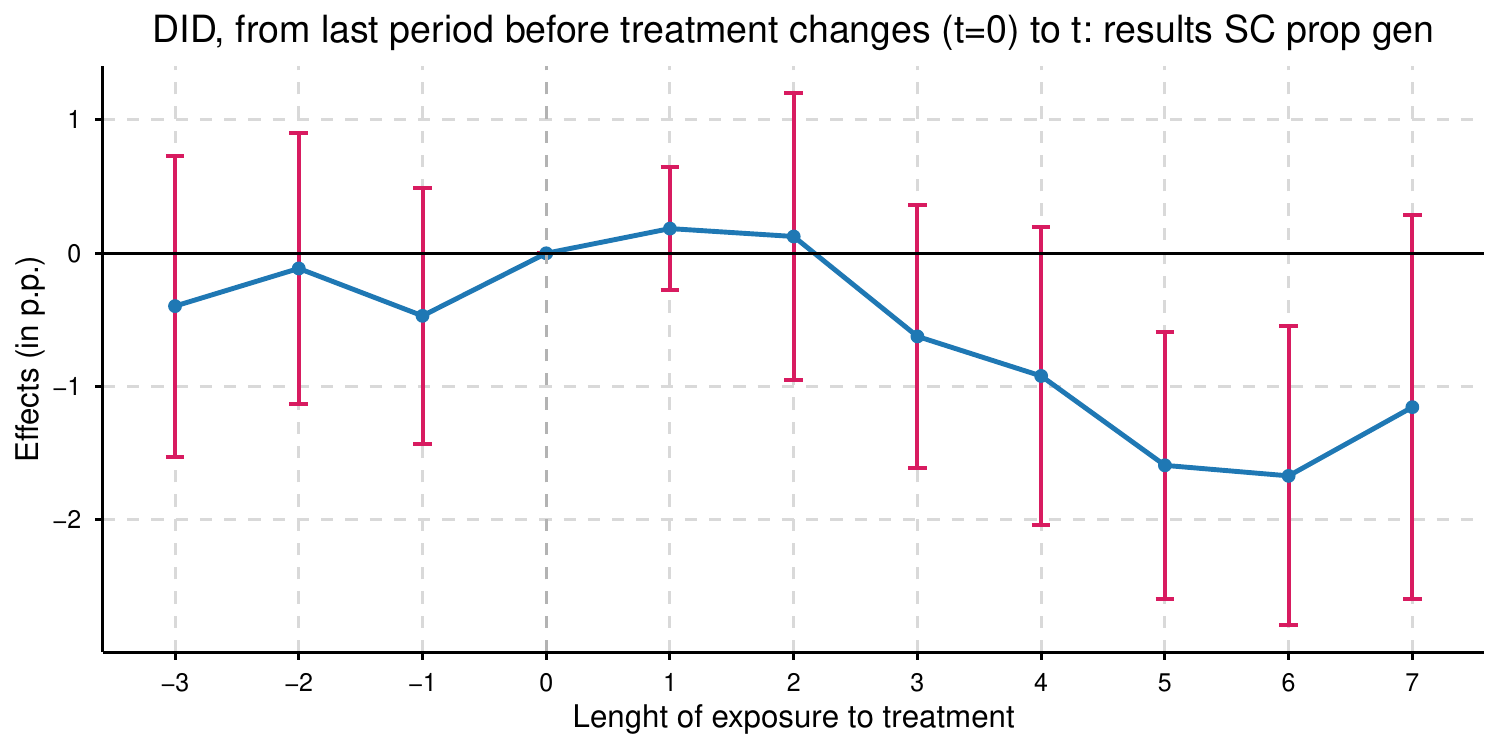}
        \caption{Proportion of students in secondary education: general-track}
        \label{fig:gener}
    \end{subfigure}
    \vskip\baselineskip
    \begin{subfigure}[b]{0.45\textwidth}
        \centering
        \includegraphics[width=\textwidth]{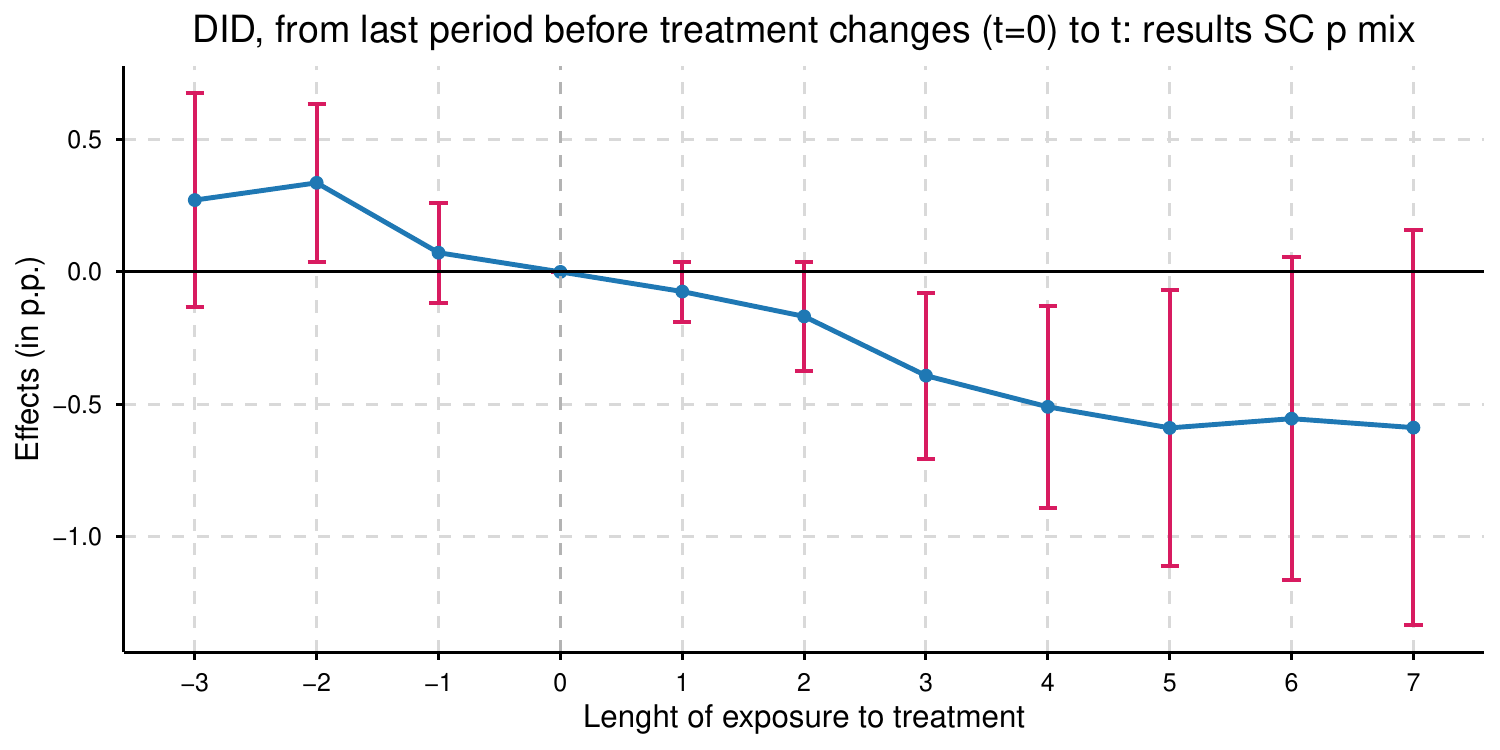}
        \caption{Proportion of mixed-use dwellings}
        \label{fig:SC_mix}
    \end{subfigure}
    \caption{Event studies: effects over time of internal migration. Lines plot estimated dynamic treatment effects ($DID_\ell$) relative to the year before treatment; red bars = 95\% CI.}
    \label{fig:SC_event}
\end{figure}

We first report the non-normalized event study estimators with three pre-trend periods in Figures \ref{fig:FA_event} and \ref{fig:SC_event}. These figures plot the dynamic treatment effects $DID_\ell$ from Equation \ref{eq:non_normalized} for relative times $\ell \in [-3, +7]$ around the first increase in migration exposure. Each point reports the average effect on the outcome $\ell$ periods after a municipality first experiences an increase in cumulative migration exposure above its baseline level, relative to comparable municipalities that have not yet increased at that time. Coefficients are normalized to zero in the year preceding treatment ($\ell = 0$), and bands denote 95\% confidence intervals. We present event graphs only for statistically significant results; estimates for non-significant outcomes are collected in  \ref{app:event_non_sign} and discussed briefly below.

\vspace{0.5em}
\noindent\textbf{International migration.} The pre-treatment coefficients in Figure \ref{fig:FA_event} are small and statistically indistinguishable from zero across all outcomes, supporting the absence of differential pre-trends before exposure changes. Post-treatment patterns differ meaningfully across dimensions. The share of the population aged 65 and above drifts gradually downward after treatment, reaching approximately $-3$ to $-4$ percentage points by $\ell = 7$. Against a pre-treatment mean of 16.9\%, this corresponds to a relative decline of roughly 18-24\%, a substantively large effect consistent with international migrants being disproportionately of working age and thereby diluting the elderly share of the local population. The proportion of single-person households declines by approximately 2 percentage points at its trough around $\ell = 2$-$3$ before stabilising; relative to the pre-treatment mean of 25.4\%, this amounts to a roughly 8\% reduction, a modest but meaningful compositional shift toward multi-person living arrangements. Three-person households move in the opposite direction, rising by approximately 2 percentage points and peaking around $\ell = 2$-$3$; against a pre-treatment mean of 15.3\%, this represents a relative increase of around 13\%, consistent with the arrival of families with one child. The share of tertiary students trends negative after treatment, with point estimates settling modestly below zero beyond $\ell = 3$. Against a pre-treatment mean of 1.7\%, even a small estimated effect in percentage points represents a meaningful relative change, but the wide confidence intervals visible in the figure caution against strong conclusions: the estimate is statistically significant on average but imprecisely identified, and the range of plausible effect sizes is large. We therefore note the negative direction of the effect while refraining from strong claims about its magnitude. Births exhibit a positive shift of roughly 20-30\% at $\ell = 1$-$2$ that persists across later horizons; against a pre-treatment average of approximately $e^{2.80} \approx 16$ births per municipality, this translates to roughly 3-5 additional births, a modest absolute number that nonetheless points to a fertility response consistent with the younger age profile of incoming international migrants. For outcomes such as two-person households, larger households, and most housing characteristics, point estimates remain close to zero throughout the post-treatment window and confidence intervals consistently include zero; their event graphs are therefore not presented here. Across all outcomes, confidence intervals widen with $\ell$, reflecting the declining number of municipalities contributing identifying variation as exposure length increases.

\vspace{0.5em}
\noindent\textbf{Internal migration.} Figure \ref{fig:SC_event} shows a similar pre-trend picture: placebo leads fluctuate around zero across outcomes, supporting the credibility of the parallel-trends assumption. After the first treatment change, the share of the population aged 65 and above declines gradually, reaching approximately $-1$ to $-2$ percentage points by $\ell = 7$. Against a pre-treatment mean of 16.9\%, this is a relative reduction of around 6-12\%, more modest than the international migration effect but equally consistent with internal movers being predominantly of working age. The number of students at residence (in logs) exhibits a clear positive jump at $\ell = 1$ of roughly 10-15\%, with coefficients remaining above zero in subsequent years. With a pre-treatment average of approximately $e^{5.69} \approx 295$ students per municipality, this effect corresponds to approximately 30-45 additional students, a substantial increase that strongly suggests internal migration inflows are driven by families with school-age children relocating within Switzerland. The compulsory-school share rises by approximately 5-7 percentage points after exposure begins and remains stable across horizons; against a pre-treatment mean of 49.6\%, this is a relative increase of 10-14\%, large enough to carry meaningful implications for local school capacity and resource planning. The general-track secondary share moves in the opposite direction, falling by roughly 1-2 percentage points, a relative decline of 11-23\% against its pre-treatment mean of 8.7\%, and the two movements together point to a compositional shift in the student body toward compulsory-age pupils. Mixed-use dwellings display a small negative response of approximately $-0.5$ percentage points, stable across $\ell$; relative to the pre-treatment mean of 19.1\%, this represents a relative decline of around 3\% and, while statistically significant, is unlikely to carry meaningful economic consequences. Several other demographic and housing outcomes, including household size distributions beyond three-person households and most dwelling-type proportions, show no statistically distinguishable response, and their event graphs are collected in the appendix. As with international migration, later-$\ell$ estimates become noisier, but the overall shape of the dynamic profiles is consistent across the exposure window.

\noindent Table \ref{tab:summary_effects} summarises the direction, approximate peak magnitude, and economic significance for all outcomes with statistically distinguishable effects.

\begin{table}[H]
\centering
\caption{Summary of non-normalized event-study results}
\label{tab:summary_effects}
\resizebox{\textwidth}{!}{%
\begin{tabular}{llcccc}
\hline
Migration type & Outcome & Dir. & Approx.\ peak effect & Pre-treat.\ mean & Relative magnitude \\
\hline
International & Pop.\ aged 65+           & $-$ & $ -3$ to $-4$ p.p.\ at $\ell=7$  & 16.9\%                  & $\sim$18-24\% of mean \\
International & Single-person HH         & $-$ & $\approx -2$ p.p.\ at $\ell=2$-$3$     & 25.4\%                  & $\sim$8\% of mean      \\
International & Three-person HH          & $+$ & $\approx +2$ p.p.\ at $\ell=2$-$3$     & 15.3\%                  & $\sim$13\% of mean     \\
International & Tertiary student share   & $-$ & Negative, imprecise                       & 1.7\%                   & Wide CI; interpret with caution \\
International & Births (log)             & $+$ & $ +20$-$30$\% at $\ell=5$  & ${\approx}16$ births    & $\sim$3-5 additional births \\
\hline
Internal      & Pop.\ aged 65+           & $-$ & $ -1$ to $-2$ p.p.\ at $\ell=7$  & 16.9\%                  & $\sim$6-12\% of mean  \\
Internal      & Students at residence    & $+$ & $ +10$-$15$\% at $\ell=7$       & ${\approx}295$ students & $\sim$30-45 students  \\
Internal      & Compulsory-school share  & $+$ & $ +2$ to $5$ p.p. at $\ell=5$     & 49.6\%                  & $\sim$4-10\% of mean \\
Internal      & General-track share      & $-$ & $ -1$ to $-2$ p.p.    at $\ell=5-6$     & 8.7\%                   & $\sim$11-23\% of mean \\
Internal      & Mixed-use dwelling share & $-$ & $\approx -0.5$ p.p.             & 19.1\%                  & $\sim$3\% of mean      \\
\hline
\end{tabular}}
\end{table}
\vspace{0.5em}
\noindent

\subsubsection{Normalized treatment effects}

\begin{figure}[H]
    \centering
    \begin{subfigure}[b]{0.45\textwidth}
        \centering
        \includegraphics[width=\textwidth]{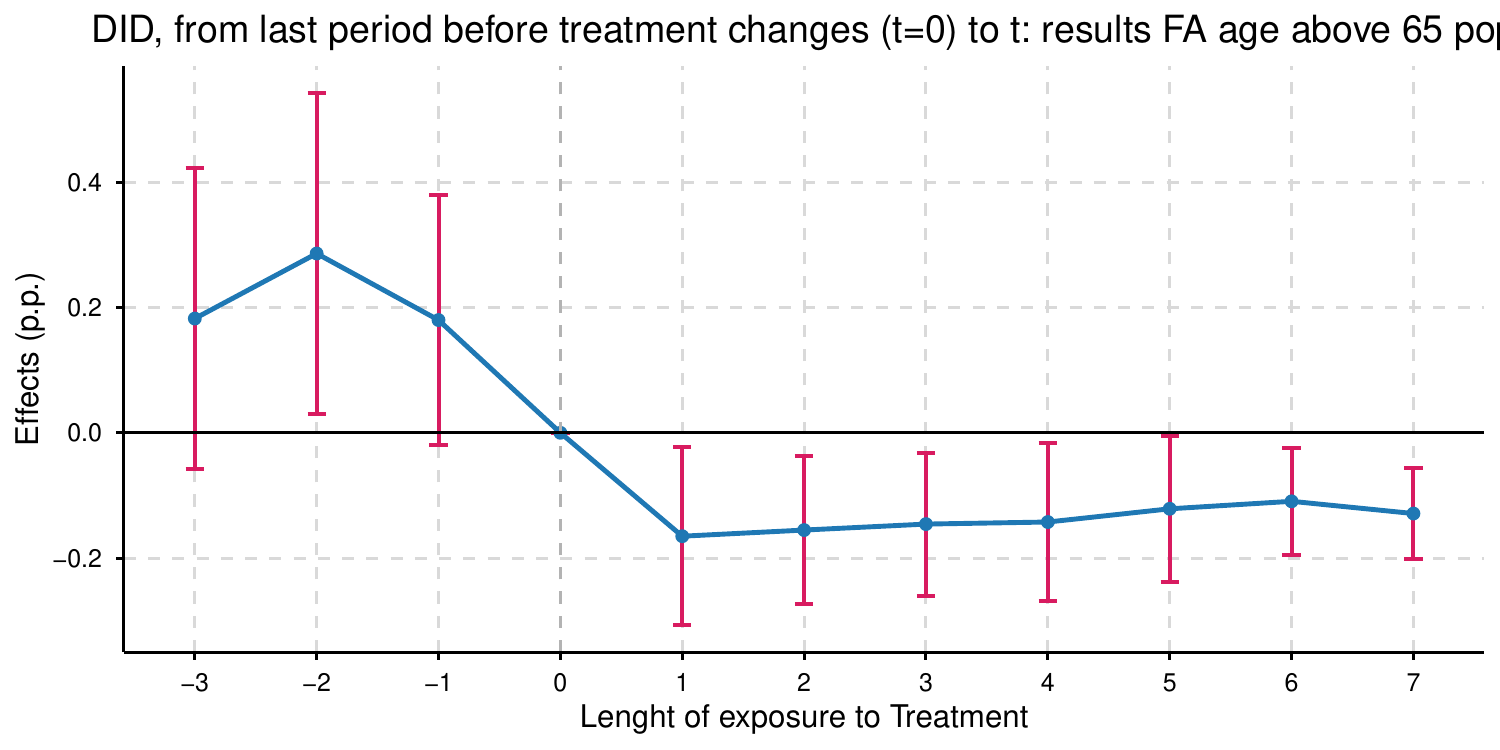}
        \caption{Proportion of population aged $>65$}
        \label{fig:n_age_65_FA}
    \end{subfigure}
    \hfill
    \begin{subfigure}[b]{0.45\textwidth}
        \centering
        \includegraphics[width=\textwidth]{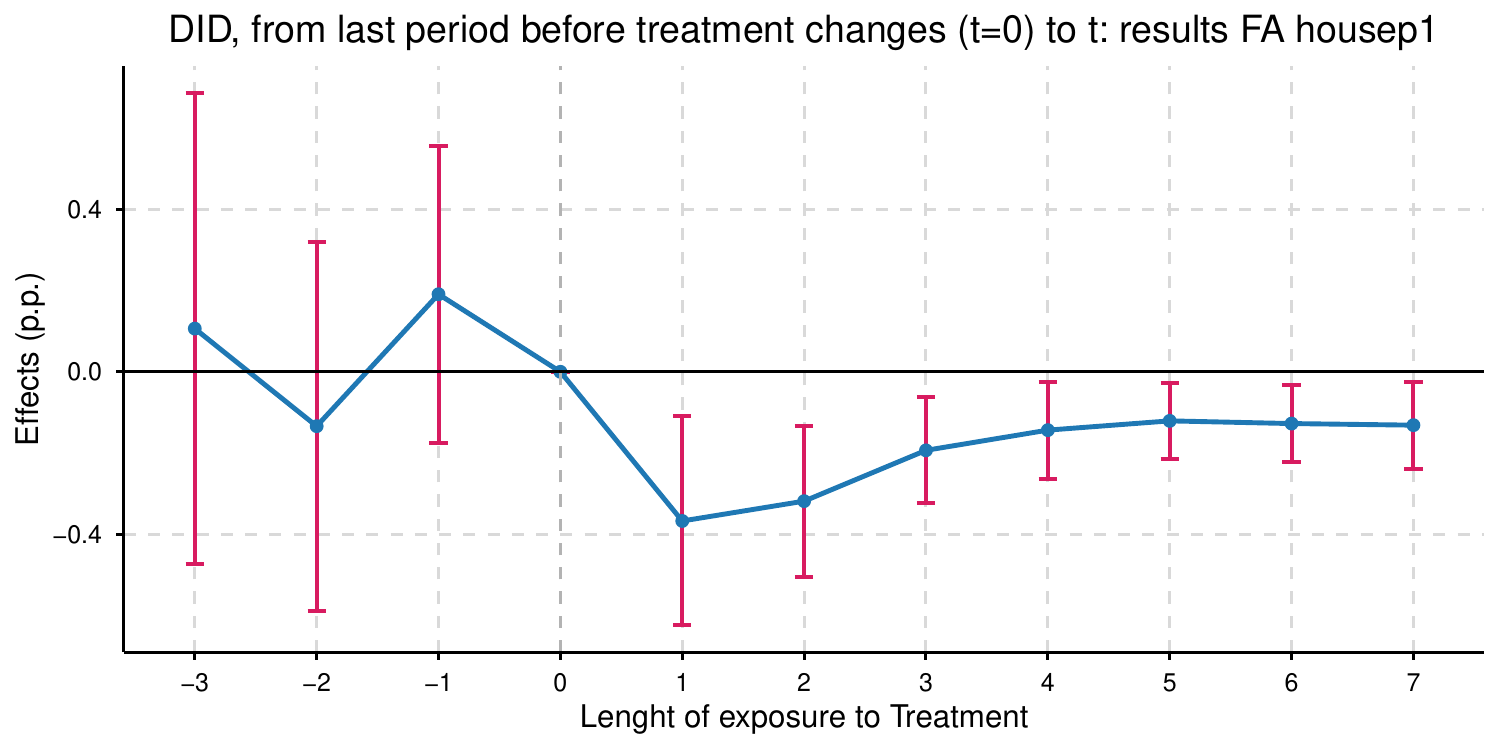}
        \caption{Proportion of single households}
        \label{fig:n_housep1}
    \end{subfigure}
    \vskip\baselineskip
    \begin{subfigure}[b]{0.45\textwidth}
        \centering
        \includegraphics[width=\textwidth]{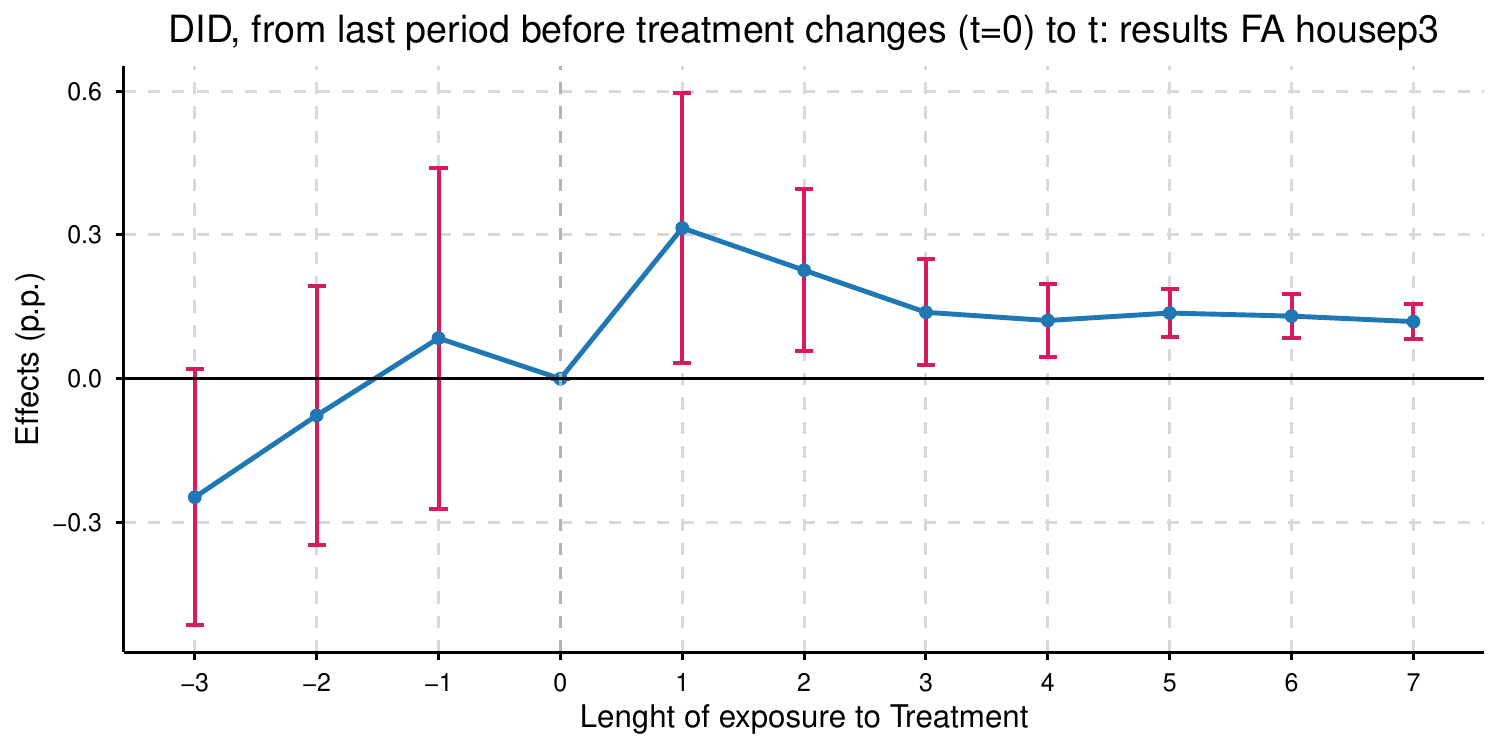}
        \caption{Proportion of 3-people households}
        \label{fig:n_housep3}
    \end{subfigure}
    \hfill
    \begin{subfigure}[b]{0.45\textwidth}
        \centering
        \includegraphics[width=\textwidth]{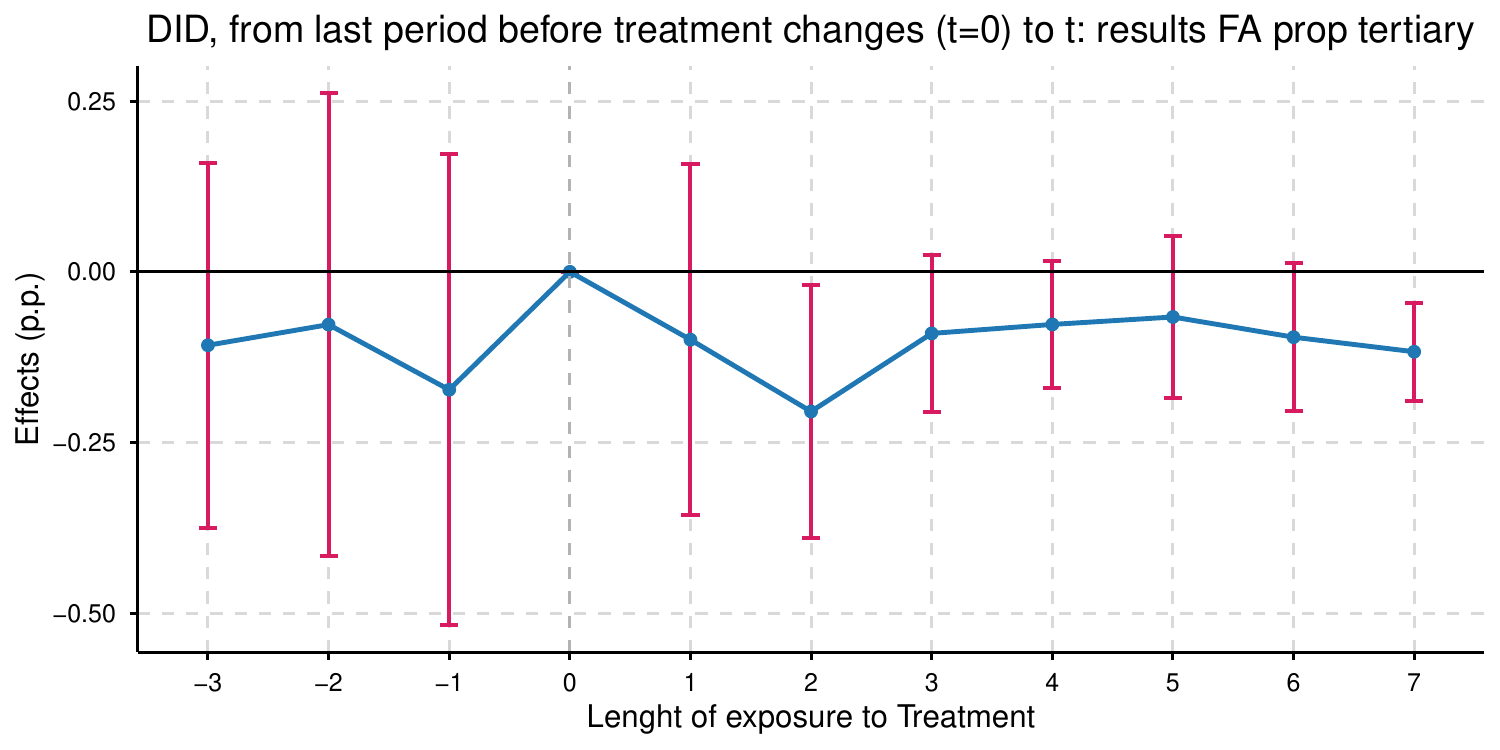}
        \caption{Proportion of students in tertiary education}
        \label{fig:n_prop_tertiary}
    \end{subfigure}
    \vskip\baselineskip
    \begin{subfigure}[b]{0.45\textwidth}
        \centering
        \includegraphics[width=\textwidth]{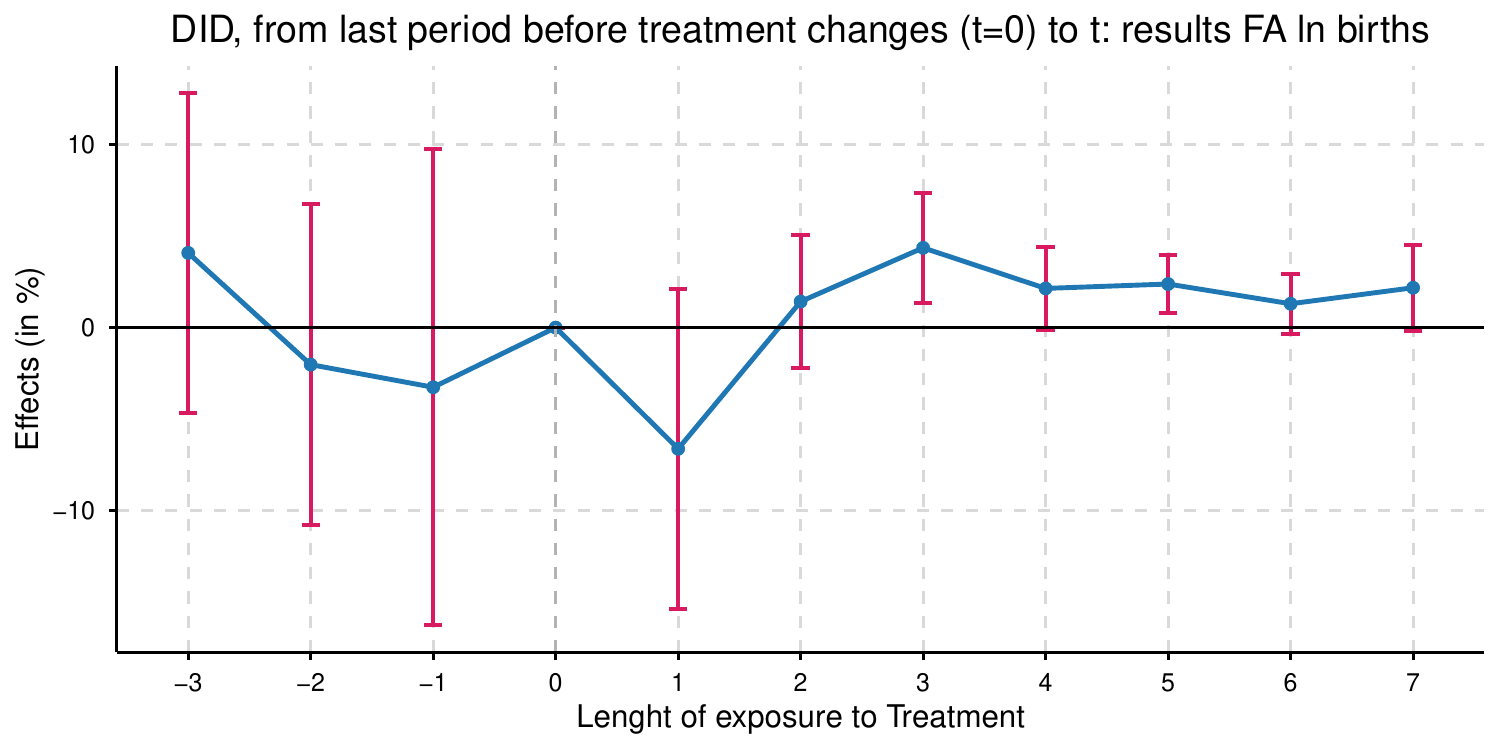}
        \caption{Total births }
        \label{fig:n_FA_births}
    \end{subfigure}
    \caption{Event studies: effects over time of international migration. Lines plot estimated dynamic treatment effects ($DID_\ell^n$) relative to the year before treatment; red bars = 95\% CI.}
    \label{fig:FA_event_normalized}
\end{figure}

\begin{figure}[H]
    \centering
    \begin{subfigure}[b]{0.45\textwidth}
        \centering
        \includegraphics[width=\textwidth]{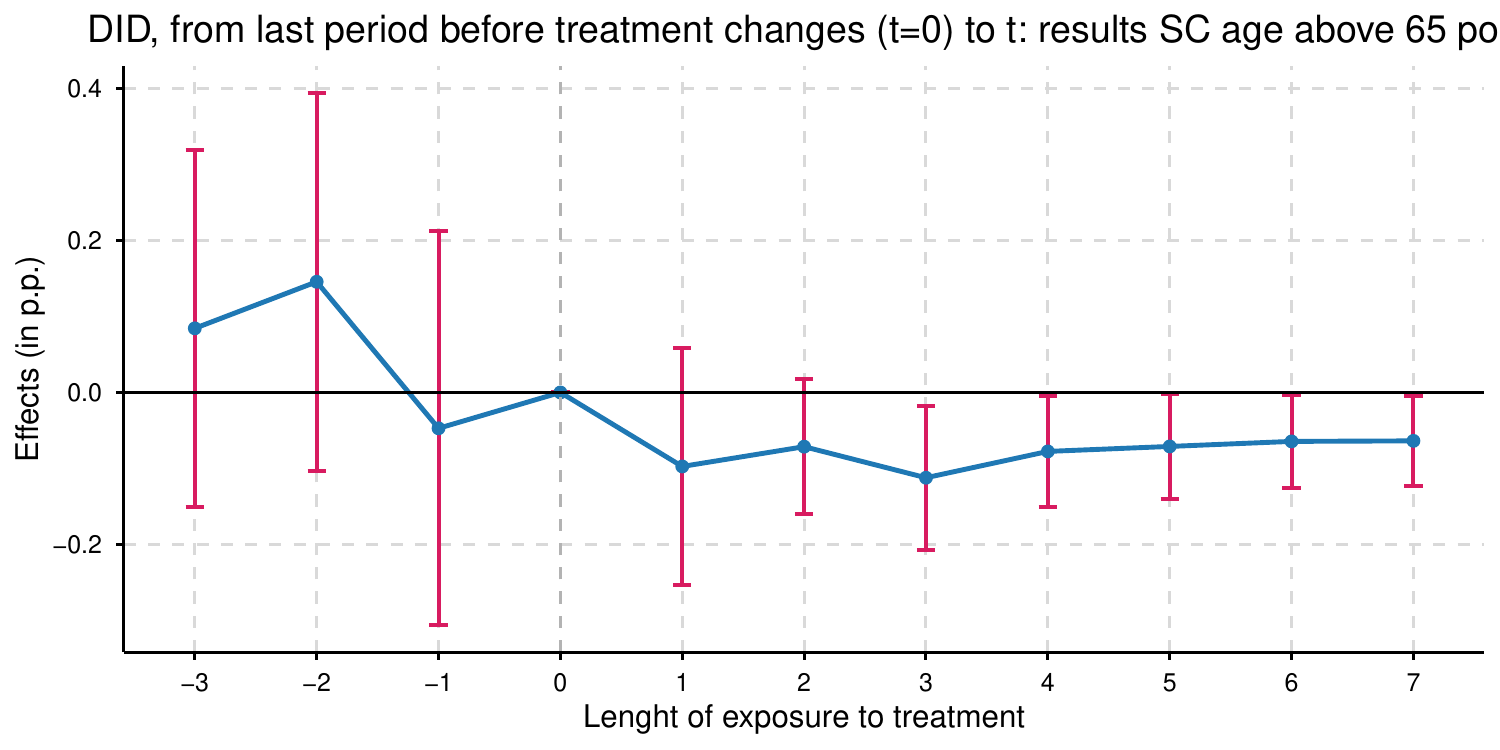}
        \caption{Proportion of population aged $>$ 65}
    \end{subfigure}
    \hfill
    \begin{subfigure}[b]{0.45\textwidth}
        \centering
        \includegraphics[width=\textwidth]{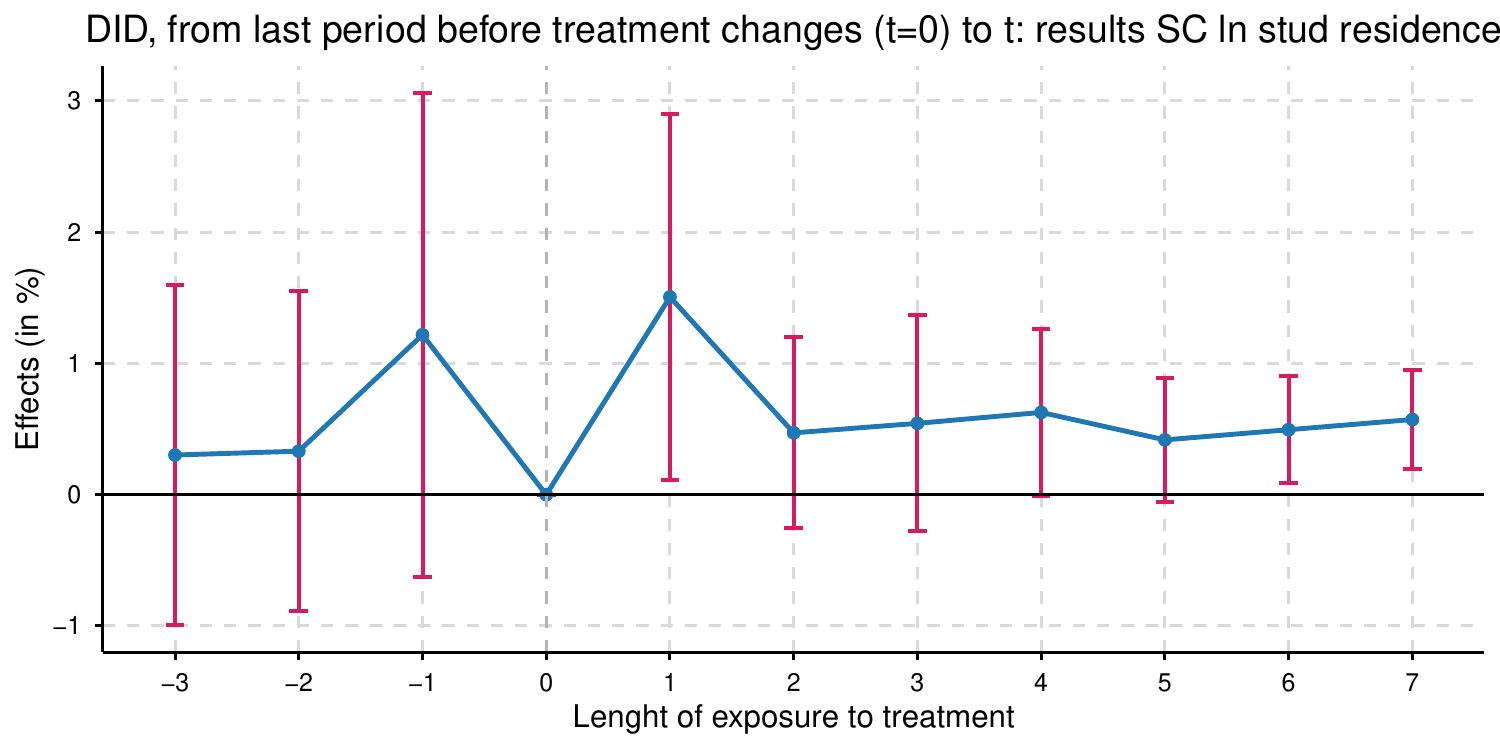}
        \caption{Total students at residence}
        \label{fig:SC_stud_n}
    \end{subfigure}
    \vskip\baselineskip
    \begin{subfigure}[b]{0.45\textwidth}
        \centering
        \includegraphics[width=\textwidth]{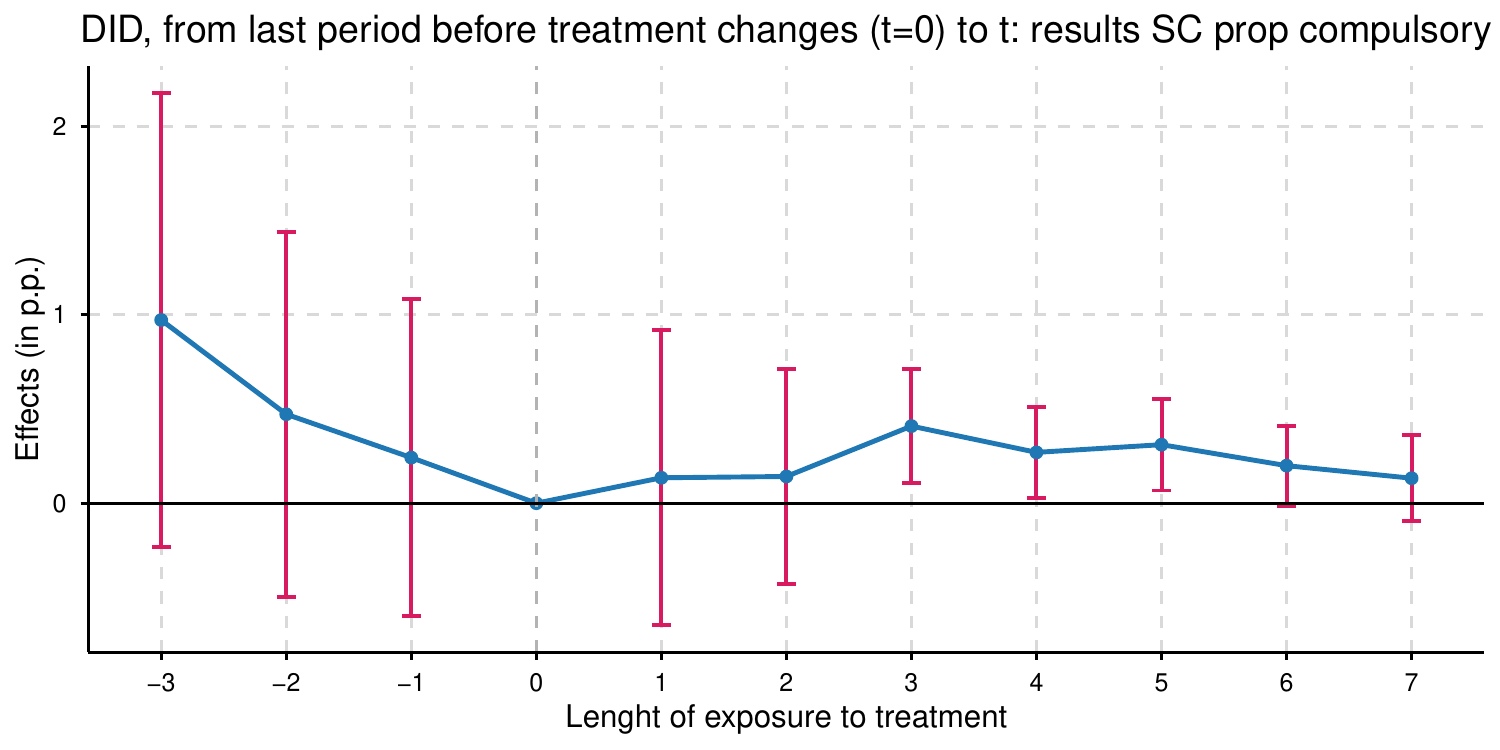}
        \caption{Proportion of students in compulsory education}
        \label{fig:n_compuls}
    \end{subfigure}
    \hfill
    \begin{subfigure}[b]{0.45\textwidth}
        \centering
        \includegraphics[width=\textwidth]{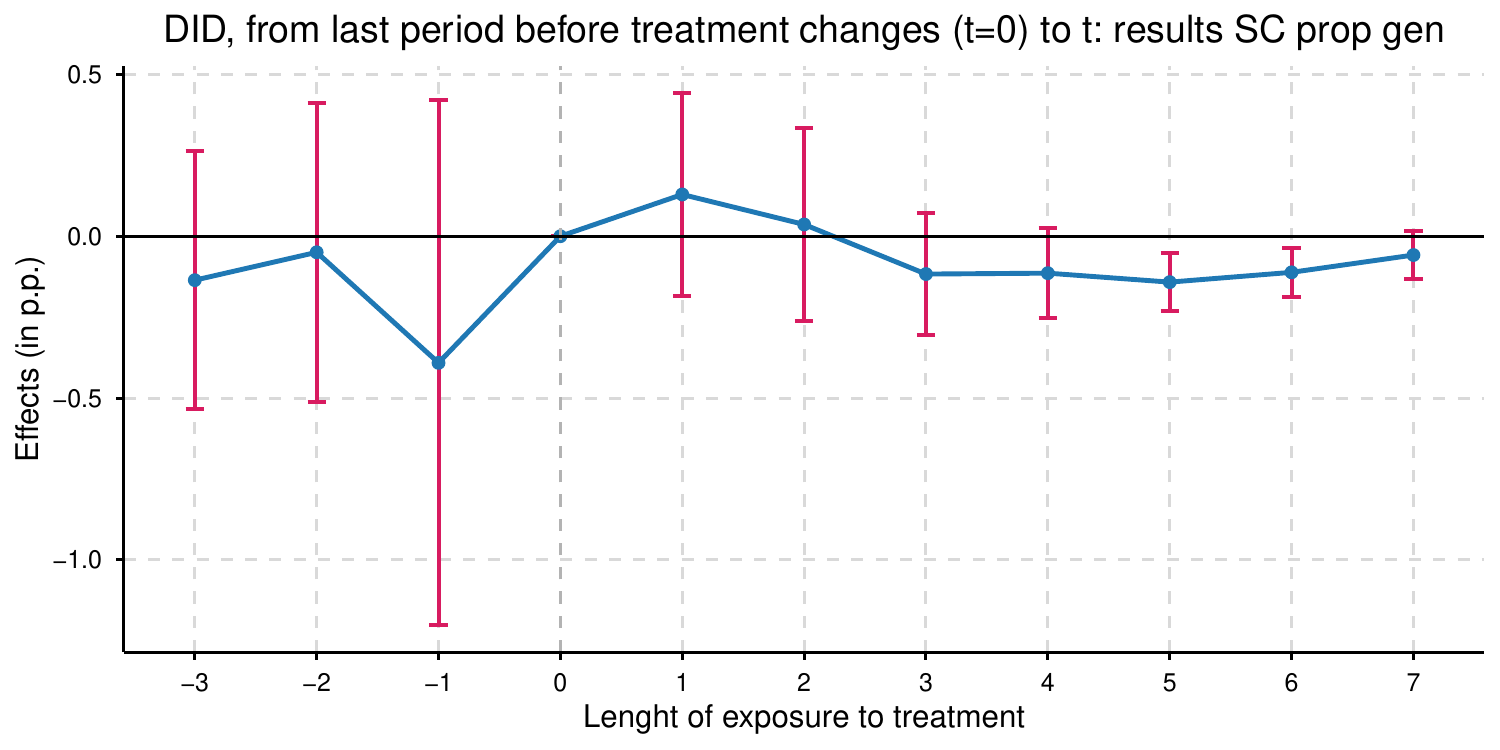}
        \caption{Proportion of students in secondary education: general-track}
        \label{fig:n_ger}
    \end{subfigure}
    \vskip\baselineskip
    \begin{subfigure}[b]{0.45\textwidth}
        \centering
        \includegraphics[width=\textwidth]{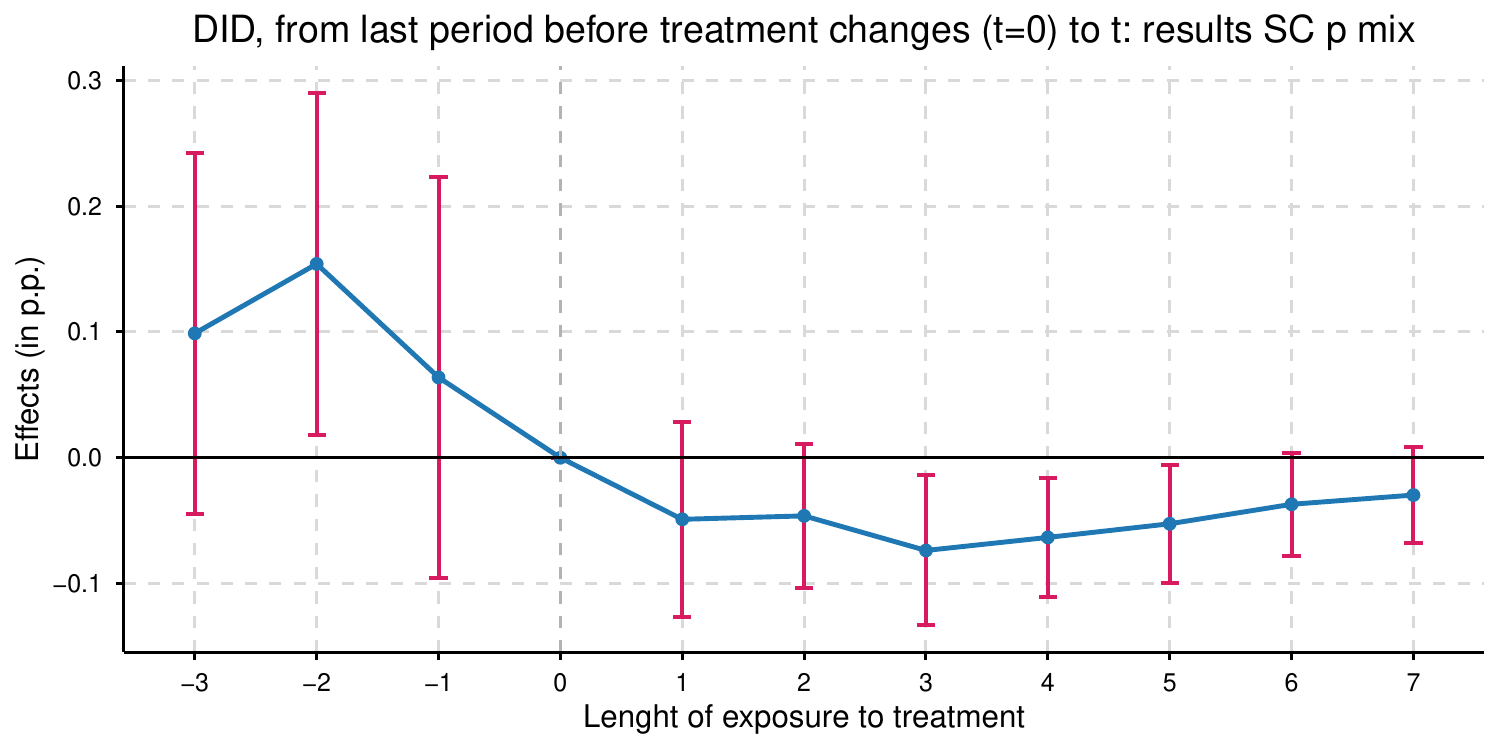}
        \caption{Proportion of mixed-use dwellings}
        \label{fig:n_SC_mix}
    \end{subfigure}
    \caption{Event studies: effects over time of internal migration. Lines plot estimated dynamic treatment effects ($DID_\ell^n$) relative to the year before treatment; red bars = 95\% CI.}
    \label{fig:SC_event_normalized}
\end{figure}

Figures \ref{fig:FA_event_normalized} and \ref{fig:SC_event_normalized} report the normalized event studies, plotting the dynamic treatment effects $DID^{n}_\ell$ from Equation \ref{eq:normalized eff} for relative times $\ell \in [-3, +7]$. These coefficients measure the weighted average of the contemporaneous treatment at $\ell$ and its $\ell - 1$ first lags on the contemporaneous outcomes.\footnote{Relevant weights are reported in Tables \ref{tab_app:weights_international} and \ref{tab_app:weights_internal}. Non-normalized estimates describe how outcomes evolve after $\ell$ periods of higher migration exposure. Normalized estimates convert the $\ell$-th effect into a weighted average of contemporaneous and lagged exposure effects, allowing direct assessment of whether lagged migration exposure matters independently. Comparing the two helps characterize the temporal structure of treatment effects.} As before, we report event graphs only for statistically significant results.

\vspace{0.5em}
\noindent\textbf{International migration.} The normalized estimates in Figure \ref{fig:FA_event_normalized} are considerably attenuated relative to the non-normalized ones, which is expected given that normalized coefficients represent per-unit exposure averages rather than cumulative effects. Peak estimates fall to approximately $-0.1$ to $-0.2$ percentage points for the elderly share, roughly 1\% of its pre-treatment mean, around $-0.3$ percentage points for single-person households, approximately $+0.3$ to $+0.5$ percentage points for three-person households, and roughly $-0.2$ to $-0.3$ percentage points for the tertiary student share; births retain a positive normalized effect on the order of $+5$ to $+8$\%. Crucially, signs and the relative ordering of outcomes are fully preserved across both sets of estimates. Across all outcomes, the normalized profiles are nearly flat as $\ell$ increases, indicating that contemporaneous and lagged exposure contributions are broadly similar in magnitude. This absence of a systematic slope implies that neither a strongly front-loaded impact, which would produce declining normalized effects as $\ell$ grows, nor a delayed-accumulation dynamic, which would produce rising effects, is driving the results. Instead, most of the impact materialises within the first few periods after exposure begins and then persists at a roughly constant level.

\vspace{0.5em}
\noindent\textbf{Internal migration.} The normalized profiles in Figure \ref{fig:SC_event_normalized} confirm the patterns established in the non-normalized analysis. The negative relationship between internal migration exposure and the elderly share is preserved at approximately $-0.1$ to $-0.15$ percentage points, though the confidence interval extends into positive territory, indicating that the per-period lagged contribution is estimated less precisely than the cumulative non-normalized effect. The positive effect on students at residence remains robust at approximately $+1$-$2$\%, equivalent to roughly 3-6 additional students per municipality per period of exposure;  modest in absolute terms but consistent and stable across the exposure window. The compulsory-school share retains a positive normalized effect of approximately $+1$-$2$ percentage points, while the general-track share shows a corresponding decline of roughly $-0.5$ to $-0.8$ percentage points; both profiles are stable across $\ell$, reinforcing the compositional interpretation from the non-normalized results. The small negative effect on mixed-use dwellings persists at around $-0.05$ to $-0.1$ percentage points. As in the international case, the flatness of the profiles across $\ell$ indicates that contemporaneous and lagged exposure contribute in broadly similar ways to the overall impact.

\vspace{0.5em}
\noindent For both migration types, the normalized event-study curves closely track the shapes of the non-normalized ones in terms of sign, timing, and relative ordering. The primary difference is one of scale, attributable to the normalization procedure itself. The absence of any systematic trend in normalized effects across $\ell$ indicates that the results are not driven by the accumulation of lagged exposure over time, but rather reflect a persistent, roughly time-invariant relationship between migration inflows and the outcomes studied.

\subsection{Average cumulative effects per treatment unit}
Finally, we report the average total effect per unit of treatment ($\hat{\delta}$ from Equation \ref{eq: Av tot effects}, which summarizes the overall impact of a 1-percentage-point increase in migration exposure. When exposure rises, that change can affect outcomes not only in the year it occurs but also in subsequent years. The estimator adds up all these immediate and lagged effects generated by a single treatment increment and then averages them across all municipalities that experience a change in exposure. In our data, these effects typically unfold over about three years, so the reported coefficient captures the average combined impact of one treatment increase over this window.
All outcomes are expressed in percentage
points (p.p.) of the corresponding population share or housing proportion,
except for \emph{number of students at residence} and  \emph{number of births} , which are in natural logarithms.
A coefficient of $0.004$ means that a $1$\,p.p.\ increase in cumulative
migration exposure raises the outcome by $0.4$\,p.p for the proportion, or by $0.4\%$ for the number of births and number of students at residence. 
\vspace{0.5em}
\noindent
A separate column reports the $p$-value of the joint test for the nullity of all
placebo coefficients (no pre-trend).  High placebo $p$-values ($>0.7$) across
outcomes confirm that pre-treatment trajectories are statistically flat, lending
credibility to a causal interpretation.

\begin{table}[H]
\caption{Average total effect of a 1-percentage-point increase in cumulative international migration exposure on demographic, educational, and housing outcomes, 2010-2021. For each 1-percentage-point increase in exposure, the estimator sums the immediate and subsequent effects generated by that treatment change and averages them across switching municipalities. 95\% confidence intervals in columns 4-5.}
\centering
\resizebox{\textwidth}{!}{%
\begin{tabular}{|l|l|l|l|l|l|}
\hline
Outcome &Average Total Effect&$p$-value (ATE)&$p$-value (placebos)&    CI Lower&    CI Upper\\
\hline
Below 20-year-olds relative to total population                                       & 0.00107                         & 0.44809                        & 0.83439                        & -0.00170                        & 0.00385                         \\

{\color[HTML]{3531FF} Population aged above 64 relative to total population}          & {\color[HTML]{3531FF} -0.00425} & {\color[HTML]{3531FF} 0.00274} & {\color[HTML]{3531FF} 0.10289} & {\color[HTML]{3531FF} -0.00702} & {\color[HTML]{3531FF} -0.00147} \\
Proportion of students  in compulsory school                                          & 0.00625                         & 0.43674                        & 0.86147                        & -0.00950                        & 0.02200                         \\
Proportion of students in vocational  secondary education                             & -0.01178                        & 0.06998                        & 0.62410                        & -0.02452                        & 0.00096                         \\
Proportion of students in general secondary education                                 & 0.00324                         & 0.12718                        & 0.63394                        & -0.00092                        & 0.00740                         \\
{\color[HTML]{3531FF} Proportion of students in  tertiary education}                  & {\color[HTML]{3531FF} -0.00322} & {\color[HTML]{3531FF} 0.02835} & {\color[HTML]{3531FF} 0.62377} & {\color[HTML]{3531FF} -0.00609} & {\color[HTML]{3531FF} -0.00034} \\
{\color[HTML]{3531FF} Proportion of 1-people households relative to total households} & {\color[HTML]{3531FF} -0.00503} & {\color[HTML]{3531FF} 0.00081} & {\color[HTML]{3531FF} 0.12782} & {\color[HTML]{3531FF} -0.00797} & {\color[HTML]{3531FF} -0.00209} \\
Proportion of 2-people households relative to total households                        & 0.00015                         & 0.92642                        & 0.87087                        & -0.00294                        & 0.00323                         \\
{\color[HTML]{3531FF} Proportion of 3-people households relative to total households} & {\color[HTML]{3531FF} 0.00456}  & {\color[HTML]{3531FF} 0.00000} & {\color[HTML]{3531FF} 0.34222} & {\color[HTML]{3531FF} 0.00300}  & {\color[HTML]{3531FF} 0.00613}  \\
Proportion of above 4-people households relative to total households                  & 0.00032                         & 0.76368                        & 0.17043                        & -0.00179                        & 0.00243                         \\
Proportion of 1/2 bedroom apartments over total housing                               & 0.00109                         & 0.51984                        & 0.31406                        & -0.00222                        & 0.00440                         \\
Proportion of small family apartments over total housing                              & 0.00158                         & 0.29773                        & 0.56248                        & -0.00139                        & 0.00454                         \\
Proportion of large family apartments over total housing                              & 0.00155                         & 0.26625                        & 0.73617                        & -0.00118                        & 0.00427                         \\
Proportion of mixed-use dwellings over total housing                                  & -0.00259                        & 0.12910                        & 0.15607                        & -0.00592                        & 0.00075                         \\
Proportion of births relative to total population                                     & 0.00040                         & 0.09390                        & 0.69619                        & -0.00007                        & 0.00086                         \\
{\color[HTML]{3531FF} Total number of births in the municipality (logged)}            & {\color[HTML]{3531FF} 0.06496}  & {\color[HTML]{3531FF} 0.01995} & {\color[HTML]{3531FF} 0.47659} & {\color[HTML]{3531FF} 0.01025}  & {\color[HTML]{3531FF} 0.11967} \\

\hline
\end{tabular}%
}
\label{tab:1pcinternational}
\end{table}

\begin{table}[H]
\caption{Average total effect of a 1-percentage-point increase in cumulative internal migration exposure on demographic, educational, and housing outcomes, 2010-2021. For each 1-percentage-point increase in exposure, the estimator sums the immediate and subsequent effects generated by that treatment change and averages them across switching municipalities 95\% confidence intervals in columns 4-5.}
\centering
\resizebox{\textwidth}{!}{%
\begin{tabular}{|l|l|l|l|l|l|}
\hline
  Outcome &Average Total Effect& $p$-value (ATE) &$p$-value (placebos)&    CI Lower&    CI Upper\\
\hline
Below 20-year-olds relative to total population                                  & 0.00086                         & 0.46277                        & 0.55361                        & -0.00143                        & 0.00315                         \\
Population aged 20 to 64 relative to total population                            & 0.00081                         & 0.54932                        & 0.83010                        & -0.00185                        & 0.00348                         \\
{\color[HTML]{3531FF} Population aged above 64 relative to total population}     & {\color[HTML]{3531FF} -0.00230} & {\color[HTML]{3531FF} 0.01874} & {\color[HTML]{3531FF} 0.55694} & {\color[HTML]{3531FF} -0.00422} & {\color[HTML]{3531FF} -0.00038} \\
{\color[HTML]{3531FF} Total students resident in the municipality (logged)}      & {\color[HTML]{3531FF} 0.01757}  & {\color[HTML]{3531FF} 0.01311} & {\color[HTML]{3531FF} 0.64367} & {\color[HTML]{3531FF} 0.00369}  & {\color[HTML]{3531FF} 0.03145}  \\
{\color[HTML]{3531FF} Proportion of students  in compulsory school}              & {\color[HTML]{3531FF} 0.00700}  & {\color[HTML]{3531FF} 0.04542} & {\color[HTML]{3531FF} 0.46627} & {\color[HTML]{3531FF} 0.00014}  & {\color[HTML]{3531FF} 0.01386}  \\
Proportion of students in vocational  secondary education & -0.00509 & 0.05939 & 0.07569 & -0.01038 &  0.00020 \\
{\color[HTML]{3531FF} Proportion of students in general secondary education}     & {\color[HTML]{3531FF} -0.00267} & {\color[HTML]{3531FF} 0.04592} & {\color[HTML]{3531FF} 0.73714} & {\color[HTML]{3531FF} -0.00529} & {\color[HTML]{3531FF} -0.00005} \\
Proportion of students in  tertiary education                                    & 0.00142                         & 0.18029                        & 0.39130                        & -0.00066                        & 0.00349                         \\
Proportion of 1-people households relative to total households                   & -0.00221                        & 0.09555                        & 0.95501                        & -0.00480                        & 0.00039                         \\
Proportion of 2-people households relative to total households                   & 0.00005                         & 0.97632                        & 0.80390                        & -0.00355                        & 0.00366                         \\
Proportion of 3-people households relative to total households                   & 0.00156                         & 0.24588                        & 0.06783                        & -0.00107                        & 0.00419                         \\
Proportion of above 4-people households relative to total households             & 0.00060                         & 0.67332                        & 0.70695                        & -0.00218                        & 0.00337                         \\
Proportion of 1/2 bedroom apartments over total housing                          & -0.00002                        & 0.97965                        & 0.38113                        & -0.00133                        & 0.00129                         \\
Proportion of small family apartments over total housing                         & 0.00181                         & 0.09160                        & 0.22986                        & -0.00029                        & 0.00391                         \\
Proportion of large family apartments over total housing                         & 0.00007                         & 0.94196                        & 0.47395                        & -0.00179                        & 0.00193                         \\
Proportion of single-detached dwellings over total housing                       & -0.00041                        & 0.46143                        & 0.82589                        & -0.00150                        & 0.00068                         \\
{\color[HTML]{3531FF} Proportion of mixed-use dwellings over total housing}      & {\color[HTML]{3531FF} -0.00145} & {\color[HTML]{3531FF} 0.02354} & {\color[HTML]{3531FF} 0.17684} & {\color[HTML]{3531FF} -0.00271} & {\color[HTML]{3531FF} -0.00020} \\
Proportion of births relative to total population                                & 0.00015                         & 0.56338                        & 0.28683                        & -0.00036                        & 0.00067 \\        
Total number of births in the municipality (logged)                              & 0.01993                         & 0.50546                        & 0.95968                        & -0.03874                        & 0.07861                         \\
 \hline
\end{tabular}%
}
\label{tab:1pcinternal}
\end{table}
\vspace{0.5em}
\noindent
\textbf{Demographic effects.}
International migration exposure generates several measurable demographic shifts. A 1-percentage-point (p.p.) increase in exposure is associated with a 6.5\% increase in births (even though with a large confidence interval ranging from 1.0\% to 12.0\%). Given this wide confidence interval, and the fact that absolute birth counts are small in many of the predominantly rural municipalities in our sample, this result should be interpreted with some caution. At the same time, exposure is linked to a 0.42 p.p. decline in the share of residents aged 65 and over (-0.00425; p = 0.003). For internal migration, the estimated demographic responses move in a similar direction but with smaller magnitudes: a 1 p.p. increase in exposure is associated with a 0.23 p.p. reduction in the population aged 65+ (-0.0023; p = 0.019). By contrast, effects on younger age groups are not statistically significant for either migration type, indicating that the demographic response is concentrated at older ages rather than reflecting a broad reshaping of the age distribution.

\vspace{0.5em}
\noindent
\textbf{Educational effects.}
The education-related outcomes also exhibit systematic responses to migration exposure. For international migration, a 1 p.p. increase in exposure leads to a 0.32 p.p. decline in the tertiary-education share (with the 95\% confidence interval ranging from $-0.0061$ to $-0.0003$).
Internal migration exposure mainly affects the school-aged population at compulsory and secondary levels. A 1 p.p. increase in internal exposure corresponds to a 0.70 p.p. increase in the compulsory-school share (0.00699; p $\simeq$ 0.046) and a 0.26 p.p. decline in the general-track secondary share ($-0.00264$; p $\simeq$ 0.048). In addition, the log number of students at residence rises by 1.76\% (0.0176; p $\simeq$ 0.013). Other education-related outcomes, such as the tertiary share under internal migration and vocational-track enrollment under international migration, do not exhibit statistically significant cumulative effects, suggesting that migration exposure reallocates students across educational tracks rather than expanding overall participation at higher levels.

\vspace{0.5em}
\noindent
\textbf{Housing and household composition.}
International migration exposure is associated with notable adjustments in household structure. A 1 p.p. rise in exposure leads to a 0.46 p.p. increase in three-person households (0.00456; p $<$ 0.0001), while the share of single-person households falls by 0.50 p.p. ($-0.00503$; p $\simeq$ 0.0008). These effects are tightly estimated, with confidence intervals comfortably excluding zero.
For internal migration, exposure changes correlate with small but significant shifts in the housing stock: the share of mixed-use dwellings decreases by 0.14 p.p. ($-0.00144$; p $\simeq$ 0.025).
For most other housing categories, including apartment size distributions and detached housing, estimated effects are small and statistically indistinguishable from zero, implying that migration exposure, in particular to international migration, affects household composition more strongly than the physical housing stock.

\section{Discussion and Policy Relevance}
\label{discussion}
The results presented above suggest that migration (both international and internal) acts as a counterweight to population aging in the canton of Fribourg. 

While several outcomes exhibit statistically precise effects, many others remain statistically indistinguishable from zero, suggesting that migration exposure reshapes a subset of demographic and socio-economic margins rather than producing broad-based changes across all indicators. This section offers an interpretation and discussion on these findings and outlines potential implications for cantonal and municipal policy.

\subsection{Demographic mechanisms}
Across both types of migration, the results point to a clear rejuvenating effect on municipalities' demographic structure. A 1-percentage-point (p.p.) increase in international migration exposure reduces the share of residents aged 65 and over by about 0.42 p.p., while an equivalent increase in internal migration exposure lowers it by roughly 0.23 p.p. These declines reflect that migration inflows tend to introduce comparatively younger population groups into municipalities.
International migration also coincides with a notable rise in local fertility: a 1 p.p. increase in exposure is associated with a 6.5\% increase in births, whereas municipalities with unchanged exposure show no comparable change. This suggests that international inflows reshape not only the age distribution but also the population growth, through higher birth counts.
At the same time, effects on some related demographic margins (e.g., youth shares) are not statistically significant, indicating that the demographic response is concentrated rather than uniform across the age distribution.

These average effects are consistent with the dynamic profiles shown in the event-study graphs. Pre-treatment coefficients remain close to zero, indicating stable pre-trends, while post-treatment effects appear immediately after the first exposure change and maintain a fairly stable magnitude over the observed horizons. There is little evidence of reversals or attenuation, and the estimates do not drift markedly as additional years of exposure accumulate. This pattern supports the view that the demographic adjustments (both the reduction in the elderly share and the increase in births) are persistent rather than short-lived.
Taken together, the decline in the elderly share and the rise in births point to a robust even if smaller in magnitude demographic pattern: migration inflows, particularly international ones, lead to a younger population profile, with effects that appear early and remain relatively stable over time.

\subsection{Education implications}
The educational outcomes also display systematic responses to migration exposure, with clear and stable patterns in the event-study profiles. Internal migration exposure is associated with increases in the school-age population: a 1-percentage-point (p.p.) increase raises the log number of students at residence by 1.76\%, and the share of students in compulsory schooling by 0.70 p.p. At the same time, the share of students in the general-track secondary system declines by 0.26 p.p. These effects suggest compositional adjustments in the local student population following internal inflows.
International migration shows a different structure of adjustments, with a modest decline in the tertiary-education share (-0.32 p.p.). The data do not allow us to distinguish between two possible mechanisms behind this decline: incoming migrants' children may join compulsory schools and thereby mechanically dilute the tertiary share, or the arrival of international migrants may coincide with an outflow of highly educated incumbent residents. These mechanisms have different policy implications and distinguishing between them would require individual-level data beyond the scope of this study.

Other education outcomes show weaker or statistically inconclusive responses, which is consistent with migration affecting the composition of enrollment across tracks more than generating broad changes in overall participation rates.

The dynamic graphs support the credibility and persistence of these effects. Placebo (pre-treatment) coefficients remain close to zero across all school-related outcomes, indicating no differential trends before exposure changes occur. Post-treatment, the effects emerge immediately or within the first two periods and remain relatively stable over the available exposure window. 

\subsection{Housing-market adjustments}

Migration exposure also induces measurable adjustments in household structure and the composition of the housing stock. The most precisely estimated housing-related responses concern household composition, while several housing-stock categories exhibit small and statistically inconclusive effects.
For international migration, a 1-p.p. increase in exposure leads to a 0.46 p.p. increase in three-person households and a 0.50 p.p. decline in single-person households, indicating a shift toward larger household types. For internal migration, the share of mixed-use dwellings decreases by 0.14 p.p., reflecting changes in how residential and hybrid spaces are allocated in municipalities experiencing sustained inflows.
The event-study graphs shed light on the timing and persistence of these changes. Pre-treatment coefficients cluster around zero, suggesting stable pre-trends in housing and household outcomes. For international migration, the expansion of three-person households and the decline in single-person households show smooth, sustained profiles with little evidence of oscillation or attenuation. Internal migration's impact on mixed-use dwellings is similarly stable, with the event-study estimates showing a consistent negative pattern that does not reverse as additional years of exposure accumulate.
This combination of minimal pre-trend movement, immediate onset, and stable post-treatment trajectories suggests that housing-market and household-structure adjustments are persistent responses to migration exposure rather than short-term fluctuations. The magnitude of the effects is modest in absolute terms but systematic across outcomes, reinforcing the idea that migration incrementally but durably reshapes residential demand and the composition of occupied housing units.
\subsection{Policy relevance for the canton of Fribourg}
The evidence points to several policy domains in which sustained migration exposure has direct relevance for the canton of Fribourg. Migration—both internal and international—appears to contribute to a younger population structure, modest shifts in the composition of student cohorts, and gradual adjustments in housing demand. These patterns unfold relatively quickly after exposure changes and remain stable over time, suggesting that municipalities may need to plan for persistent demographic and structural adjustments rather than temporary fluctuations. Three areas stand out:

\begin{enumerate}
    \item \textbf{Demographic sustainability.}  
The canton faces long-term demographic pressures linked to population aging, and the results indicate that migration flows help counteract this trend. Municipalities experiencing increases in migration exposure, especially international inflows, see reductions in the elderly share and increases in births. While the magnitudes are modest at any single point in time, their persistence and consistency suggest that migration contributes to stabilizing the age distribution. For municipalities with shrinking or aging populations, these inflows may ease pressures on social services, local labor supply, and community viability, although the uneven spatial distribution of migrants implies that not all localities benefit equally.
    \item \textbf{Education investments and integration.}  
Internal migration inflows expand the compulsory-school population over relatively short horizons, while international migration subtly shifts the distribution of students across post-compulsory tracks. These effects do not dissipate quickly, implying that municipalities and the canton may need to anticipate continued demand for compulsory-school capacity, orientation services, and integration support. The rise in student numbers associated with internal migration suggests that growing municipalities may face renewed pressure to adjust classroom capacity, staffing, and infrastructure, whereas municipalities seeing increased international exposure may need to tailor support programs for students arriving at different stages of the educational path.
    \item \textbf{Housing and spatial planning.}  
Migration exposure reshapes household composition and affects the use of local housing stock. International migration shifts demand toward larger household types, while internal migration reduces the prevalence of mixed-use dwellings. These adjustments appear steadily over time and remain stable, suggesting that municipalities may need to account for changes in housing needs—both in terms of unit size and the balance between residential and mixed-use zoning. For urbanizing areas or municipalities undergoing redevelopment, these patterns may inform decisions on building density, land-use allocation, and the planning of family-oriented housing.   Under the cantonal \textit{LATeC} (\href{https://bdlf.fr.ch/app/fr/texts_of_law/710.1}{Loi sur l'aménagement du territoire et les constructions}), communes retain zoning autonomy within cantonal guidelines. 
Thus, encouraging mid-size rental construction and adapting local plans is administratively feasible.
\end{enumerate}

\subsection{Broader implications}

Beyond Fribourg, these findings illustrate how cumulative migration exposure can
reshape local communities in a decentralized federal context.  Because effects
emerge gradually yet persistently, local governments benefit from forward-looking
planning that recognizes migration as a structural driver of demographic and
social change rather than a transient shock. The intertemporal difference-in-differences design used here captures precisely this long-run dimension, showing that migration alters the composition and spatial organization of local societies.
In sum, migration in the canton of Fribourg has acted less as a disruptive
impulse and more as a slow-moving engine of demographic renewal.  Recognizing
this cumulative nature is key for evidence-based policies in housing, education,
and demographic sustainability.
It is important to note that all estimates in this paper are identified from municipalities experiencing net in-migration (switchers-in), as this is the case for most municipalities in the Canton of Fribourg. The dynamics in declining or emigration-affected municipalities may differ substantially, and the policy conclusions drawn here should not be extrapolated to that context without further evidence.

\section{Conclusions}
\label{conclusions}
This study quantifies how cumulative migration exposure shapes demographic, educational, and housing outcomes across municipalities in the canton of Fribourg between 2010 and 2021. Using the intertemporal Difference-in-Differences estimator of \textcite{de_chaisemartin_difference--differences_2024}, we estimate the average effect of a one-percentage-point increase in cumulative migration balance relative to baseline population.
Across the broader set of outcomes considered, many estimated effects are small and not statistically distinguishable from zero, implying that migration exposure does not generate uniform changes across all demographic, education, and housing indicators.
For both international and internal migration exposure, the main demographic finding is a modest but systematic reduction in the share of elderly residents. A 1 p.p.\ increase in exposure lowers the population aged 65+ by roughly 0.42 p.p.\ for international migration and 0.23 p.p.\ for internal migration. For international migration, we also detect a positive effect on local births: a 1 p.p.\ increase in exposure raises the total number of births by about 6.5\%, with confidence intervals indicating statistically significant but imprecisely estimate. These demographic shifts are consistent with migration-driven compositional renewal rather than broad-based changes across all demographic margins.

Education outcomes respond differently to internal and international flows. Internal migration increases the number of resident students and raises the share in compulsory schooling, while reducing the share enrolled in general-track upper secondary programs. International migration is instead associated with a small decline in the tertiary-education share. These effects emerge quickly after the first treatment change and persist over longer exposure horizons, suggesting that migration affects not only the size of local student populations but also their composition.

Housing and household structure also adjust in measurable ways. International migration increases the prevalence of three-person households and reduces single-person households, indicating a shift toward larger household types. Internal migration leads to a reduction in the share of mixed-use dwellings, consistent with gradual reallocation of the housing stock in growing municipalities. As in the demographic and education domains, these adjustments are modest in magnitude but stable across exposure durations.

Taken together, the results show that migration in the canton of Fribourg acts less as a short-term shock and more as a slow-moving mechanism of demographic and social adjustment. Although yearly changes are small, their cumulative impact leads to persistent shifts in age structure, school enrollment, and housing demand. The observed effects are estimated over a 12-year window and therefore capture short- to medium-run dynamics; longer-run reversals, for instance as migrant cohorts age in place and themselves contribute to the elderly share, cannot be excluded and would require extended data to assess. For municipal and cantonal policymakers, these findings underscore the importance of incorporating migration exposure into long-run planning for education, spatial development, and demographic sustainability. Methodologically, the study demonstrates the usefulness of intertemporal DiD estimators for analysing cumulative, non-binary treatments in decentralized settings such as Switzerland.

\section{Acknowledgments}
The author thanks Thomas Christin (Head of the Statistics and Data Service (SSD) at the Department of Economy, Employment, and Vocational Training for the Etat de Fribourg) for his help in acquiring and processing the data, Sarina Joy Oberhaensli  for her initial contribution to the data retrieval and Martin Huber for the continuous support and supervision. 

Finally, the author made use of an AI tool (Claude, developed by Anthropic) 
to support her work. The tool was not employed in any creative capacity, 
but solely as an aid for scientific writing and coding. Specifically, 
it was used to rephrase and improve the clarity of author-written text, 
and to comment and optimize author-written code. All intellectual content, 
arguments, and results remain entirely the author's own. 
\printbibliography

\appendix

\section{Robustness checks}
\subsection{Results of Average Total Effects for different model specification}
\label{app:model specifications}
For brevity, each table reports only outcomes that reach statistical significance under that specification; the full set of outcomes is available from the author upon request.

\paragraph{Basic model: no covariates, no clustering}
\begin{table}[H]\centering
\caption{Average total effect of a 1-percentage-point increase in cumulative international migration exposure on demographic, educational, and housing outcomes, 2010-2021. Model specification excluding covariates and without clustering.}

\resizebox{\textwidth}{!}{%
\begin{tabular}{l*{5}{c}}
\hline\hline
            &Average Total Effect& P-Value ATE&P-Value Placebo&    CI Lower&    CI Upper\\
\hline
Total number of births in the municipality (logged)&    .0701582&    .0264669&    .3911687&    .0081969&    .1321195\\
Proportion of students in tertiary education&   -.0033289&    .0227334&    .7425608&   -.0061933&   -.0004646\\
Proportion of 3-people households relative to total households&    .0042861&    .00000033&     .485025&    .0026409&    .0059313\\
Proportion of 1-people households relative to total households&   -.0050843&    .0026216&    .1212438&   -.0083962&   -.0017724\\
\hline\hline
\end{tabular}}
\end{table}
\begin{table}[H]\centering
\caption{Average total effect of a 1-percentage-point increase in cumulative internal migration exposure on demographic, educational, and housing outcomes, 2010-2021.  Model specification excluding covariates and without clustering.}
\resizebox{\textwidth}{!}{%
\begin{tabular}{l*{5}{c}}
\hline\hline
            &Average Total Effect& P-Value ATE&P-Value Placebo&    CI Lower&    CI Upper\\
\hline
Population aged above 65 relative to total population&   -.0020548&    .0144231&    .7393683&   -.0037009&   -.0004086\\
Total students resident in the municipality (logged)&    .0130234&    .0249194&    .7658219&    .0016414&    .0244054\\
Proportion of mixed-use dwellings over total housing   &   -.0013635&    .0164438&    .3669046&   -.0024775&   -.0002495\\
\hline\hline
\end{tabular}}
\end{table}

\paragraph{Including within migration as covariate}
\begin{table}[H]\centering
\caption{Average total effect of a 1-percentage-point increase in cumulative international migration exposure on demographic, educational, and housing outcomes, 2010-2021. Model specification including controlling for within-migration and without clustering.}\resizebox{\textwidth}{!}{%
\begin{tabular}{l*{5}{c}}
\hline\hline
            &Average Total Effect& P-Value ATE&P-Value Placebo&    CI Lower&    CI Upper\\
\hline
Total number of births in the municipality (logged)&    .0694943&    .0311967&    .2998506&    .0062753&    .1327133\\
Proportion of students in tertiary education&   -.0033395&     .024105&     .735672&   -.0062416&   -.0004375\\
Proportion of 3-people households relative to total households&    .0043803&    .0000004&    .4447305&    .0026801&    .0060804\\
Proportion of 1-people households relative to total households&   -.0050989&    .0016145&    .2863476&   -.0082682&   -.0019295\\
\hline\hline
\end{tabular}}
\end{table}

\begin{table}[H]\centering
\caption{Average total effect of a 1-percentage-point increase in cumulative internal migration exposure on demographic, educational, and housing outcomes, 2010-2021. Model specification including controlling for within-migration and without clustering.}\resizebox{\textwidth}{!}{%
\begin{tabular}{l*{5}{c}}
\hline\hline
            &Average Total Effect& P-Value ATE&P-Value Placebo&    CI Lower&    CI Upper\\
\hline
Population aged above 65 relative to total population&   -.0020185&    .0162493&    .6729924&   -.0036647&   -.0003723\\
Total students resident in the municipality (logged)&    .0122989&    .0302211&    .5403546&    .0011758&     .023422\\
Proportion of mixed-use dwellings over total housing   &   -.0014029&    .0136518&    .3782824&   -.0025178&    -.000288\\
\hline\hline
\end{tabular}}
\end{table}

\paragraph{Including clustering and quadratic covariate}
\begin{table}[H]\centering
\caption{Average total effect of a 1-percentage-point increase in cumulative international migration exposure on demographic, educational, and housing outcomes, 2010-2021. Model specification with quadratic covariates.}
\resizebox{\textwidth}{!}{%
\begin{tabular}{l*{5}{c}}
\hline\hline
            &Average Total Effect& P-Value ATE&P-Value Placebo&    CI Lower&    CI Upper\\
\hline
Total number of births in the municipality (logged)& 0.065191321 & 1.700923e-02 & 0.54891133& 0.011650757 & 0.1187318861\\
Proportion of students in tertiary education & -0.003181336 & 2.999594e-02 & 0.58049059& -0.006054610 & -0.0003080618 \\
Proportion of 3-people households relative to total households& 0.004504507 & 1.196017e-08 & 0.38478452 & 0.002955669 & 0.0060533462 \\
Population aged above 65 relative to total population& -0.004280320 & 1.781810e-03 & 0.07002388 & -0.006965471 & -0.0015951694\\
\hline\hline
\end{tabular}}
\end{table}
\begin{table}[H]\centering
\caption{Average total effect of a 1-percentage-point increase in cumulative internal migration exposure on demographic, educational, and housing outcomes, 2010-2021.  Model specification with quadratic covariates.}
\resizebox{\textwidth}{!}{%
\begin{tabular}{l*{5}{c}}
\hline\hline
            &Average Total Effect& P-Value ATE&P-Value Placebo&    CI Lower&    CI Upper\\
\hline
Population aged above 65 relative to total population & -0.002261422 & 0.02339722 & 0.52858645 & -0.0042166975 & -3.061459e-04\\
Total students resident in the municipality (logged)& 0.018273689& 0.01280376& 0.64336532 & 0.0038848505& 3.266253e-02\\
Proportion of students in compulsory school& 0.007760458& 0.03317098& 0.45024335& 0.0006194049& 1.490151e-02\\
Proportion of students in general secondary education& -0.003301101& 0.01870247& 0.73705554& -0.0060527097& -5.494918e-04\\
Proportion of students in vocational secondary education& -0.005394181& 0.04696162& 0.06703363& -0.0107160117& -7.235015e-05\\
Proportion of mixed-use dwellings over total housing  & -0.001437657& 0.02451730& 0.17223944& -0.0026906154& -1.846991e-04 \\
\hline\hline
\end{tabular}}
\end{table}
\subsection{Results of Average Total Effects for alternative treatment binning}
\label{app:alt_binning}

A key step in our empirical design is the discretization of cumulative migration exposure into bins of equal width. 
In the baseline specification, we use 1-percentage-point (p.p.) bins, so that each treatment ``increment'' corresponds to crossing a 1 p.p.\ threshold in cumulative exposure.
While this definition is transparent and easy to interpret, it raises two natural concerns.
First, estimated effects might be sensitive to the choice of bin width rather than reflecting underlying treatment dynamics.
Second, municipalities may differ substantially in how quickly they accumulate exposure, so that the timing and size of treatment increments depend on the binning choice.

To assess the robustness of our results to these issues, we re-estimate the average total effect for a range of alternative bin widths:
\[
\text{Binwidth} \in \{1, 2, 5, 10\}\,\text{p.p.}
\]
Tables \ref{tab:bin_internal} and \ref{tab:bin_international} report the resulting estimates for internal and international migration exposure, respectively.

Changing the bin width affects the structure of the treatment variable in systematic ways. 
Finer binning (e.g.\ 1 or 2 p.p.) makes the treatment more responsive to small year-to-year changes in migration exposure and increases the number of observed treatment increments.
Coarser binning (5 or 10 p.p.) aggregates several increments into a single threshold crossing and therefore targets the effect of larger cumulative changes in exposure, but with fewer switching events.
Naturally, coarser bins also reduce the number of municipalities that contribute to each estimate, which can widen confidence intervals and alter which outcomes are precisely estimated.

Robustness of the results across different bin widths therefore strengthens the interpretation that the estimated effects reflect underlying demographic, educational, and housing adjustments rather than artifacts of the binning scheme.

\subsubsection*{Internal migration exposure}

For internal migration exposure, the estimates are remarkably consistent across bin widths.
The age structure responds similarly in all specifications: higher exposure reduces the share of residents aged 65+, with effect sizes ranging from about $-0.0023$ (1 p.p.\ bins) to $-0.0007$ (10 p.p.\ bins).
The student population also reacts in a stable way.
The effect on log students at residence is positive at all bin widths, although its magnitude naturally declines as the bin size increases (from roughly 0.0176 with 1 p.p.\ bins to 0.0032 with 10 p.p.\ bins), reflecting the fact that a larger bin corresponds to a larger treatment increment.

Patterns in household composition remain stable as well.
Three-person households tend to increase with internal migration exposure, while mixed-use dwellings decline.
At coarser binning (5 and 10 p.p.), additional outcomes become precisely estimated (e.g.\ vocational-track students, detached houses, transient (1 or 2 rooms) apartments), but their signs and orders of magnitude are consistent with the baseline specification.

Importantly, placebo $p$-values remain large across all bin widths, indicating no evidence of problematic pre-trends even under alternative discretizations.

\subsubsection*{International migration exposure}

The international migration results exhibit the same pattern of stability.
The rejuvenating demographic effects are present under all bin widths: the effect on the elderly share is negative and of similar magnitude across specifications (around $-0.0043$ with 1 p.p.\ bins and $-0.0036$ with 2 p.p.\ bins, and still negative with 10 p.p.\ bins).
The fertility response, a central result of the paper, is also robust.
The estimated average total effect on total births is about 0.065 under 1 p.p.\ bins and about 0.063 under 2 p.p.\ bins, with overlapping confidence intervals.

Household composition effects are likewise stable.
International migration is associated with fewer single-person households and more three-person households in the 1 p.p.\ specification, and the signs and relative magnitudes of these effects are preserved when moving to coarser bins.
At 2 and 5 p.p.\ binning, additional outcomes (e.g.\ vocational-track students, transient units, the birth rate relative to population) become precisely estimated, but their estimated effects are consistent with the baseline patterns.

Again, placebo $p$-values are comfortably large across all specifications, suggesting stable pre-treatment trajectories for the outcomes considered.

\subsubsection*{Overall interpretation}

Across both internal and international migration, the results are highly robust to alternative discretizations of cumulative exposure.
First, the signs of the estimated average total effects remain unchanged across all bin widths.
Second, effect magnitudes scale in a plausible way with the size of the bin: when treatment increments correspond to larger changes in exposure, the coefficient per 1 p.p.\ of exposure naturally becomes smaller.
Third, placebo tests consistently fail to reject the null of no pre-trends, even under alternative binning rules.

Taken together, these robustness checks indicate that the main conclusions of the paper, namely, that migration exposure rejuvenates the population, increases births, and induces systematic adjustments in student and household composition, are not artifacts of the specific binning choice used in the baseline specification.
Instead, they appear to reflect underlying responses to sustained changes in cumulative migration exposure.

\begin{table}[H]
\centering
\caption{Internal migration exposure: results for different binning}
\centering
\resizebox{\ifdim\width>\linewidth\linewidth\else\width\fi}{!}{
\begin{tabular}[t]{lcccccc}
\toprule
\multicolumn{1}{c}{ } & \multicolumn{6}{c}{} \\
\cmidrule(l{3pt}r{3pt}){2-7}
Outcome & Average Total Effect & P-Value ATE & P-Value Placebo & CI Lower & CI Upper & Binwidth\\
\midrule
Population aged above 65 relative to total population & -0.00230 & 0.01857 & 0.55649 & -0.00422 & -0.00039 & 1\\
Total students resident in the municipality (logged) & 0.01764 & 0.01273 & 0.63457 & 0.00376 & 0.03151 & 1\\
Proportion of students in compulsory school & 0.00700 & 0.04553 & 0.46613 & 0.00014 & 0.01386 & 1\\
Proportion of students in general secondary education & -0.00264 & 0.04802 & 0.74828 & -0.00526 & -0.00002 & 1\\
Proportion of mixed-use dwellings over total housing & -0.00144 & 0.02462 & 0.17554 & -0.00269 & -0.00018 & 1\\
\addlinespace
Total students resident in the municipality (logged) & 0.01150 & 0.00223 & 0.18168 & 0.00413 & 0.01888 & 2\\
Proportion of small family apartments over total housing & 0.00158 & 0.00887 & 0.53379 & 0.00040 & 0.00277 & 2\\
Proportion of mixed-use dwellings over total housing & -0.00140 & 0.00011 & 0.14601 & -0.00211 & -0.00069 & 2\\
\addlinespace
Below 20-year-olds relative to total population & 0.00055 & 0.02597 & 0.56429 & 0.00007 & 0.00104 & 5\\
Proportion of students in vocational secondary education & -0.00189 & 0.00308 & 0.10903 & -0.00315 & -0.00064 & 5\\
Proportion of 3-people households relative to total households & 0.00099 & 0.00091 & 0.31021 & 0.00040 & 0.00157 & 5\\
Proportion of 2-people households relative to total households & -0.00132 & 0.00260 & 0.15970 & -0.00217 & -0.00046 & 5\\
Proportion of single-detached dwellings over total housing & -0.00058 & 0.00831 & 0.86419 & -0.00102 & -0.00015 & 5\\
Proportion of mixed-use dwellings over total housing & -0.00051 & 0.02954 & 0.12828 & -0.00098 & -0.00005 & 5\\
\addlinespace
Population aged 20 to 64 relative to total population & 0.00052 & 0.01417 & 0.58922 & 0.00010 & 0.00094 & 10\\
Population aged above 65 relative to total population & -0.00069 & 0.00112 & 0.11394 & -0.00110 & -0.00027 & 10\\
Total students resident in the municipality (logged) & 0.00323 & 0.01190 & 0.07116 & 0.00071 & 0.00574 & 10\\
Proportion of 3-people households relative to total households & 0.00066 & 0.01947 & 0.58646 & 0.00011 & 0.00121 & 10\\
Proportion of 2-people households relative to total households & -0.00071 & 0.00446 & 0.25814 & -0.00120 & -0.00022 & 10\\
Proportion of 1/2 bedroom apartments over total housing & 0.00030 & 0.01318 & 0.43443 & 0.00006 & 0.00054 & 10\\
\bottomrule
\end{tabular}
}
\label{tab:bin_internal}
\end{table}
\begin{table}[H]
\centering
\caption{International migration exposure: results for different binning}
\centering
\resizebox{\ifdim\width>\linewidth\linewidth\else\width\fi}{!}{
\begin{tabular}[t]{lcccccc}
\toprule
\multicolumn{1}{c}{ } & \multicolumn{6}{c}{} \\
\cmidrule(l{3pt}r{3pt}){2-7}
Outcome & Average Total Effect & P-Value ATE & P-Value Placebo & CI Lower & CI Upper & Binwidth\\
\midrule
Population aged above 65 relative to total population & -0.00425 & 0.00274 & 0.10289 & -0.00702 & -0.00147 & 1\\
Proportion of students in tertiary education & -0.00322 & 0.02835 & 0.62377 & -0.00609 & -0.00034 & 1\\
Proportion of 3-people households relative to total households & 0.00456 & 0.00000 & 0.34222 & 0.00300 & 0.00613 & 1\\
Proportion of 1-people households relative to total households & -0.00503 & 0.00081 & 0.12782 & -0.00797 & -0.00209 & 1\\
Total number of births in the municipality (logged) & 0.06496 & 0.01995 & 0.47659 & 0.01025 & 0.11967 & 1\\
\addlinespace
Population aged above 65 relative to total population & -0.00365 & 0.00000 & 0.06981 & -0.00518 & -0.00213 & 2\\
Total students resident in the municipality (logged) & 0.01324 & 0.04206 & 0.49344 & 0.00048 & 0.02601 & 2\\
Proportion of students in vocational secondary education & -0.00484 & 0.02940 & 0.30353 & -0.00919 & -0.00048 & 2\\
Proportion of students in tertiary education& -0.00226 & 0.00270 & 0.47403 & -0.00373 & -0.00078 & 2\\
Proportion of 1/2 bedroom apartments over total housing & 0.00112 & 0.00710 & 0.13406 & 0.00030 & 0.00193 & 2\\
Total number of births in the municipality (logged) & 0.06312 & 0.00012 & 0.87655 & 0.03087 & 0.09536 & 2\\
Proportion of births relative to total population & 0.00051 & 0.00195 & 0.92613 & 0.00019 & 0.00084 & 2\\
\addlinespace
Below 20-year-olds relative to total population & 0.00071 & 0.04899 & 0.28248 & 0.00000 & 0.00141 & 5\\
Total students resident in the municipality (logged) & 0.00727 & 0.01753 & 0.21437 & 0.00127 & 0.01326 & 5\\
\bottomrule
\end{tabular}}
\label{tab:bin_international}
\end{table}

\subsection{Results of Effects for Selection on Observables  method}
\label{app:soo}

A standard Selection on Observables  comparison using never-treated municipalities as controls is not feasible here, since virtually all municipalities experienced at least one upward shift in migration exposure over 2010-2021. We therefore compare newly-treated municipalities against not-yet-treated ones at the moment of first switching, controlling for the lagged outcome, pre-2009 municipal characteristics, and cohort fixed effects. Standard errors are clustered at the municipality level. The resulting estimate captures the average cross-sectional difference at first switching  (a different parameter than the cumulative dynamic effect in the main analysis) so the appropriate comparison is one of sign and direction only, not magnitude. Results are reported in Table~\ref{tab:soo_results}. For international migration, five of the six significant main results are directionally confirmed, and four reach conventional significance thresholds: the share of residents aged 65+ ($\hat{\beta}= - 0.00345,p<0.001$), single-person households ($\hat{\beta} = - 0.00432 , p=0.021$), three-person households ($\hat{\beta} = 0.00384, 
p=0.031$), and tertiary students ($\hat{\beta} = - 0.00324, p=0.081$). Total births is the only exception, returning a negative and insignificant estimate; given the small absolute birth counts in predominantly rural municipalities and the wide confidence interval already noted in the main text, we do not view this as evidence against the DiD result. For internal migration, the Selection on Observables estimates are generally weaker but directionally consistent: total students at residence ($\hat{\beta} = 0.016, 
p=0.062$) and the compulsory school share ($\hat{\beta} = 0.010,
p=0.015$) are both confirmed in sign and significance. Overall, this robustness check supports the main findings, particularly for the demographic and household composition outcomes at the core of the paper.

\begin{table}[H]
\caption{Selection-on-observables robustness check. OLS estimates comparing 
newly-treated municipalities against not-yet-treated municipalities at the 
moment of first upward treatment switch, controlling for the lagged outcome, 
pre-2009 covariates, and cohort fixed effects. SE clustered at municipality 
level. Outcomes significant in the main analysis are highlighted in blue.}
\centering
\resizebox{\textwidth}{!}{%
\begin{tabular}{lrrr}
\hline\hline Outcome & Average Effect & SE & $p$-value  \\
\hline
\multicolumn{4}{l}{\textit{Panel A: International migration}} \\
Below 20-year-olds relative to total population                                       & 0.00297  & 0.00125 & 0.01971 \\
Population aged 20 to 64 relative to total population                                 & -0.00003 & 0.00158 & 0.98730 \\
{\color[HTML]{3531FF} Population aged above 64 relative to total population}          & -0.00345 & 0.00096 & 0.00050 \\
Total students resident in the municipality (logged)                                  & 0.00604  & 0.01036 & 0.56089 \\
Proportion of students  in compulsory school                                          & 0.01599  & 0.00498 & 0.00173 \\
Proportion of students in vocational  secondary education                             & -0.01650 & 0.00362 & 0.00001 \\
Proportion of students in general secondary education                                 & 0.00083  & 0.00266 & 0.75689 \\
{\color[HTML]{3531FF} Proportion of students in  tertiary education}                  & -0.00324 & 0.00184 & 0.08052 \\
{\color[HTML]{3531FF} Proportion of 1-people households relative to total households} & -0.00432 & 0.00184 & 0.02059 \\
Proportion of 2-people households relative to total households                        & -0.00416 & 0.00211 & 0.05113 \\
{\color[HTML]{3531FF} Proportion of 3-people households relative to total households} & 0.00384  & 0.00176 & 0.03120 \\
Proportion of above 4-people households relative to total households                  & 0.00320  & 0.00150 & 0.03600 \\
Proportion of 1/2 bedroom apartments over total housing                               & 0.00036  & 0.00084 & 0.67007 \\
Proportion of small family apartments over total housing                              & -0.00009 & 0.00113 & 0.93479 \\
Proportion of large family apartments over total housing                              & 0.00157  & 0.00103 & 0.12881 \\
Proportion of single-detached dwellings over total housing                            & -0.00060 & 0.00057 & 0.29585 \\
Proportion of mixed-use dwellings over total housing                                  & -0.00066 & 0.00093 & 0.47645 \\
Proportion of births relative to total population                                     & 0.00067  & 0.00050 & 0.18436 \\
{\color[HTML]{3531FF} Total number of births in the municipality (logged)}            & -0.07533 & 0.05957 & 0.20880\\
\hline
\multicolumn{4}{l}{\textit{Panel B: Internal migration}} \\
Below 20-year-olds relative to total population                                                      & 0.00098        & 0.00096 & 0.30691 \\
Population aged 20 to 64 relative to total population                                                & -0.00015       & 0.00103 & 0.88466 \\
{\color[HTML]{3531FF} Population aged above 64 relative to total population}                         & -0.00135       & 0.00092 & 0.14657 \\
{\color[HTML]{3531FF} Total students resident in the municipality (logged)}                          & 0.01616        & 0.00855 & 0.06167 \\
{\color[HTML]{3531FF} Proportion of students  in compulsory school}                                  & 0.00962        & 0.00390 & 0.01523 \\
Proportion of students in vocational  secondary education                                            & -0.00150       & 0.00392 & 0.70176 \\
{\color[HTML]{3531FF} Proportion of students in general secondary education} & 0.00014        & 0.00230 & 0.95279 \\
Proportion of students in  tertiary education                                                        & -0.00109       & 0.00159 & 0.49522 \\
Proportion of 1-people households relative to total households                                       & -0.00169       & 0.00179 & 0.34673 \\
Proportion of 2-people households relative to total households                                       & 0.00154        & 0.00194 & 0.42809 \\
Proportion of 3-people households relative to total households                                       & -0.00150       & 0.00184 & 0.41554 \\
Proportion of above 4-people households relative to total households                                 & 0.00169        & 0.00171 & 0.32672 \\
Proportion of 1/2 bedroom apartments over total housing                                              & -0.00128       & 0.00071 & 0.07443 \\
Proportion of small family apartments over total housing                                             & -0.00129       & 0.00141 & 0.36451 \\
Proportion of large family apartments over total housing                                             & 0.00231        & 0.00126 & 0.07050 \\
Proportion of single-detached dwellings over total housing                                           & 0.00029        & 0.00063 & 0.64083 \\
{\color[HTML]{3531FF} Proportion of mixed-use dwellings over total housing}                          & -0.00046       & 0.00049 & 0.35425 \\
Proportion of births relative to total population                                                    & 0.00034        & 0.00043 & 0.43018 \\
Total number of births in the municipality (logged)                                                  & -0.09213       & 0.06337 & 0.14899\\
\hline\hline
\multicolumn{4}{p{20cm}}{\footnotesize \textit{Notes:} Outcomes highlighted 
in blue are those significant in the main DiD analysis (Tables 
\ref{tab:1pcinternational}-\ref{tab:1pcinternal}). The SOO estimate 
captures the average cross-sectional difference at the moment of first 
treatment switching and is not directly comparable in magnitude to the 
cumulative DiD estimates in the main tables; sign and direction are the 
appropriate basis for comparison.}
\end{tabular}}
\label{tab:soo_results}
\end{table}

\section{Additional tables and graphs}
\subsection{Event studies for non-significant effects}
\label{app:event_non_sign}

\subsubsection{International migration}
\begin{figure}[H]
    \centering
    \begin{subfigure}[b]{0.45\textwidth}
        \centering
        \includegraphics[width=\textwidth]{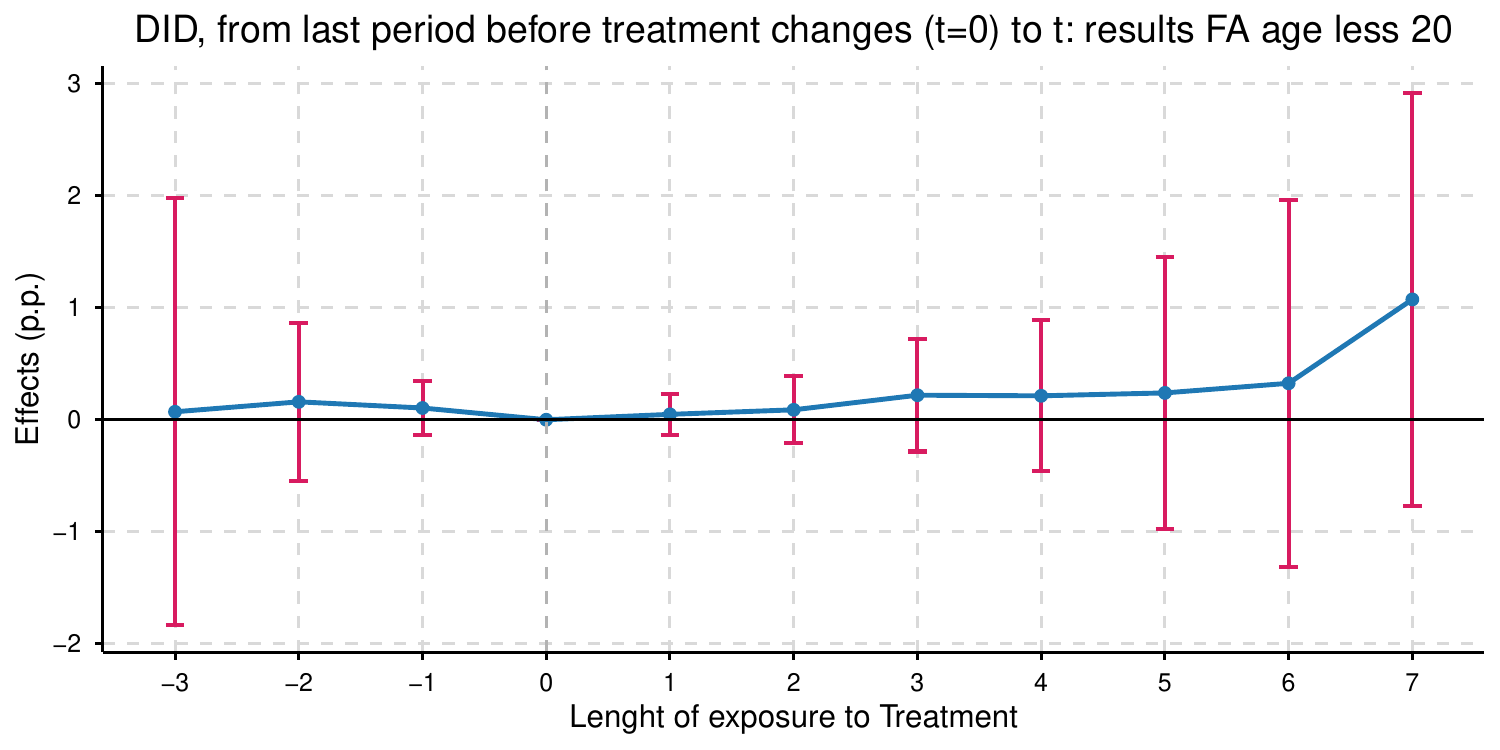}
        \caption{Proportion of population aged $<$ 20 years old}
    \end{subfigure}
    \hfill
     \begin{subfigure}[b]{0.45\textwidth}
        \centering
        \includegraphics[width=\textwidth]{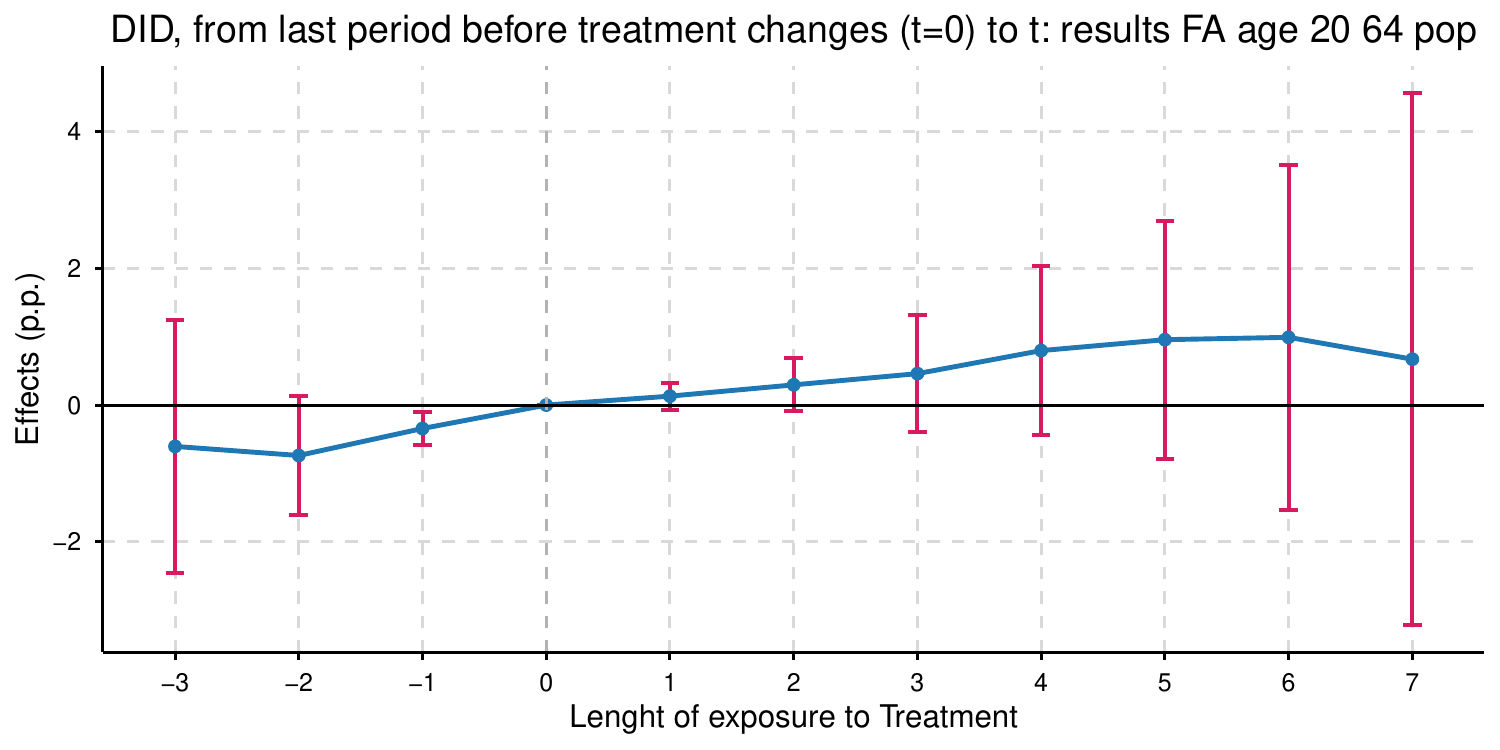}
        \caption{Population aged 20 to 64 relative to total population}
    \end{subfigure}
         \vskip\baselineskip
          \begin{subfigure}[b]{0.45\textwidth}
        \centering
        \includegraphics[width=\textwidth]{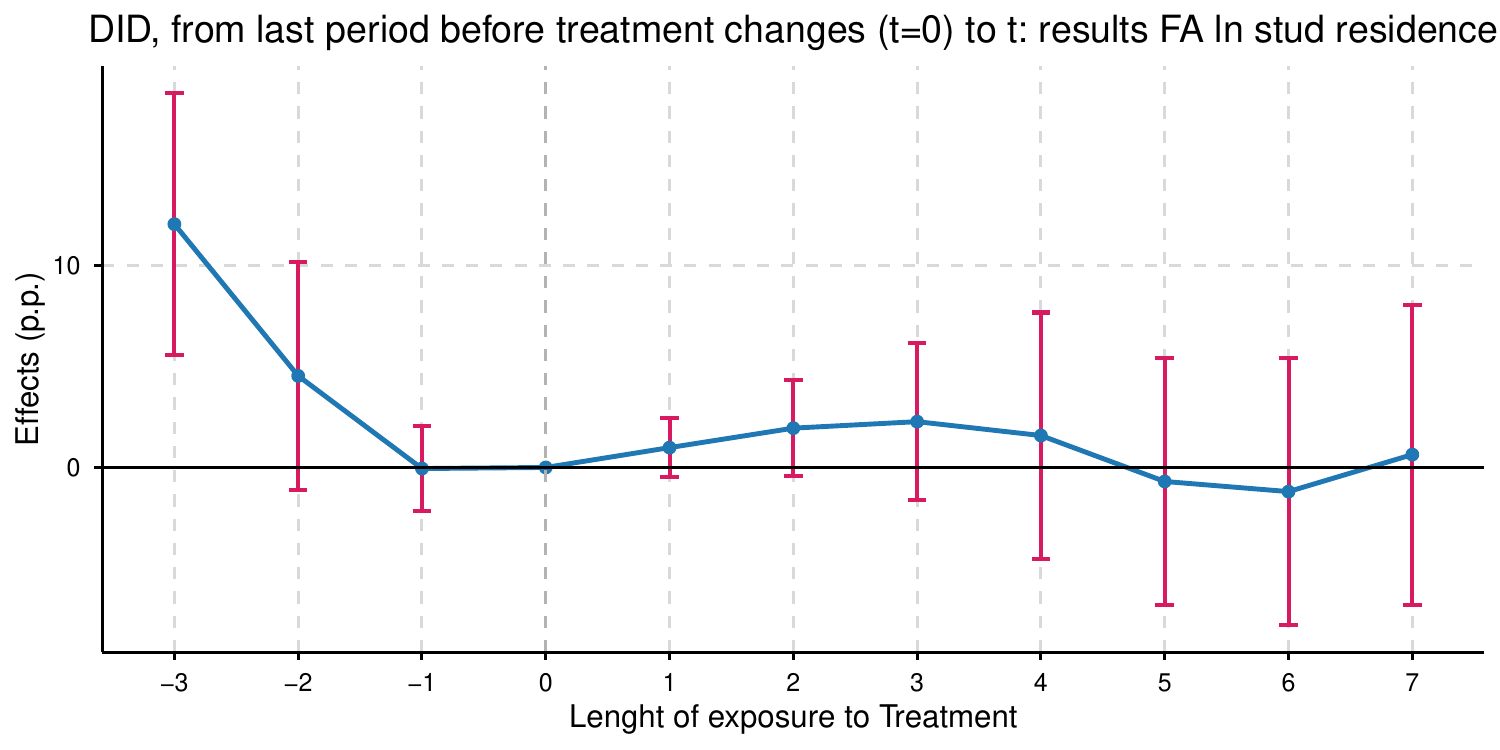}
        \caption{Total students resident in the municipality (logged)}
    \end{subfigure}
\hfill
     \begin{subfigure}[b]{0.45\textwidth}
        \centering
        \includegraphics[width=\textwidth]{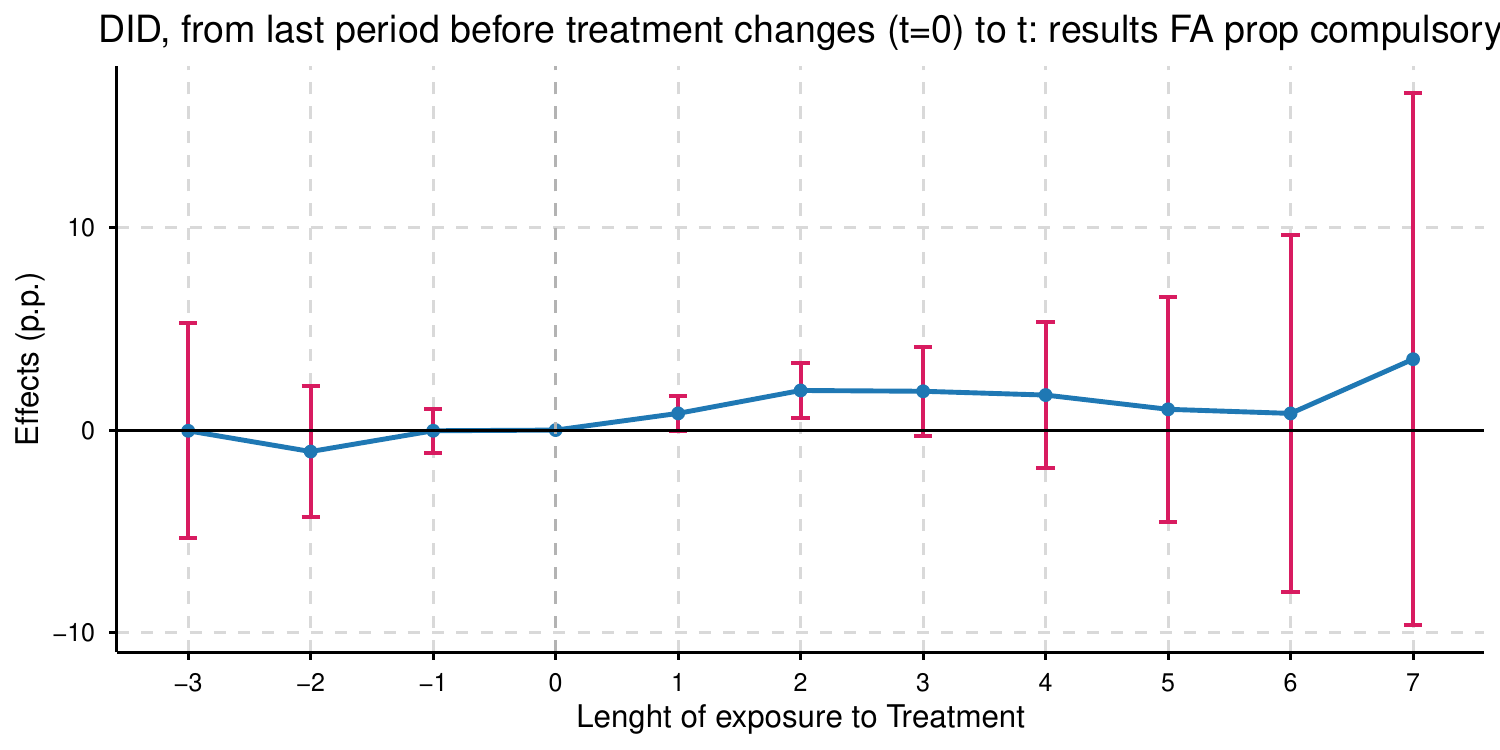}
        \caption{Proportion of students in compulsory school}
    \end{subfigure}
    
     \vskip\baselineskip
    \begin{subfigure}[b]{0.45\textwidth}
        \centering
        \includegraphics[width=\textwidth]{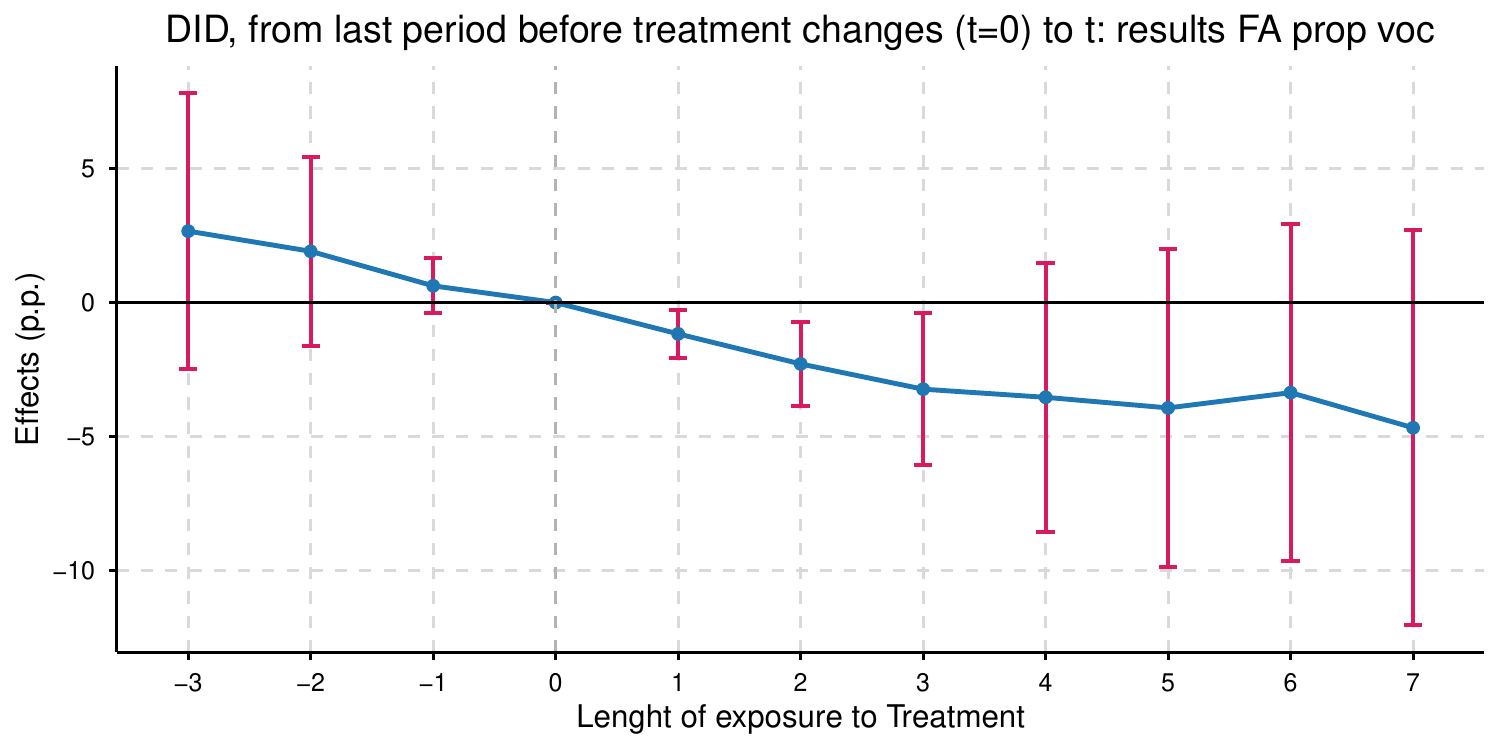}
        \caption{Proportion of students in vocational secondary education}
    \end{subfigure}
      \hfill
 \begin{subfigure}[b]{0.45\textwidth}
        \centering
        \includegraphics[width=\textwidth]{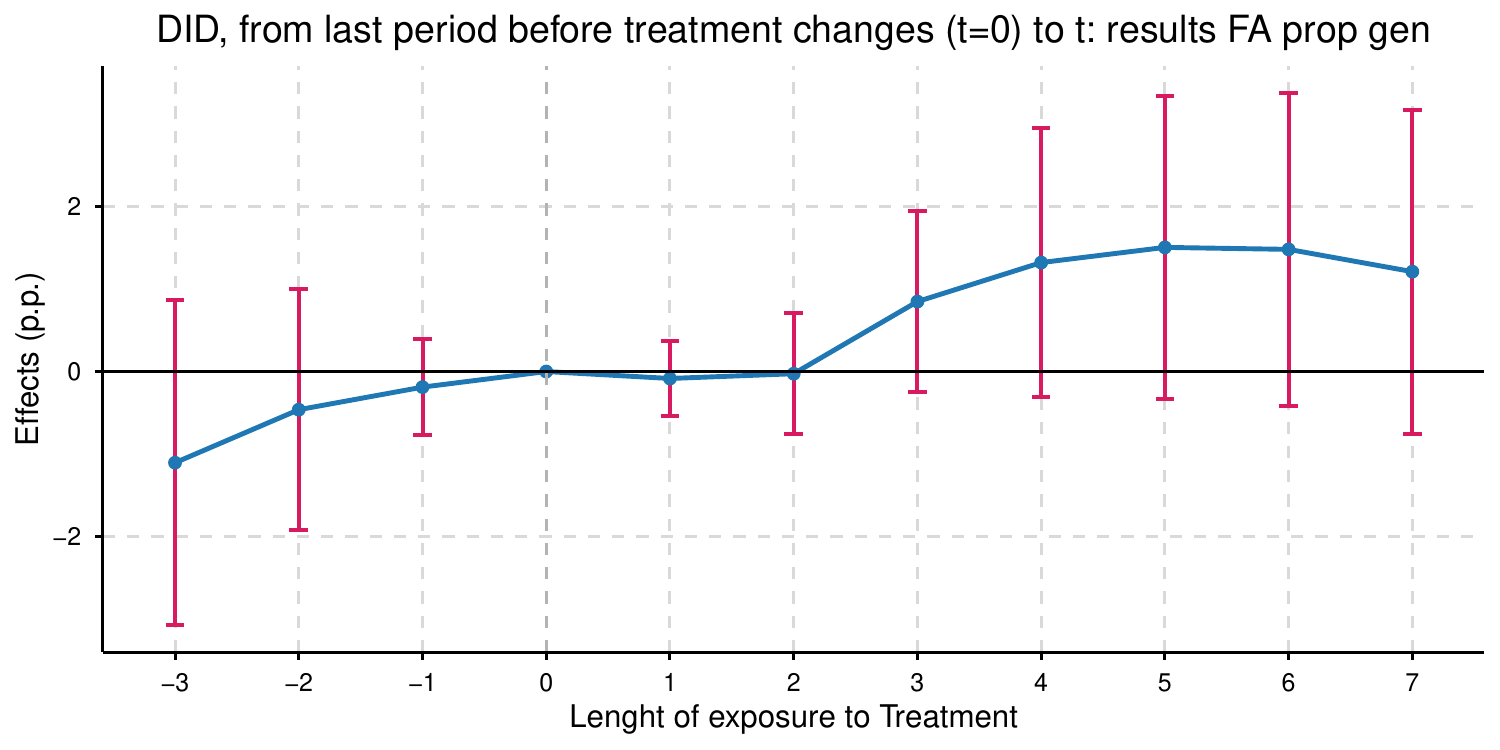}
        \caption{Proportion of students in general secondary education}
    \end{subfigure}
    \vskip\baselineskip
    \begin{subfigure}[b]{0.45\textwidth}
        \centering
        \includegraphics[width=\textwidth]{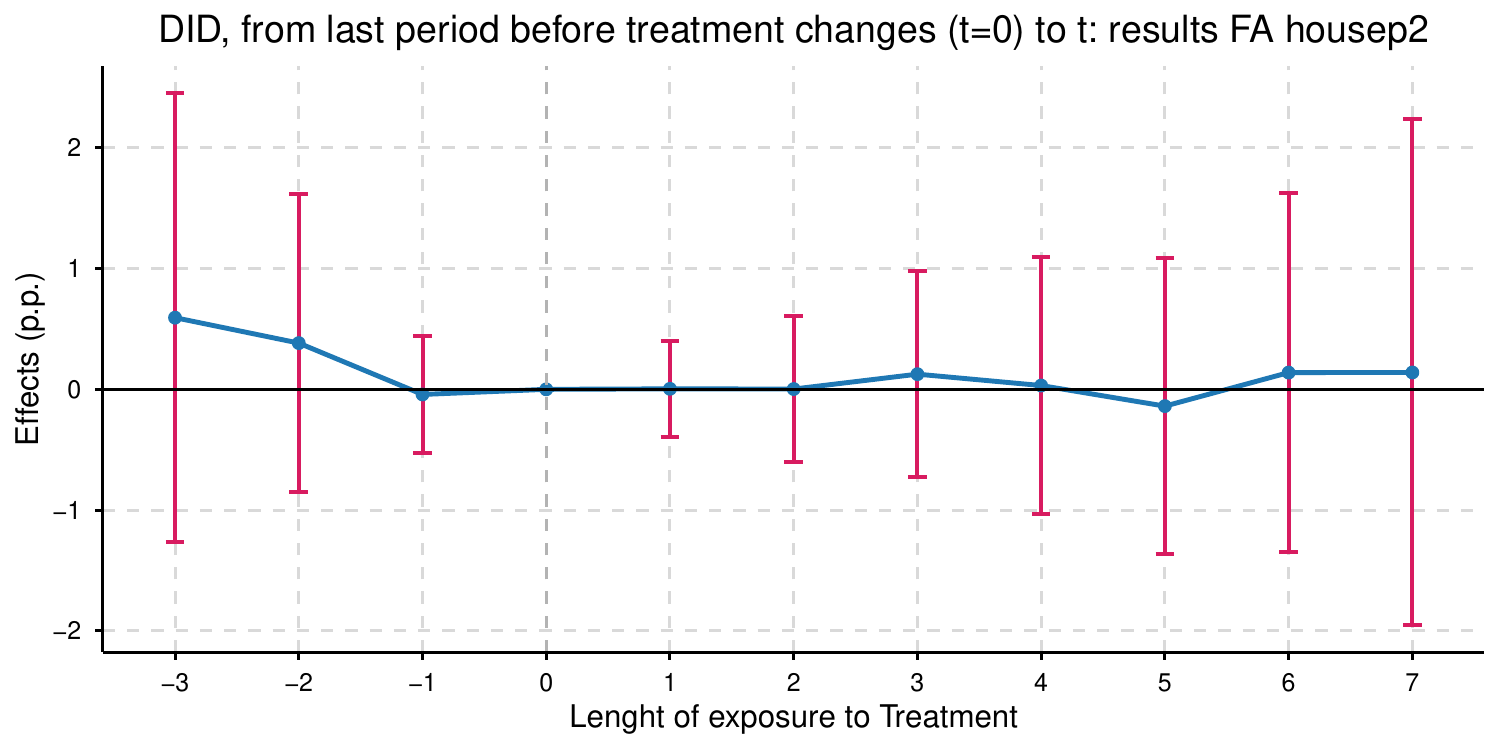}
        \caption{Proportion of 2-people households relative to total households}
    \end{subfigure}
    \hfill
    \begin{subfigure}[b]{0.45\textwidth}
        \centering
         \includegraphics[width=\textwidth]{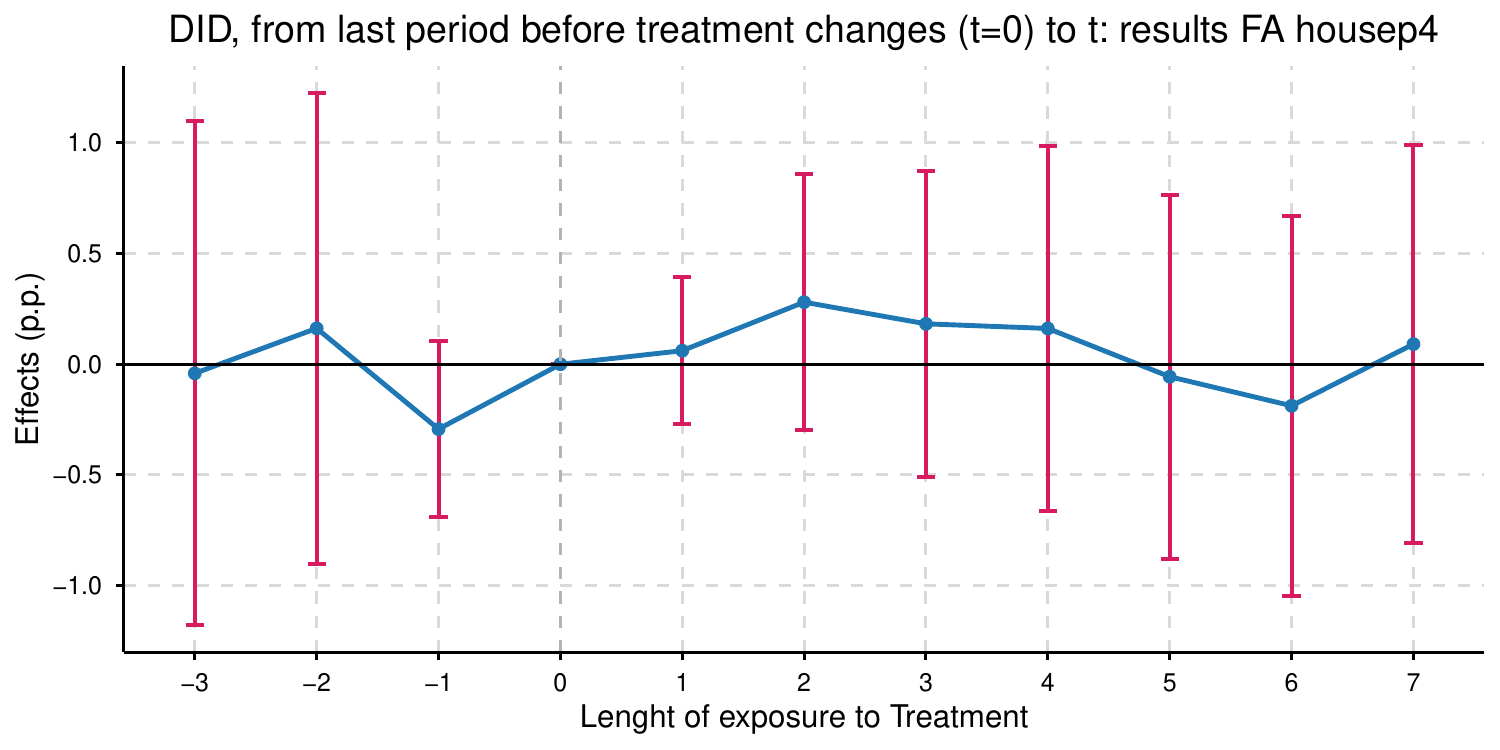}
        \caption{Proportion of above 4-people households relative to total households}
    \end{subfigure}
    \caption{Event studies: effects over time of international migration for non significant results. Lines plot estimated dynamic treatment effects ($DID_\ell$) relative to the year before treatment; red bars = 95\% CI.}
\end{figure}

\begin{figure}[H]
\ContinuedFloat
    \centering
     \begin{subfigure}[b]{0.45\textwidth}
        \centering
        \includegraphics[width=\textwidth]{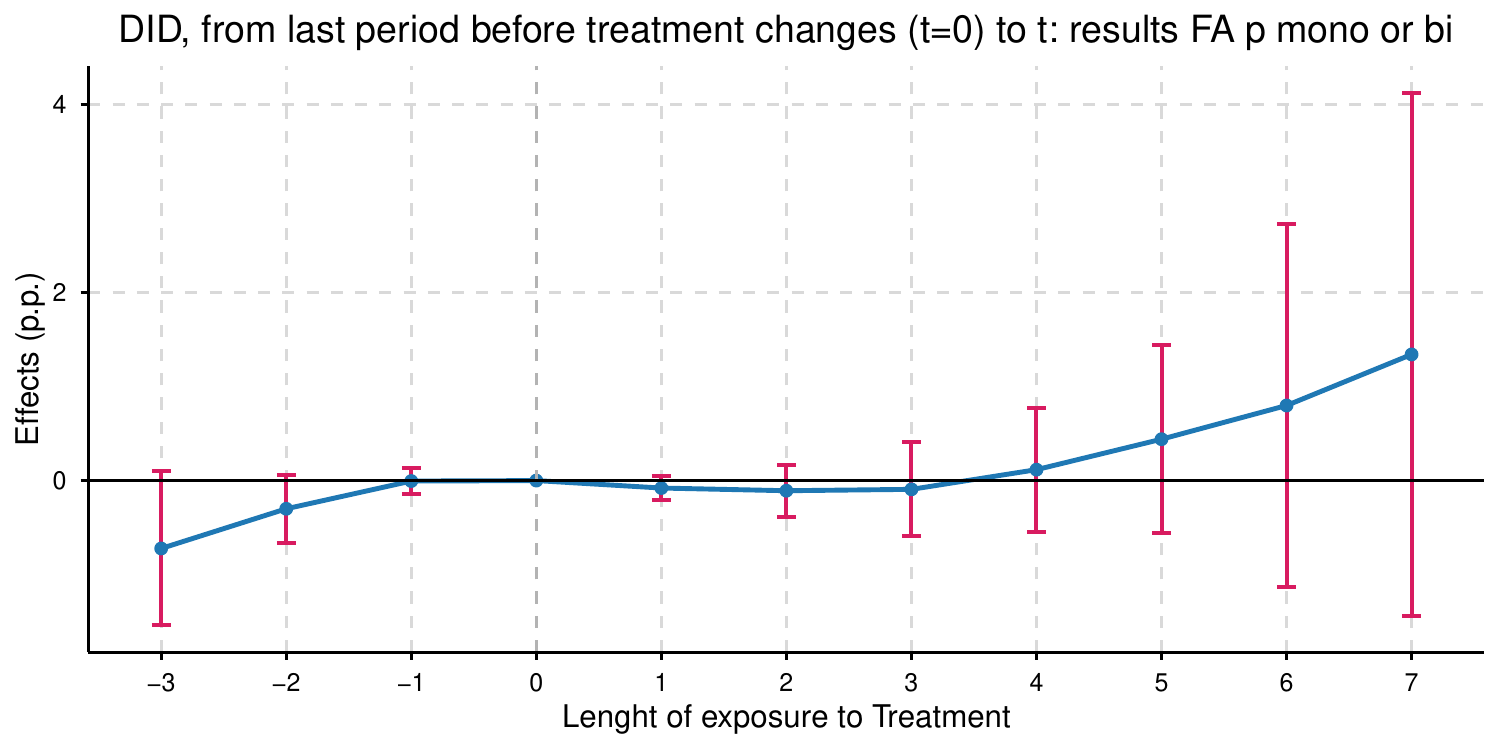}
        \caption{Proportion of 1/2 bedroom apartments over total housing}
    \end{subfigure}
    \hfill
       \begin{subfigure}[b]{0.45\textwidth}
        \centering
         \includegraphics[width=\textwidth]{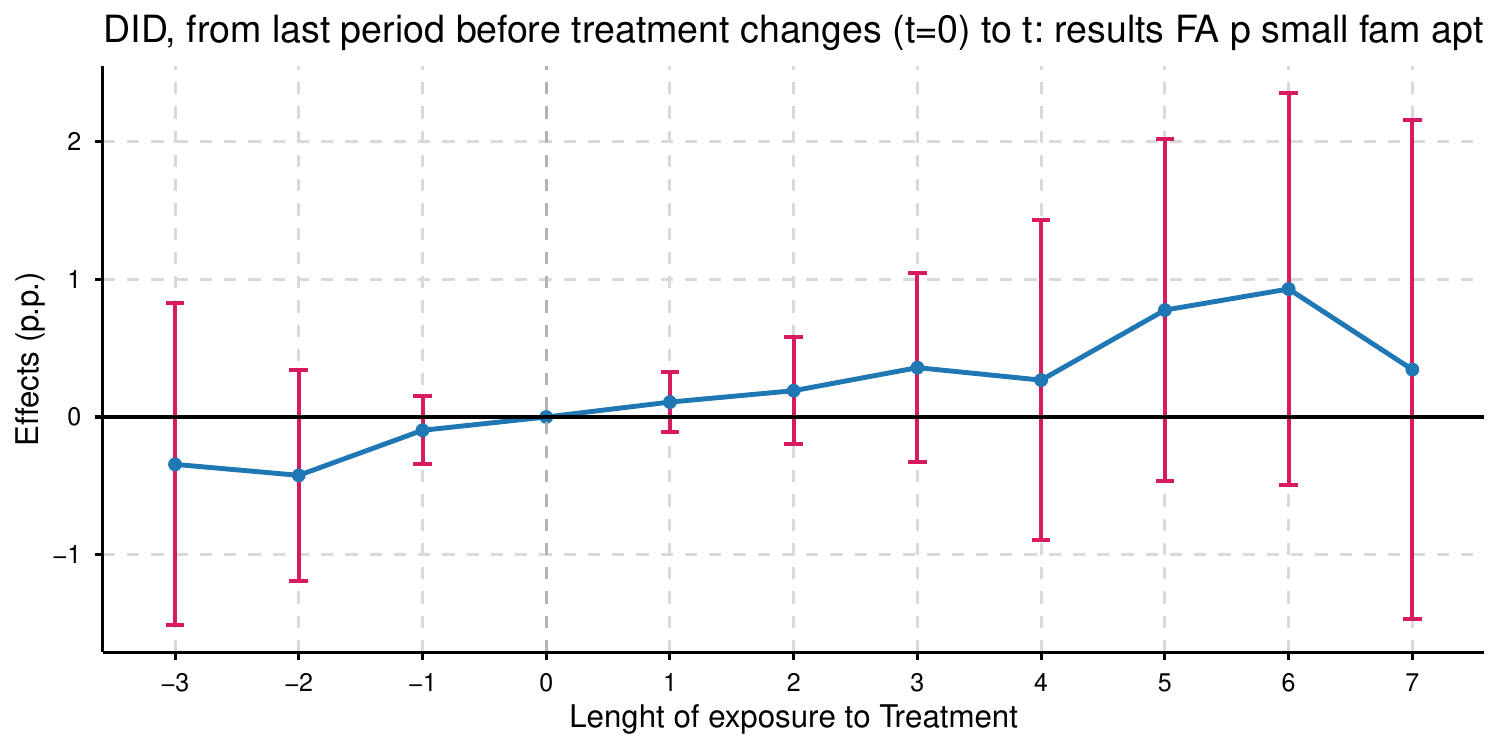}
        \caption{Proportion of small family apartments over total housing}
    \end{subfigure}
 \vskip\baselineskip
    \begin{subfigure}[b]{0.45\textwidth}
        \centering
        \includegraphics[width=\textwidth]{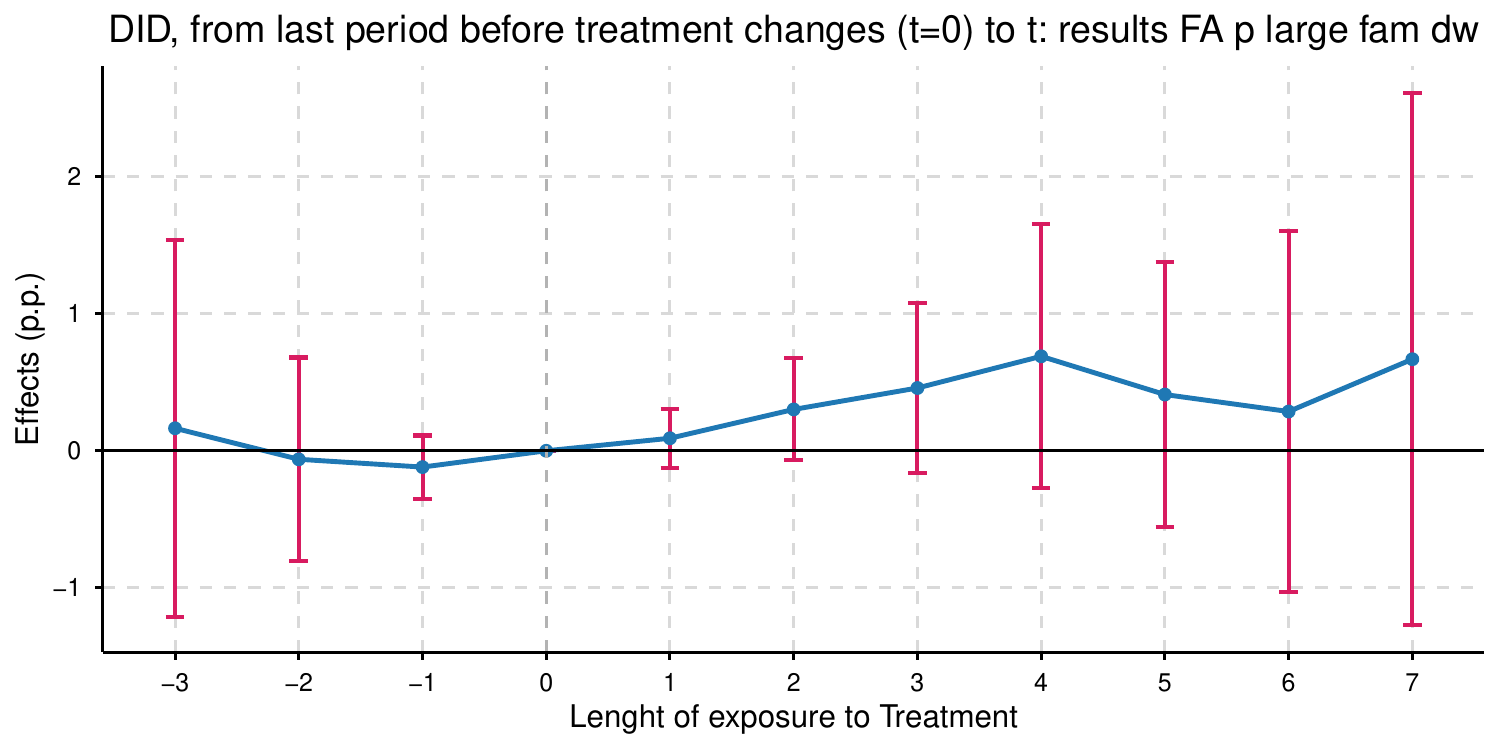}
        \caption{Proportion of large family apartments over total housing}
    \end{subfigure}
    \hfill
    \begin{subfigure}[b]{0.45\textwidth}
        \centering
         \includegraphics[width=\textwidth]{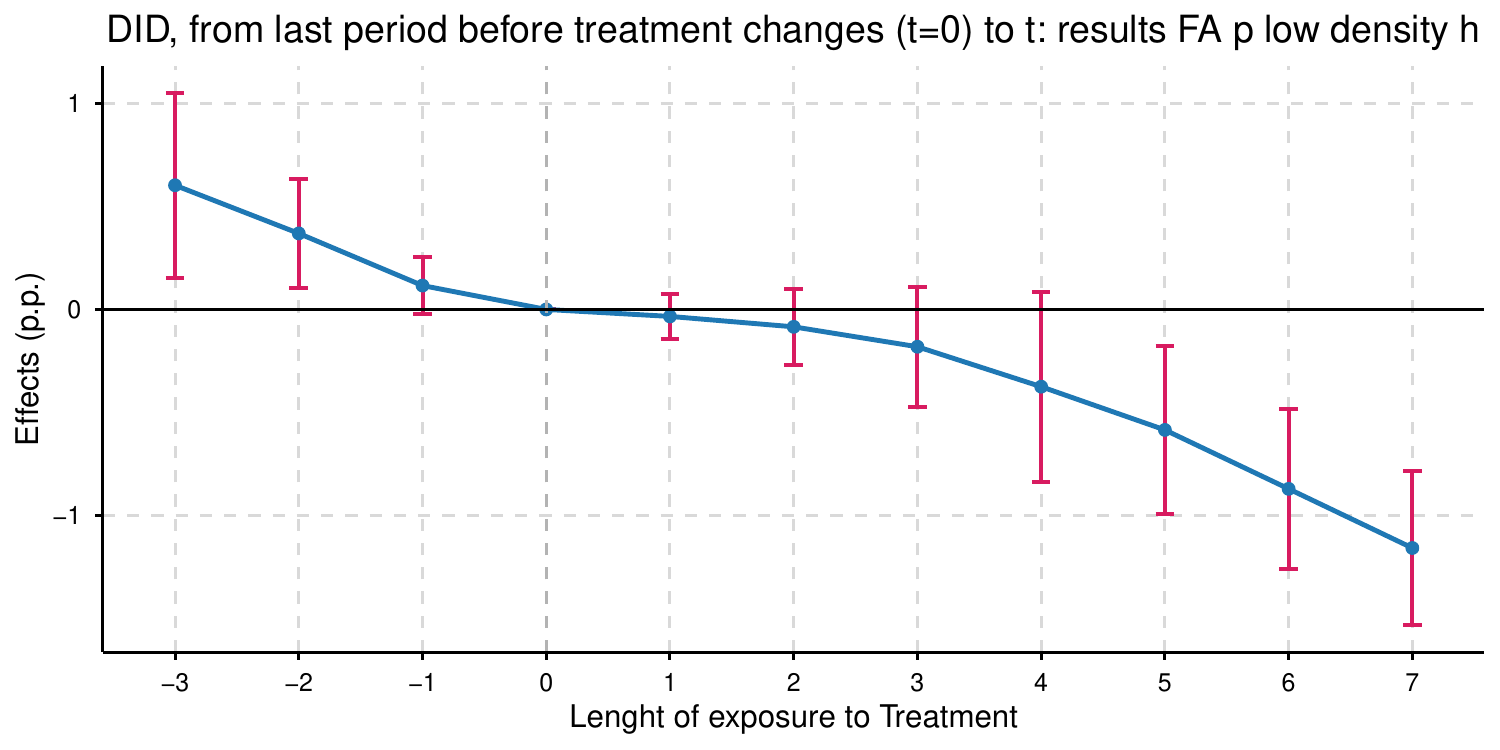}
        \caption{Proportion of single-detached dwellings over total housing}
    \end{subfigure}
    \vskip\baselineskip
       \begin{subfigure}[b]{0.45\textwidth}
        \centering
         \includegraphics[width=\textwidth]{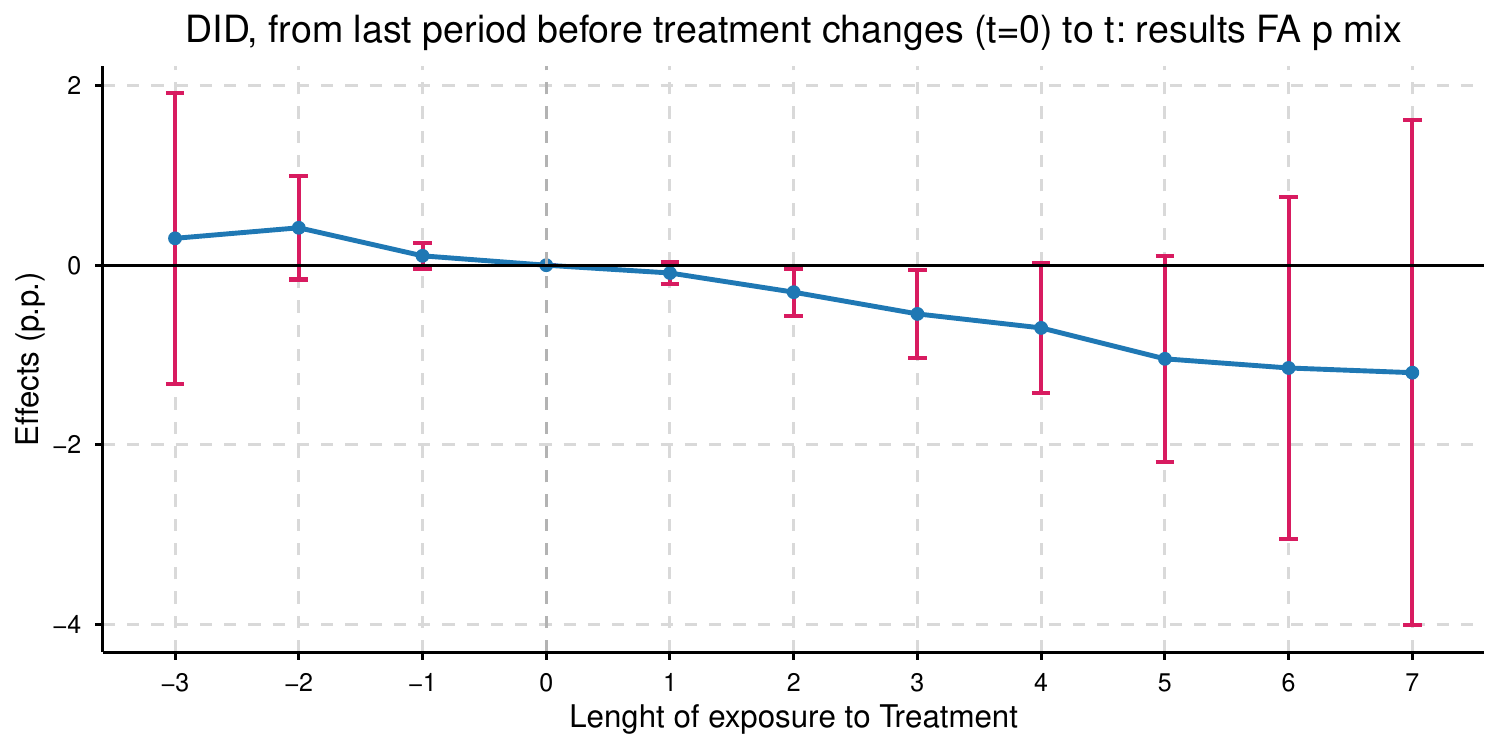}
        \caption{Proportion of mixed-use dwellings over total housing}
    \end{subfigure}
\hfill
    \begin{subfigure}[b]{0.45\textwidth}
        \centering
        \includegraphics[width=\textwidth]{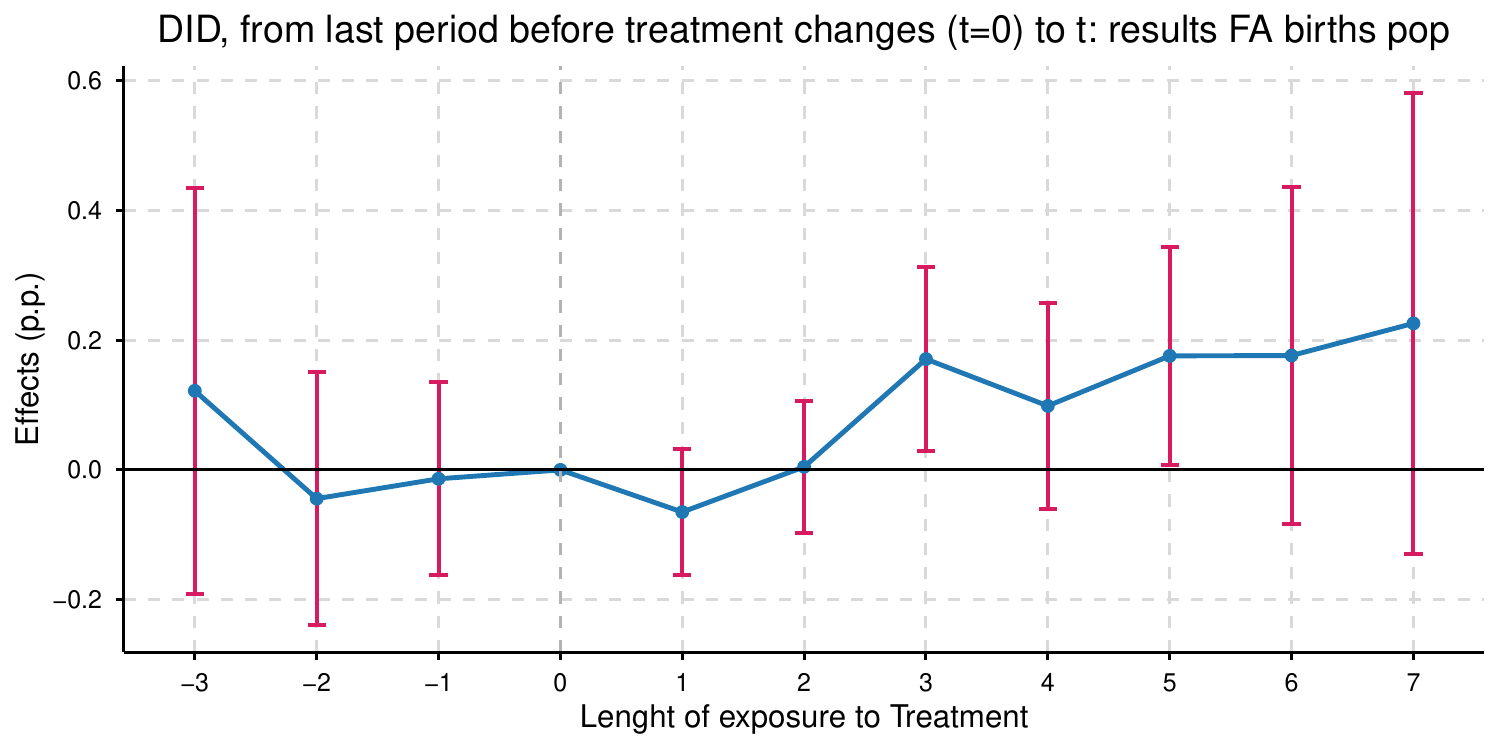}
        \caption{Proportion of births relative to total population}
    \end{subfigure}
    \caption{Event studies: effects over time of international migration for non significant results. Lines plot estimated dynamic treatment effects ($DID_\ell$) relative to the year before treatment; red bars = 95\% CI.}
\end{figure}

\subsubsection{Internal migration}
\begin{figure}[H]
    \centering
    \begin{subfigure}[b]{0.45\textwidth}
        \centering
        \includegraphics[width=\textwidth]{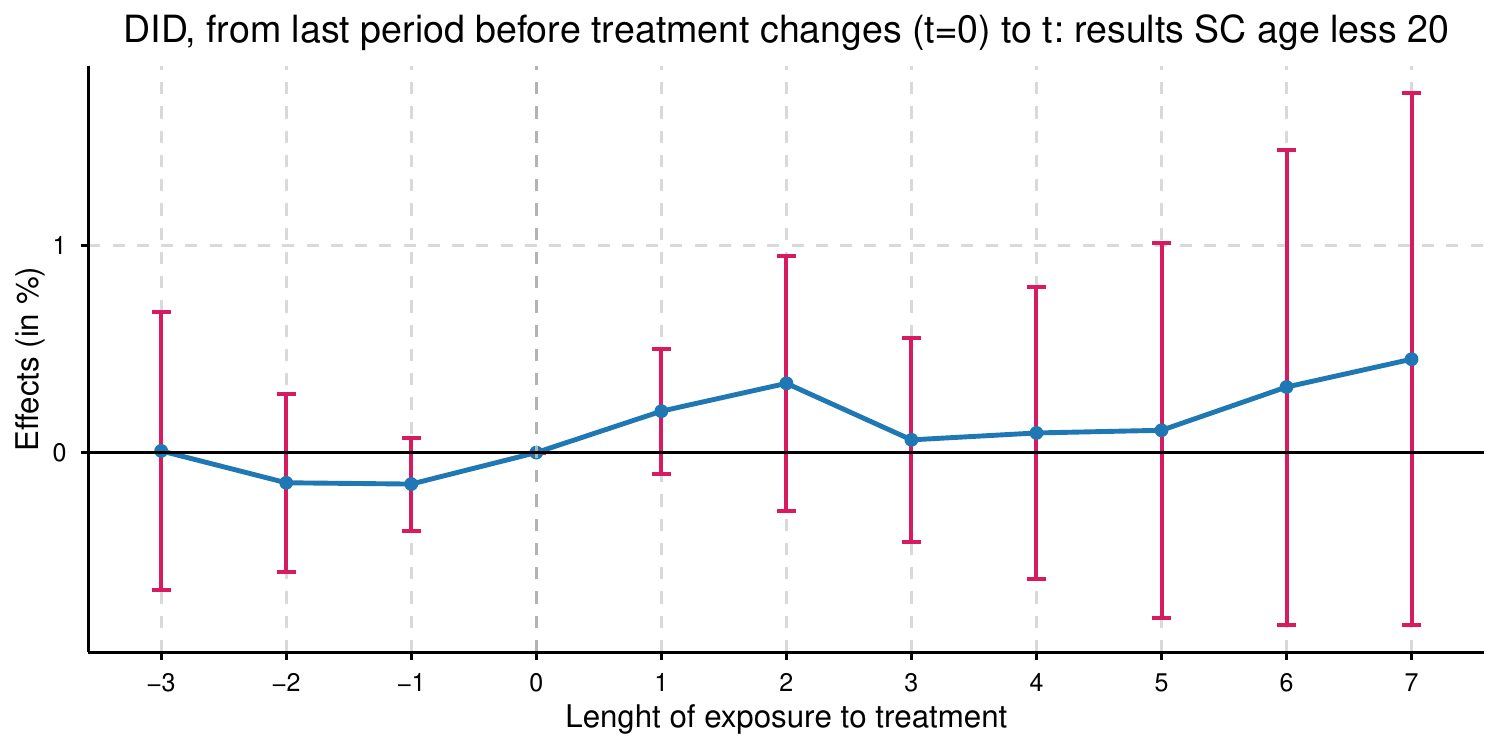}
        \caption{Proportion of population aged $<$ 20 years old}
    \end{subfigure}
    \hfill
    \begin{subfigure}[b]{0.45\textwidth}
        \centering
        \includegraphics[width=\textwidth]{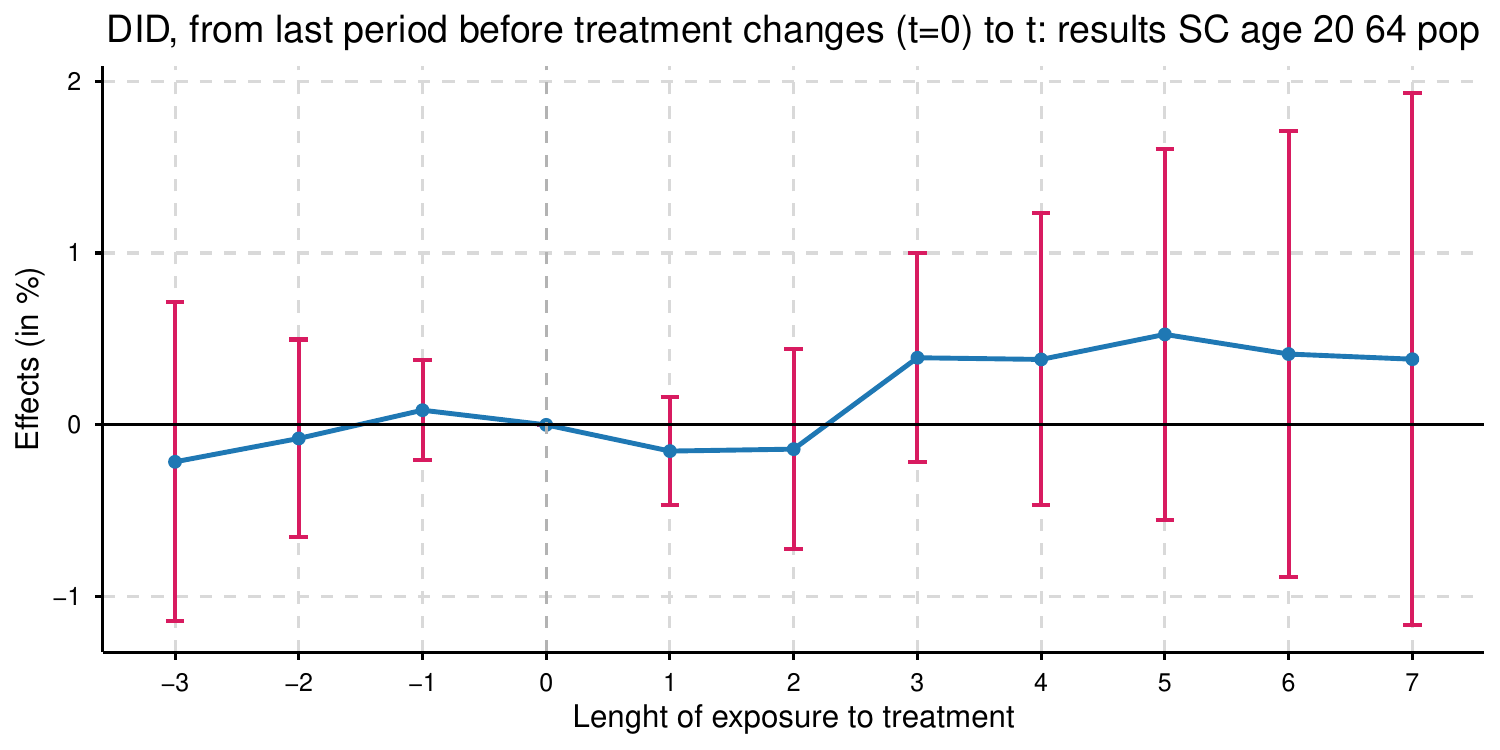}
        \caption{Proportion of population aged between 21 and 64 years old}
    \end{subfigure}
     \vskip\baselineskip
    \begin{subfigure}[b]{0.45\textwidth}
        \centering
        \includegraphics[width=\textwidth]{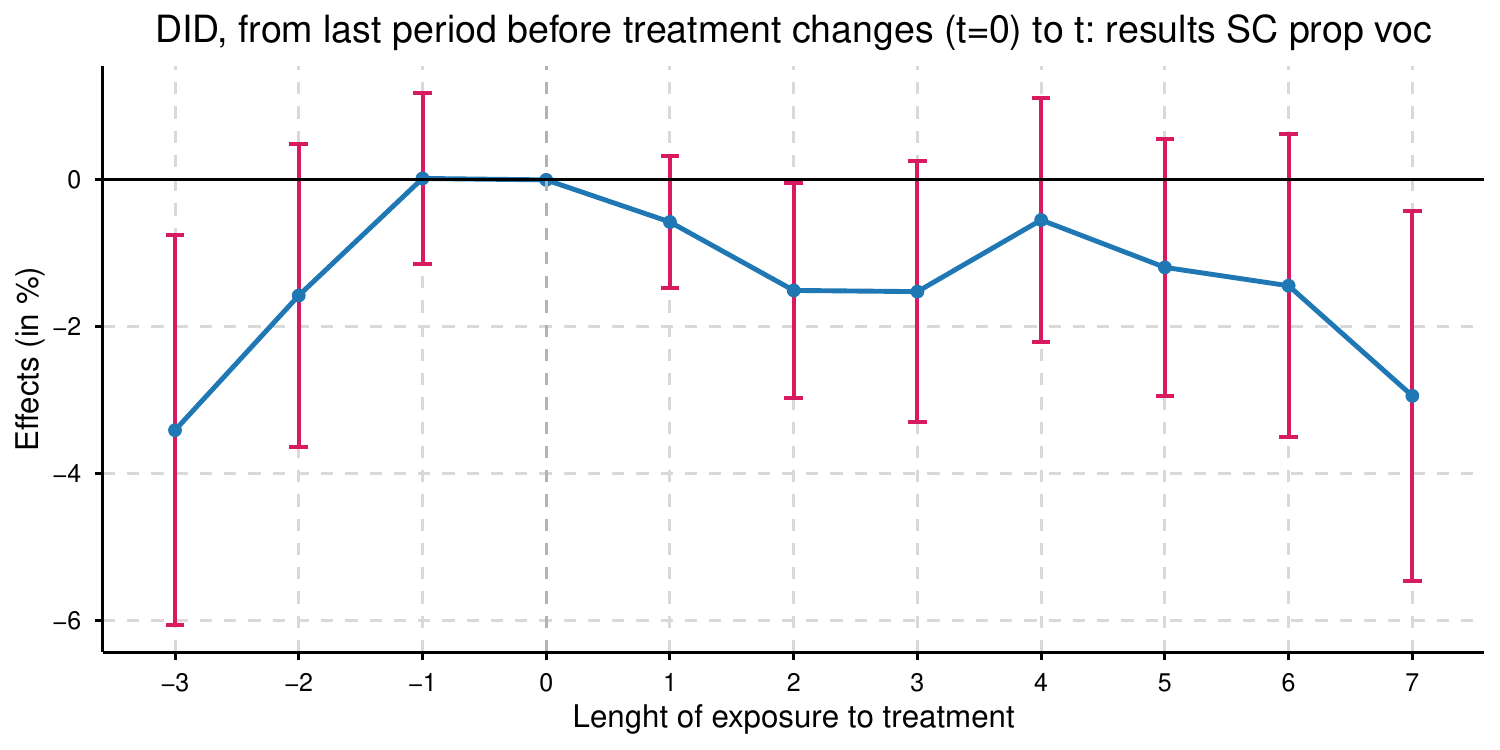}
        \caption{Proportion of students in vocational secondary education }
    \end{subfigure}
      \hfill
 \begin{subfigure}[b]{0.45\textwidth}
        \centering
        \includegraphics[width=\textwidth]{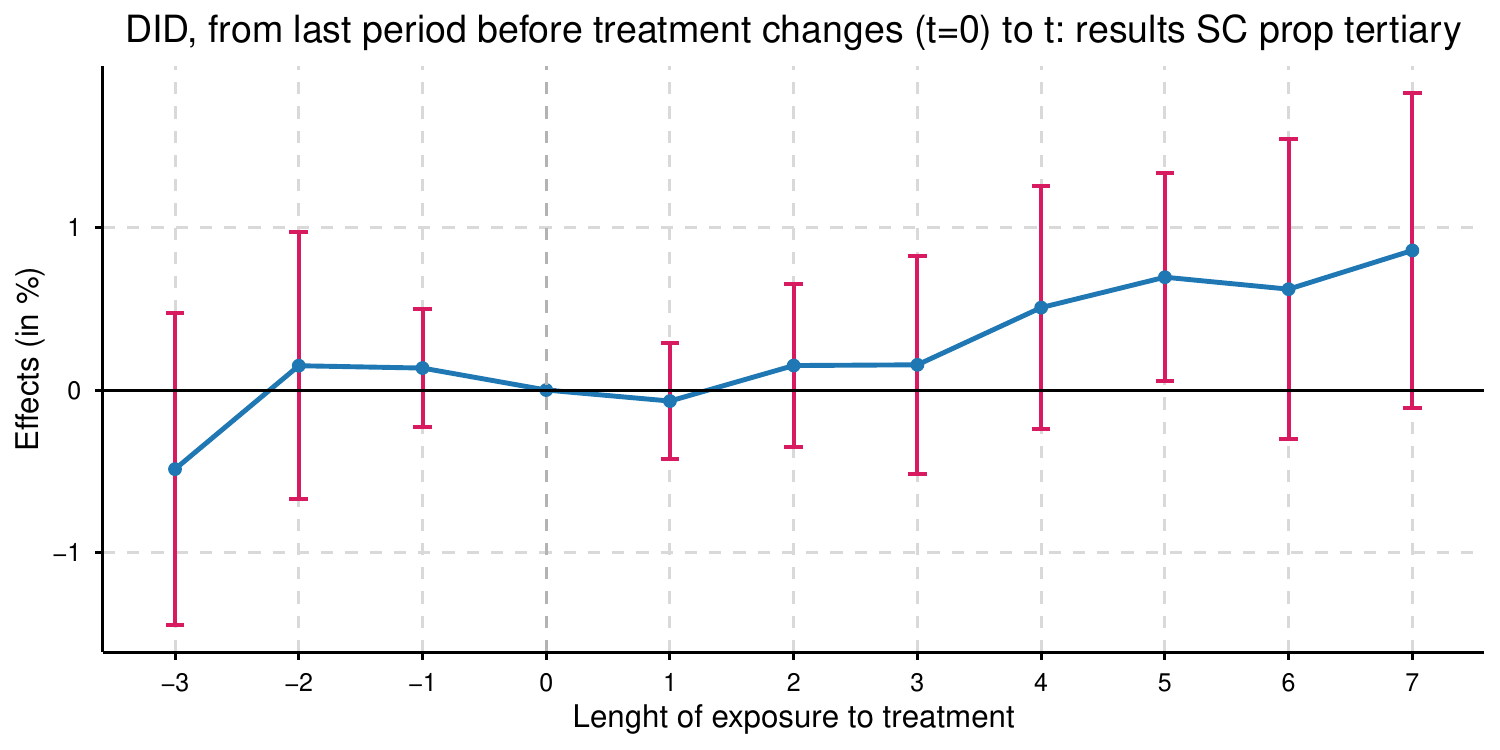}
        \caption{Proportion of students in tertiary education}
    \end{subfigure}
    \vskip\baselineskip
    \begin{subfigure}[b]{0.45\textwidth}
        \centering
        \includegraphics[width=\textwidth]{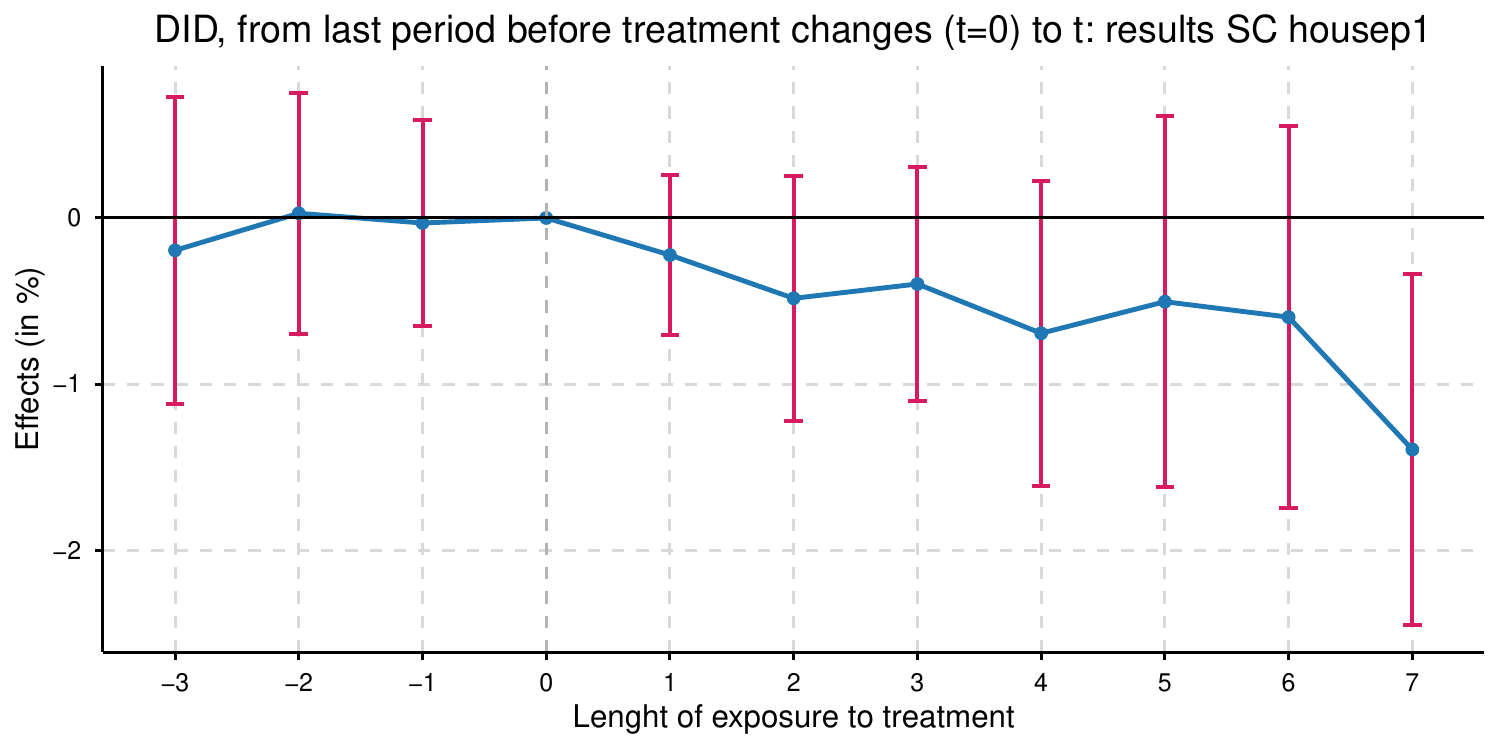}
        \caption{Proportion of 1-people households relative to total households}
    \end{subfigure}
    \hfill
    \begin{subfigure}[b]{0.45\textwidth}
        \centering
         \includegraphics[width=\textwidth]{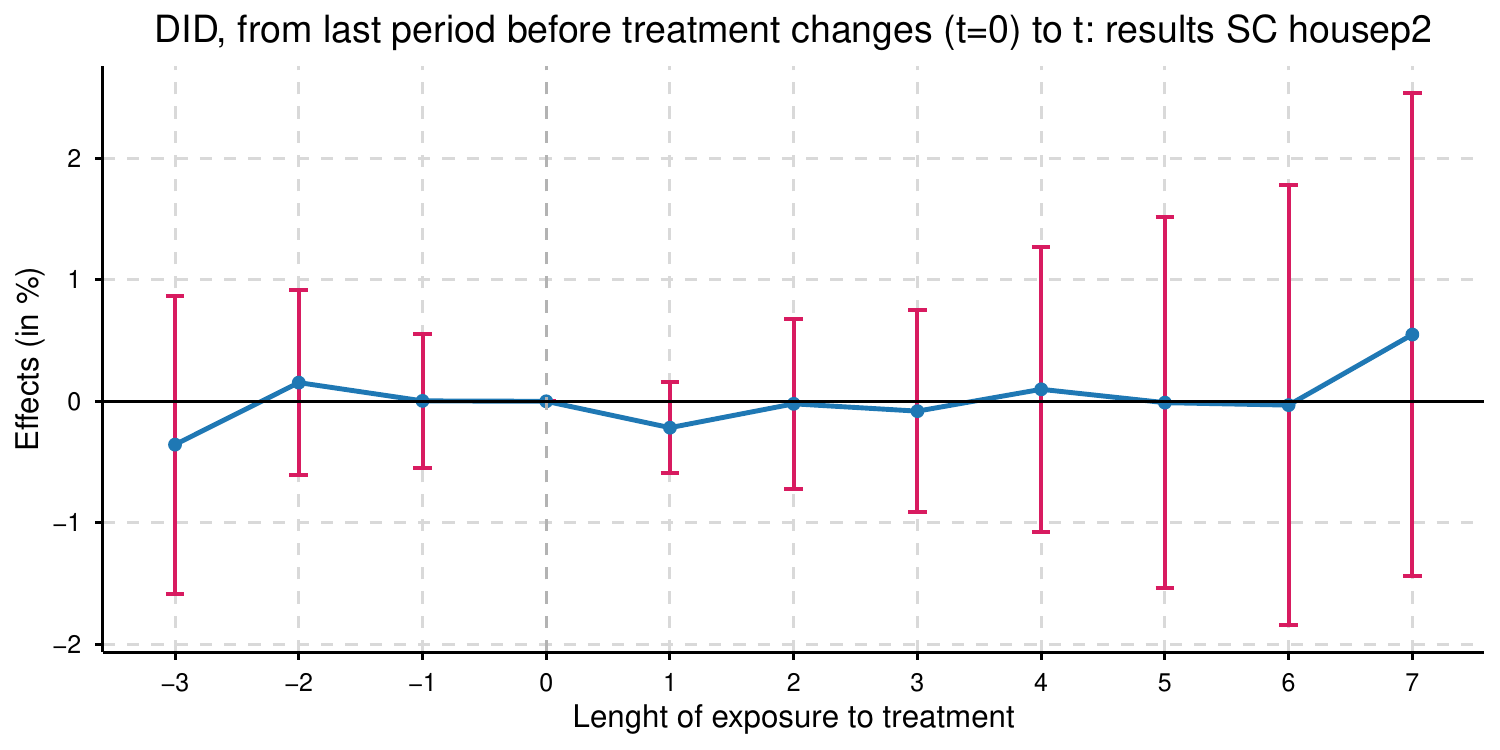}
        \caption{Proportion of 2-people households relative to total households}
    \end{subfigure}
     \vskip\baselineskip
    \begin{subfigure}[b]{0.45\textwidth}
        \centering
          \includegraphics[width=\textwidth]{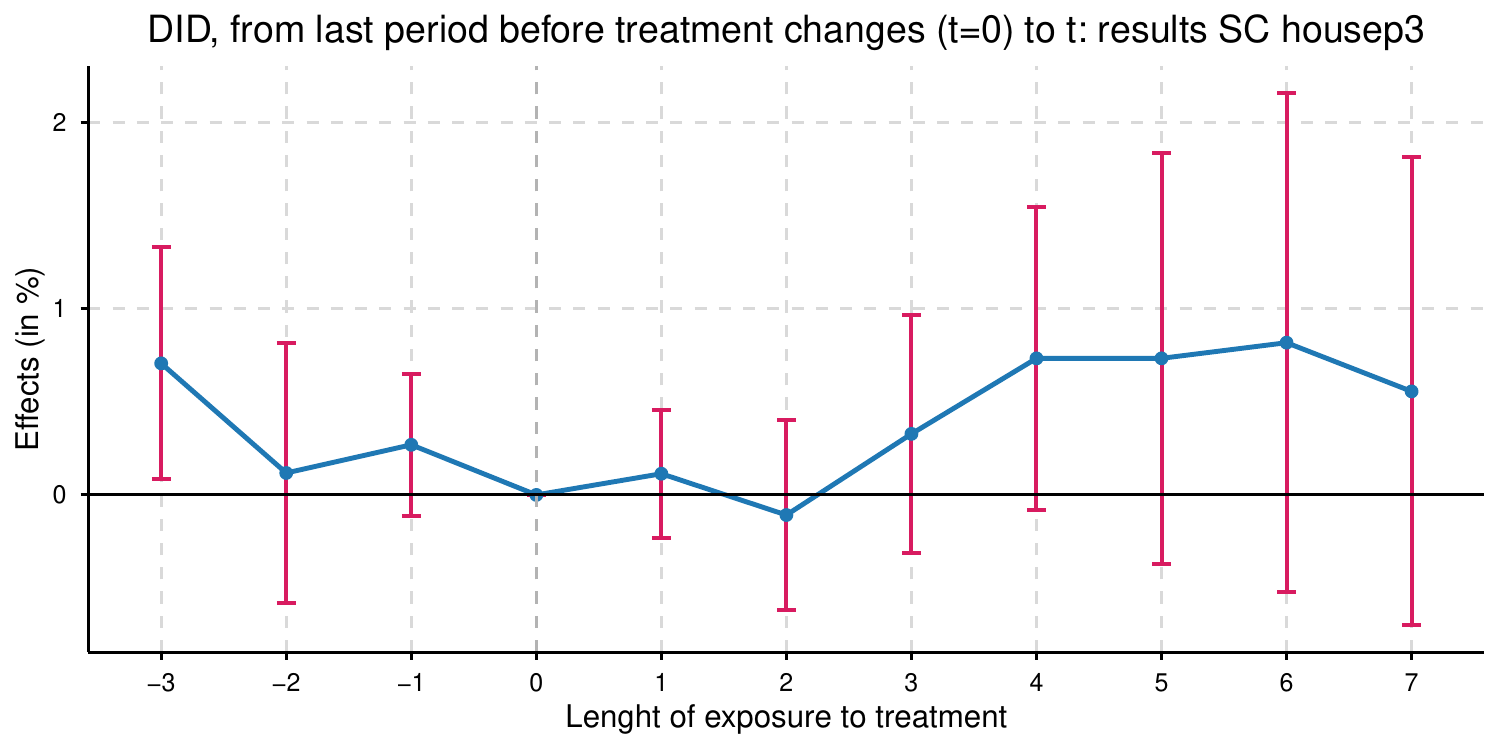}
        \caption{Proportion of 3-people households relative to total households}
    \end{subfigure}
    \hfill
    \begin{subfigure}[b]{0.45\textwidth}
        \centering
         \includegraphics[width=\textwidth]{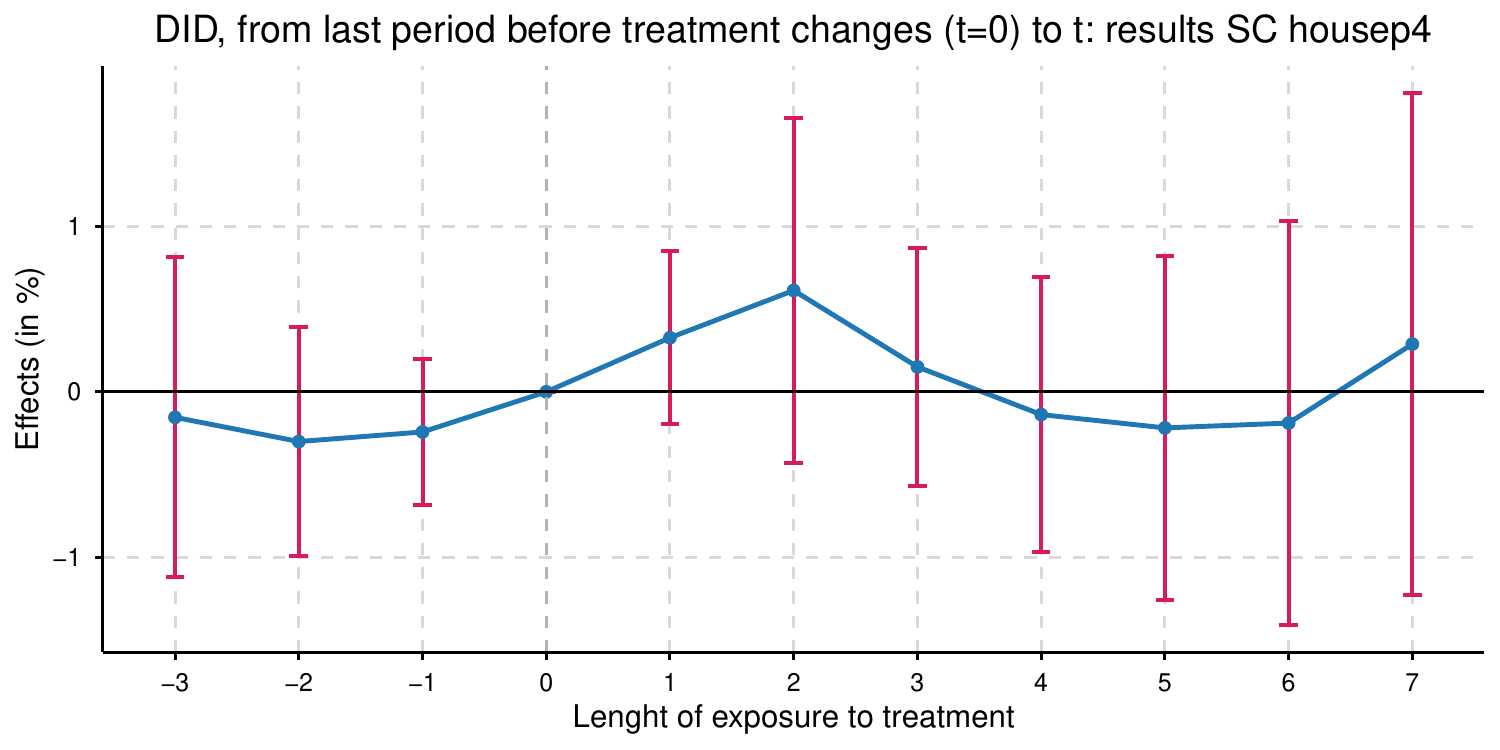}
        \caption{Proportion of above 4-people households relative to total households}
    \end{subfigure}
    \caption{Event studies: effects over time of internal migration for non significant results. Lines plot estimated dynamic treatment effects ($DID_\ell$) relative to the year before treatment; red bars = 95\% CI.}
\end{figure}

\begin{figure}[H]
\ContinuedFloat
    \centering
 \vskip\baselineskip
    \begin{subfigure}[b]{0.45\textwidth}
        \centering
        \includegraphics[width=\textwidth]{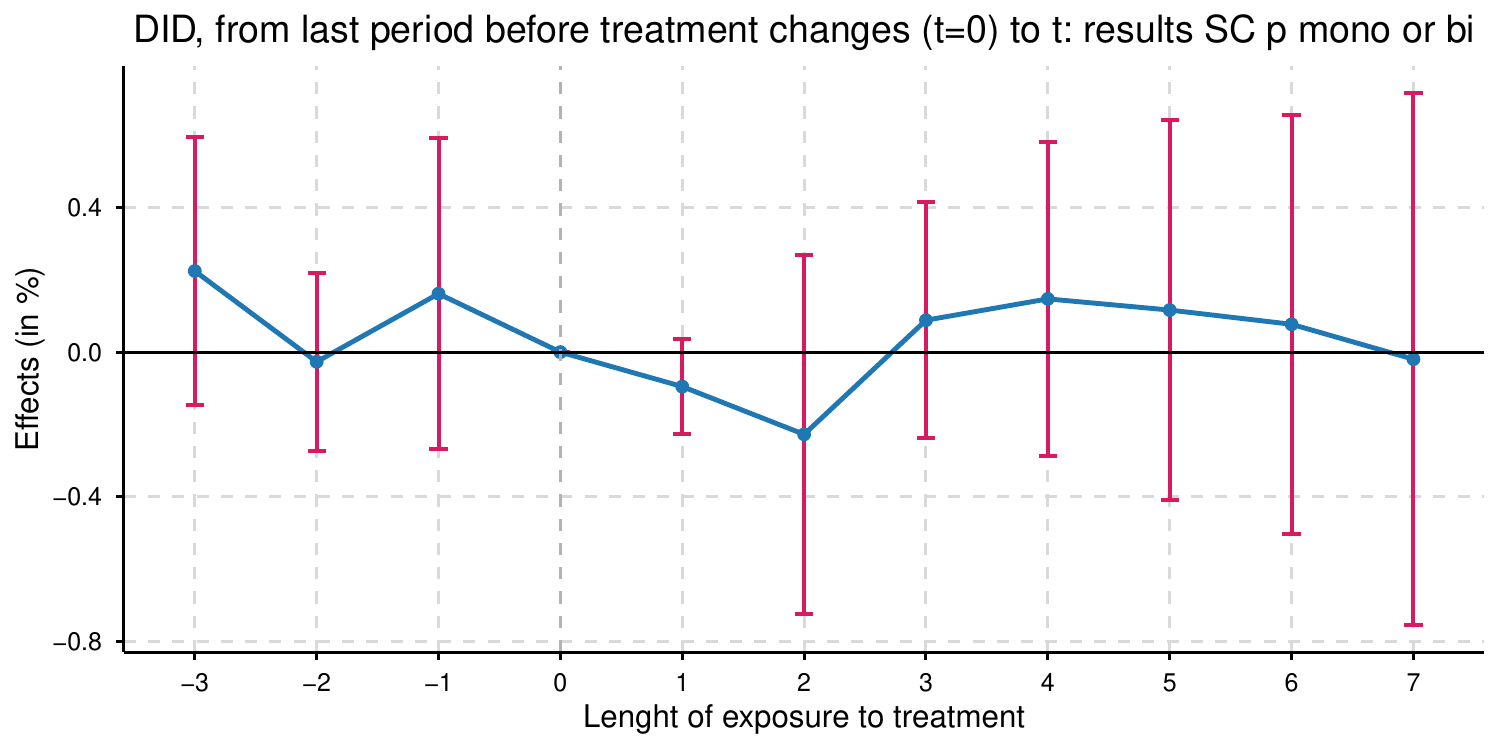}
        \caption{Proportion of 1/2 bedroom apartments over total housing}
    \end{subfigure}
    \hfill
    \begin{subfigure}[b]{0.45\textwidth}
        \centering
         \includegraphics[width=\textwidth]{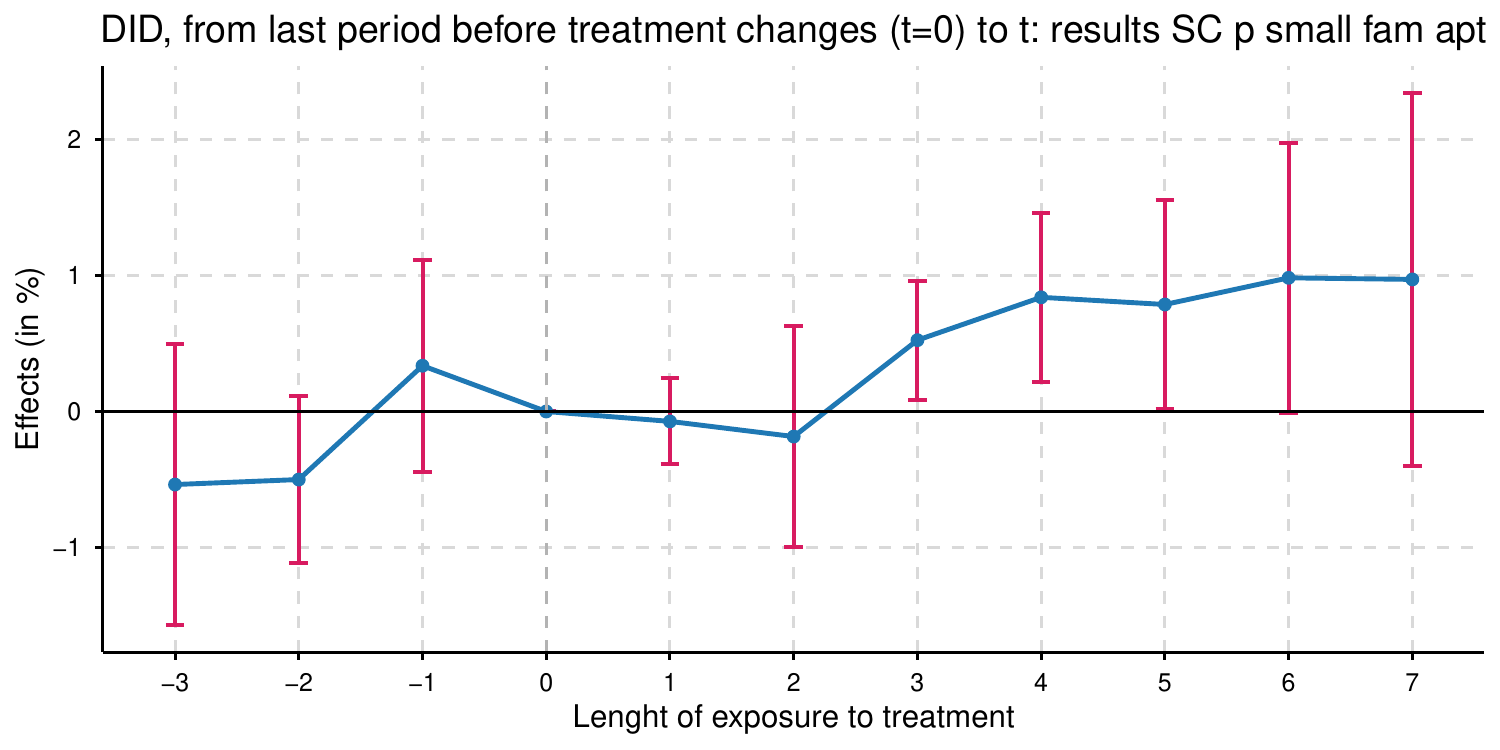}
        \caption{Proportion of small family apartments over total housing}
    \end{subfigure}
 \vskip\baselineskip
    \begin{subfigure}[b]{0.45\textwidth}
        \centering
        \includegraphics[width=\textwidth]{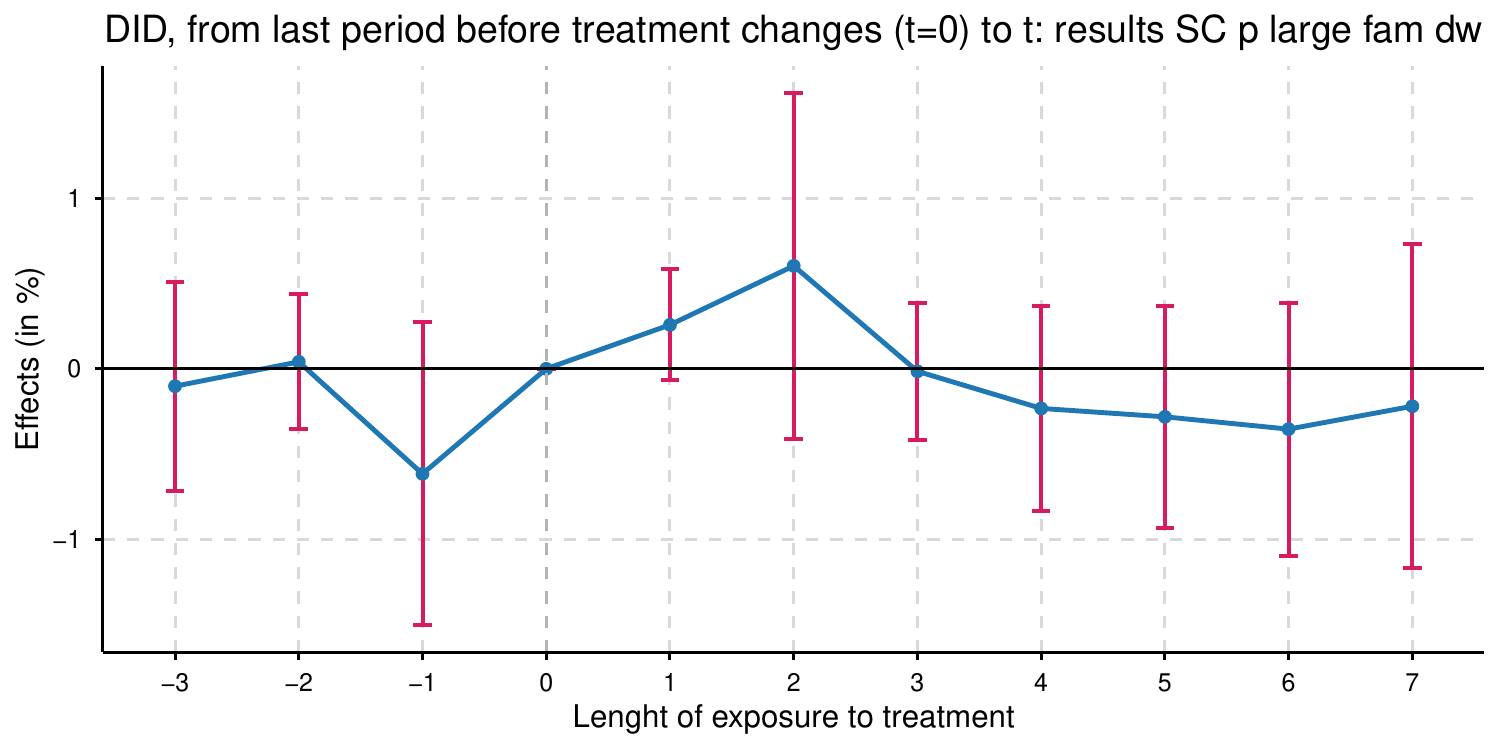}
        \caption{Proportion of large family apartments over total housing}
    \end{subfigure}
    \hfill
    \begin{subfigure}[b]{0.45\textwidth}
        \centering
        \includegraphics[width=\textwidth]{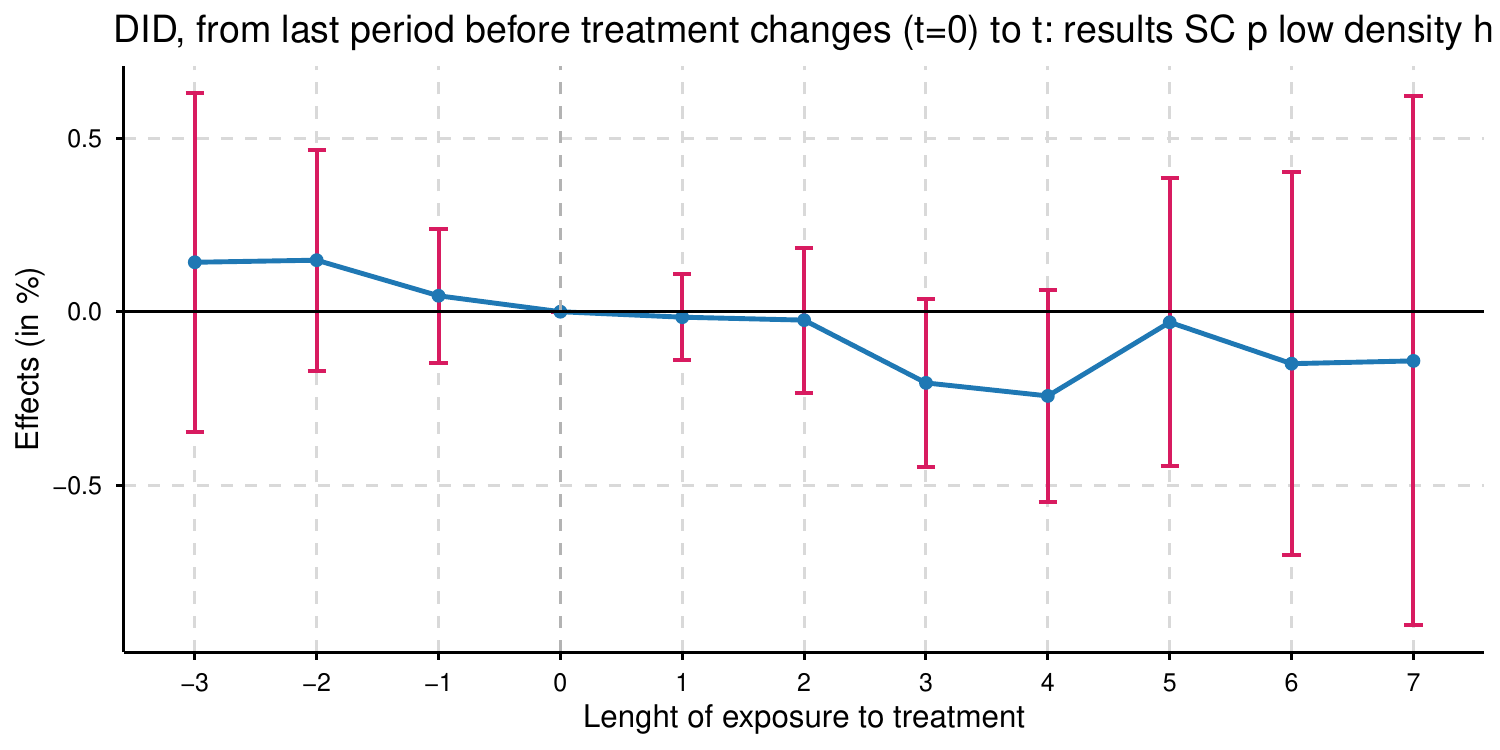}
        \caption{Proportion of single-detached dwellings over total housing}
    \end{subfigure}
    \vskip\baselineskip
    \begin{subfigure}[b]{0.45\textwidth}
        \centering
        \includegraphics[width=\textwidth]{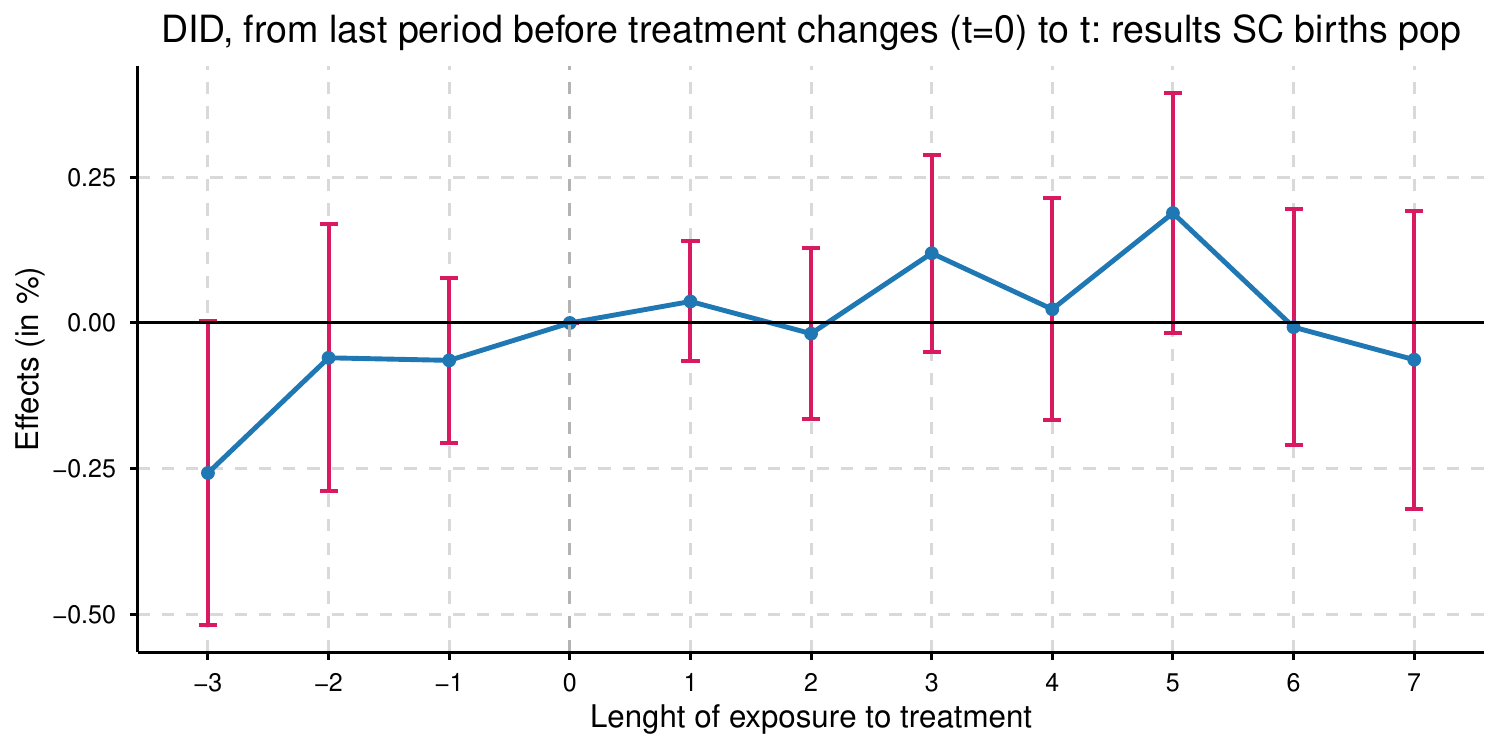}
        \caption{Proportion of births relative to total population}
    \end{subfigure}
    \hfill
    \begin{subfigure}[b]{0.45\textwidth}
        \centering
        \includegraphics[width=\textwidth]{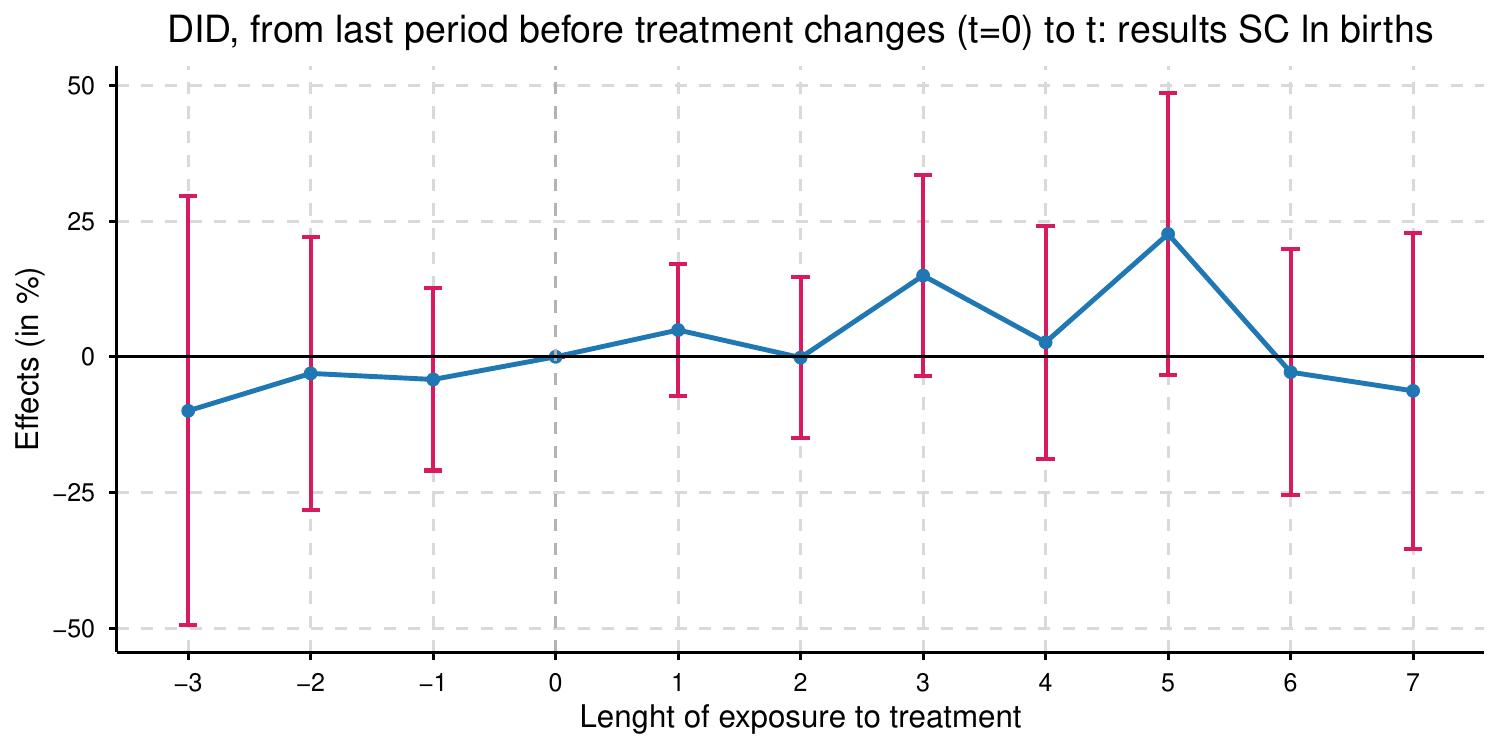}
        \caption{Total number of births in the municipality (logged)}
    \end{subfigure}
    \caption{Event studies: effects over time of internal migration for non significant results. Lines plot estimated dynamic treatment effects ($DID_\ell$) relative to the year before treatment; red bars = 95\% CI.}
\end{figure}
\subsection{Weights of normalizes event studies estimates}
We report the tables for each of our treatments that  shows the composition of the normalized event studies estimates to document which lags $k$ contribute the most to each of the estimates $DID^n_\ell$
\begin{table}[H]
\centering
\caption{Weights on treatment lags for international migration.}
\resizebox{\textwidth}{!}{%
\begin{tabular}{|l|r|r|r|r|r|r|r|}
\hline
$k^{th}$ lag (for $0 \leq k \leq \ell-1 $)  & $\ell=1$  & $\ell=2$ &$\ell=3$ &$\ell=4$&$\ell=5$&$\ell=6$&$\ell=7$ \\ \hline
 $k =0 $   &  1.0926   &  0.6084 &  0.4609 & 0.3492 & 0.2739 & 0.2250 & 0.1839 \\ 
 $k= 1$   &        .  &   0.4411   &  0.3304  & 0.3030  &  0.2556 &   0.2111   &0.1830 \\ 
 $k= 2$   &   .      &    .  &   0.2391 &    0.2179  &   0.2213   &  0.2000  & 0.1738 \\ 
 $k= 3$   &   .     &     .    &    .  &   0.1544 &    0.1590   &  0.1713  &   0.1636 \\ 
 $k= 4$   &  .       &   .     &    .   &       .   &  0.1117   &  0.1250  &   0.1396  \\ 
 $k=5$   &   .        &  .      &    .    &      .     &     .  &   0.0870  &   0.1044 \\ 
  $k=6$   &   .        &  .     &     .     &     .    &      .   &       .  &   0.0739\\  Total  & 1.0926 & 1.0494 & 1.0304  & 1.0245  & 1.0215 &  1.0194 &  1.0222 \\
  \hline
\end{tabular}%
}
\label{tab_app:weights_international}
\end{table}
\begin{table}[H]
\centering
\caption{Weights on treatment lags for internal migration.}
\resizebox{\textwidth}{!}{%
\begin{tabular}{|l|r|r|r|r|r|r|r|}
\hline
$k^{th}$ lag (for $0 \leq k \leq \ell-1 $)  & $\ell=1$  & $\ell=2$ &$\ell=3$ &$\ell=4$&$\ell=5$&$\ell=6$&$\ell=7$ \\ \hline
 $k =0 $   &   1.1455  &   0.6500 & 0.4725 & 0.3654& 0.2870 &    0.2386 &    0.2113 \\ 
 $k= 1$   &        .  & 0.4708 & 0.3590 &  0.3077  &   0.2651   &  0.2229   &  0.1882  \\ 
 $k= 2$   &   .      &    .  &     0.3004   &0.2332 &  0.2267 &  0.2043& 0.1767 \\ 
 $k= 3$   &   .     &     .    &    .  &  0.1923 & 0.1718  &  0.1743 &  0.1605  \\ 
 $k= 4$   &  .       &   .     &    .   &       .   &   0.1389&  0.1300 & 0.1386   \\ 
 $k=5$   &   .        &  .      &    .    &      .     &     .  &     0.1043 & 0.1028 \\ 
  $k=6$   &   .        &  .     &     .     &     .    &      .   &       .  &   0.0785 \\
Total &  1.1455 &1.1208 &  1.1319 & 1.0986&1.0896  &  1.0743  &  1.0566\\
\hline
\end{tabular}%
}
\label{tab_app:weights_internal}
\end{table}
\subsection{Clustering and Attractiveness}
\label{sec: app_clustering}
Here below in Figure \ref{fig:two_var_cluster} we can observe the cluster distribution of our municipalities over the two dimensions i.e. cumulative migration from 1991 until 2009 and population size in 2009. We can see that the three clusters are identified with low, medium or  high levels of both variables, with the majority of observations belonging to the first cluster where we observe very small population size and relatively low historical stock of migration. 
\begin{figure}[H]
    \centering
    \includegraphics[width=1\linewidth]{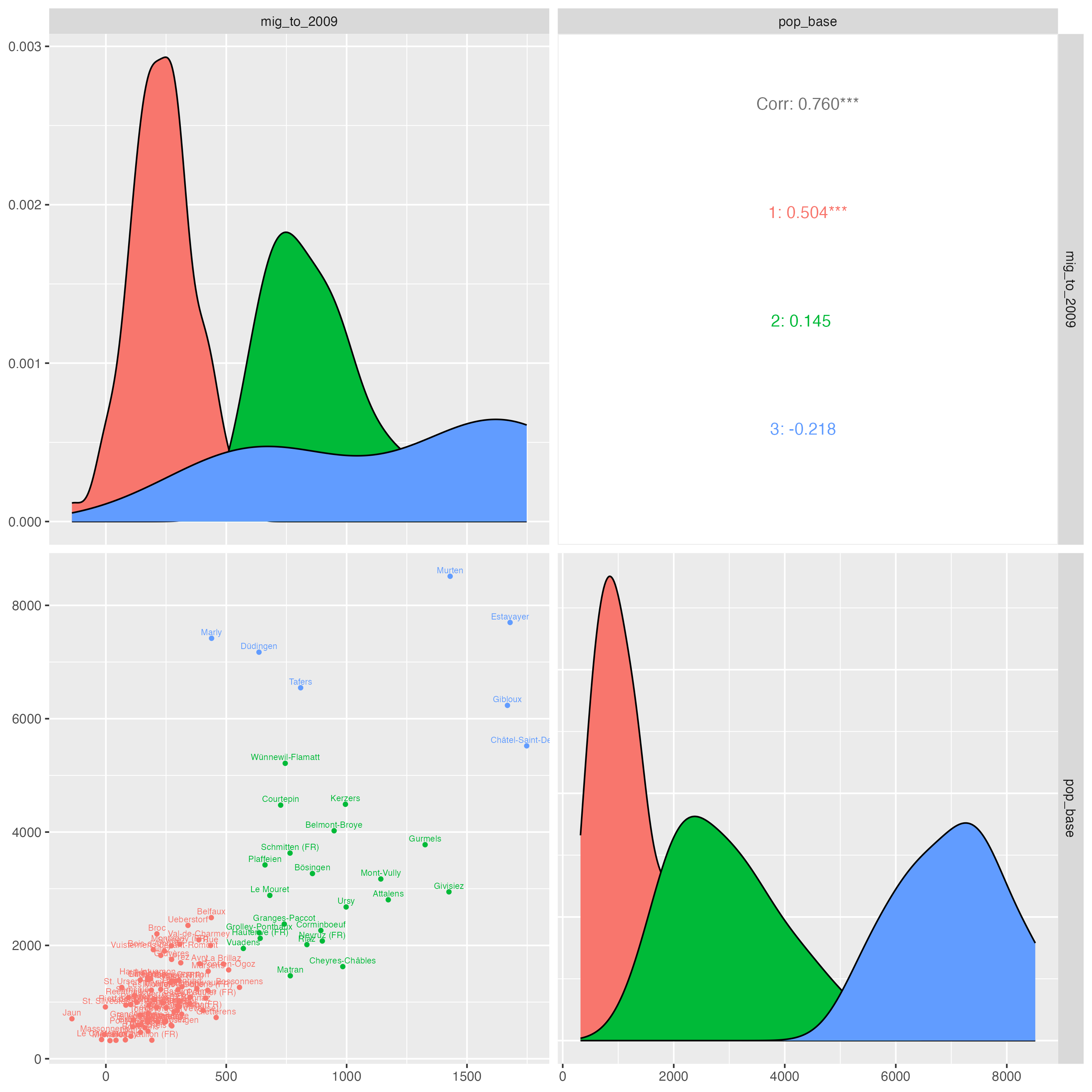}
    \caption{Clusters created on two dimensions: cumulative migration from 1991 until 2009 and population size in 2009}
    \label{fig:two_var_cluster}
\end{figure}

\begin{figure}[H]
    \centering
    \includegraphics[width=1\linewidth]{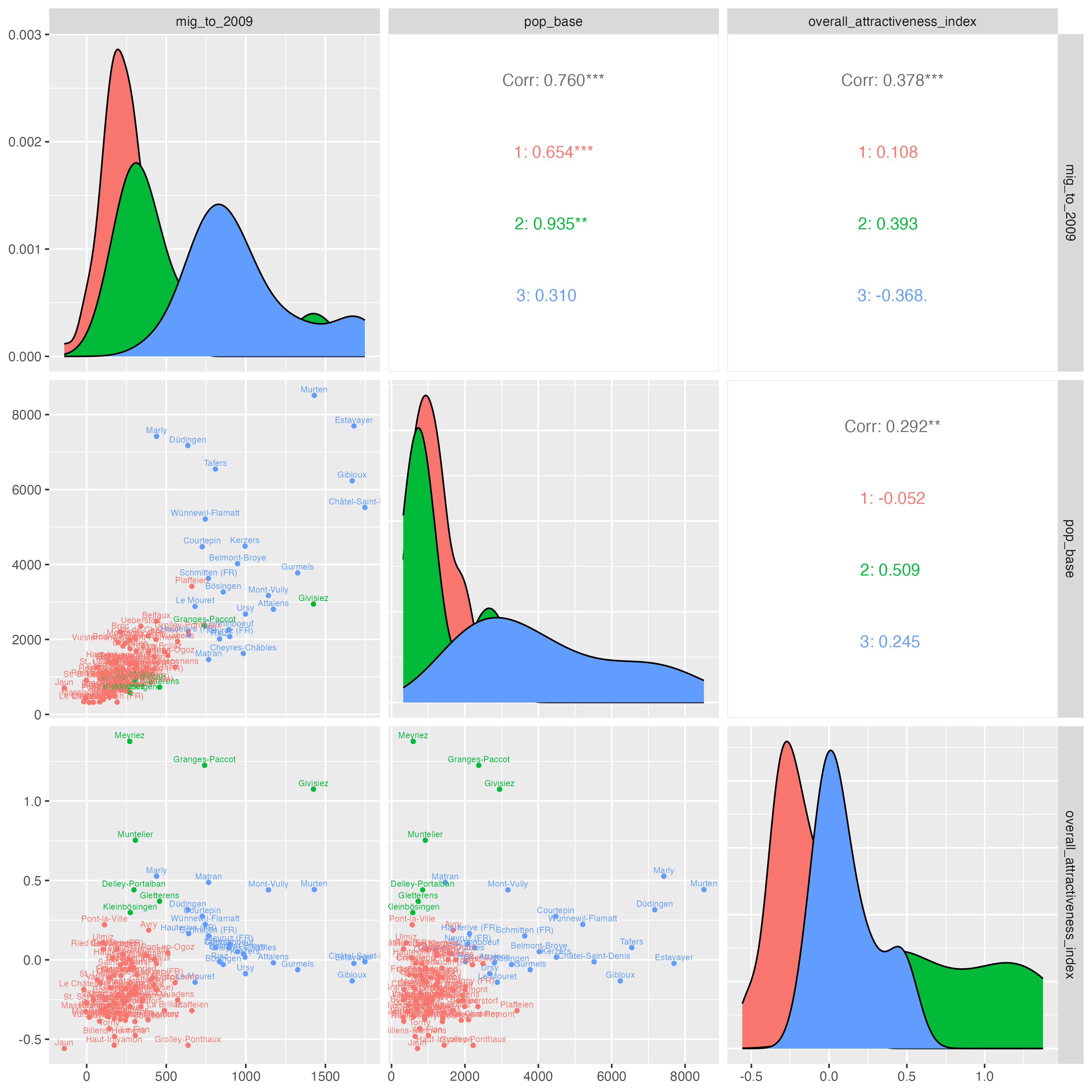}
    \caption{Clusters created on three dimensions: cumulative migration from 1991 until 2009 and population size in 2009 and an attractiveness index}
    \label{fig:three_var_cluster}
\end{figure}

In one version of our analysis we also explored the possibility of including an index on how "attractive" a certain municipality was in 2009, right before our observation period, and construct the clusters using also this dimension. This index was built to summarize five equally weighted dimensions: housing availability, economic opportunity, urban amenities, natural amenities, and fiscal conditions. Each dimension is normalized through z-scores and then averaged. The index therefore measures how appealing a municipality is relative to others when considering residential supply, labor-market conditions, built environment, natural landscape, and tax burden.
As can be observed in Figure \ref{fig:three_var_cluster}, including this dimension in our clustering procedure doesn't drastically change our clusters composition, in particular if we look at the largest cluster (with low values for both the population and the migration variable). If  we use this clustering division, our analysis yields very similar results. 

\end{document}